\documentclass[twoside,twocolumn]{article}

\usepackage{float}

\usepackage{mathptmx}
\usepackage[T1]{fontenc}
\usepackage[utf8]{inputenc}
\usepackage{amssymb,amsmath}

\usepackage{titlesec}
\usepackage{graphicx}  

\usepackage{makecell}

\usepackage{caption}
\usepackage{abstract}
\usepackage[titletoc,toc,title]{appendix}

\usepackage{fancyhdr}  
\usepackage{lastpage}

\usepackage{hyperref}
\hypersetup{unicode=true,
            pdfborder={0 0 0},
            breaklinks=true}
\usepackage{longtable,booktabs}

\makeatletter
\def\@maketitle{%
  \newpage
  \null
  \vskip 2em%
  \begin{flushleft}%
  \let \footnote \thanks
     {\normalsize\textbf{Preprint of article about an open hardware RTU device.} \par}
     {\Large \textbf{\@title} \par}%
    \vskip\baselineskip%
     { \@author \par}%
     {\normalsize Preprint submitted to 'arXiv', \@date \par}
  \end{flushleft}%
  \par
  \vskip 1.5em}
\makeatother

\usepackage{geometry}
\title{Multiple Sensor Interface by the same hardware to USB and serial connection}
\author{David Nuno G. da Silva S. Quelhas  \thanks{ Lisboa, Portugal;  E-mail: david.n.quelhas@gmail.com\\
\hspace*{1.5em}  \url{https://www.linkedin.com/in/dquelhas}\\
\hspace*{1.5em} ORCID: \href{https://orcid.org/0000-0002-0282-0972}{0000-0002-0282-0972}\\
\hspace*{1.5em} \url{https://multiple-sensor-interface.blogspot.com}}}

\titlespacing{\section}{0pc}{1pc}{0pc}
\titlespacing{\subsection}{0pc}{0.6pc}{0pc}
\titlespacing{\subsubsection}{0pc}{0.5pc}{0pc}

\titleformat{\section}
  {\normalfont\fontsize{10}{12}\bfseries}{\thesection}{1em}{}

\titleformat{\subsection}
  {\normalfont\fontsize{10}{12}\itshape}{\thesubsection}{1em}{}

\titleformat{\subsubsection}
  {\normalfont\fontsize{10}{12}\itshape}{\thesubsubsection}{1em}{}

\begin{document}

\twocolumn[
  \begin{@twocolumnfalse}

\maketitle

\setlength{\belowdisplayskip}{0.2pt} \setlength{\belowdisplayshortskip}{0.1pt}
\setlength{\abovedisplayskip}{0.2pt} \setlength{\abovedisplayshortskip}{0.1pt}

\begin{abstract}
The Multiple Sensor Interface is a simple sensor interface that works with USB, RS485 and GPIO. It allows one to make measurements using a variety of sensors based on the change of inductance, resistance, capacitance, and frequency using the same connector and same electronic interface circuit between the sensor and the microcontroller. The same device also provides some additional connectors for small voltage measurement. Any sensors used for the measurement of distinct phenomena can be used if the sensor output is based on inductance, resistance, capacitance or frequency within the measurement range of the device, obtaining a variable precision depending on the used sensor. The device presented is not meant for precise or accurate measurements. It is meant to be a reusable hardware that can be adapted/configured to a varied number of distinct situations, providing, to the user, more freedom in sensor selection as well as more options for device/system maintenance or reuse.
\end{abstract}

\begin{longtable}{@{}l@{}}
\begin{minipage}{\linewidth}\raggedright

\textbf{Keywords: }\label{SEC:Keywords}    {Sensors; Oscillators; Negative capacitance; Design aiming for reuse, repurpose, repair, customization.}

\end{minipage}\tabularnewline

\bottomrule
\end{longtable}
\setcounter{table}{0}

  \end{@twocolumnfalse}
]

\saythanks

\section{Introduction}\label{SEC:Introduction}

The electronic waste (e-waste) is a modern problem under increasing concern and awareness, there are various possible approaches to reduce and mitigate it, the most obvious is the collection and recycling of discarded devices, however the most ideal is just to make technology that lasts because not only is physically fit by quality design, production, and components; but because its design was intended to be most versatile ensuring the same device can be used and reused in various applications/contexts just by changing connections, jumpers, and firmware configurations. Some design aspects for making a device more reusable are: the use of standard connectors and protocols, think of it as a module to be part of a larger system, minimize barriers for connecting/interfacing components and devices from distinct manufacturers.

\subsection{Project objectives and trade-offs}
This article is focused on the design of a sensor interface device with USB and serial(UART,RS-485), aimed to allow the interface to many distinct 2-wire sensors based on the change of inductance, resistance, capacitance, frequency, and also small voltage; sensors that can be interchanged using the same hardware and same port of the device, thus meaning the electronics designed must also be versatile.\par
Providing a versatile device to the users will probably have its negative trade-offs, like:\newline
1- probably significant lower precision/accuracy;\newline
2- some sensor calibration must be provided/done by the end user after replacing a sensor;\newline
3- the calibration function will not be linear or 'easy' as desired for sensors and its interfaces.\par
However, for some applications the mentioned trade-offs are not necessarily a deal-breaker, such as when the user is technical and is ok with using a device that requires more setup/configuration, some users like devices that are more customizable or repairable. Also is possibly valued a device that if no longer useful for a user, it might still be useful for another user on a different application/context.\par

\subsection{License and context}
The hardware design here disclosed is distributed under "CERN Open Hardware Licence Version 2 - Weakly
Reciprocal" (\href{https://cern-ohl.web.cern.ch}{CERN-OHL-W}), its associated software/firmware under GNU licenses (GPL, LGPL). \par
This article is published under the Creative Commons license Attribution-NonCommercial-ShareAlike 4.0 International (\href{https://creativecommons.org/licenses/by-nc-sa/4.0}{CC BY-NC-SA 4.0}).\par
This article is about a 'hobby' project done by the author (David Nuno Quelhas, MSc Electronics Eng, alumni of Instituto Superior Tecnico, Portugal) with occasional 'work' between the years 2012 and 2024 on his 'free time'.

\begin{figure*}[t!]
\centering
\includegraphics[width=\textwidth,height=8cm]{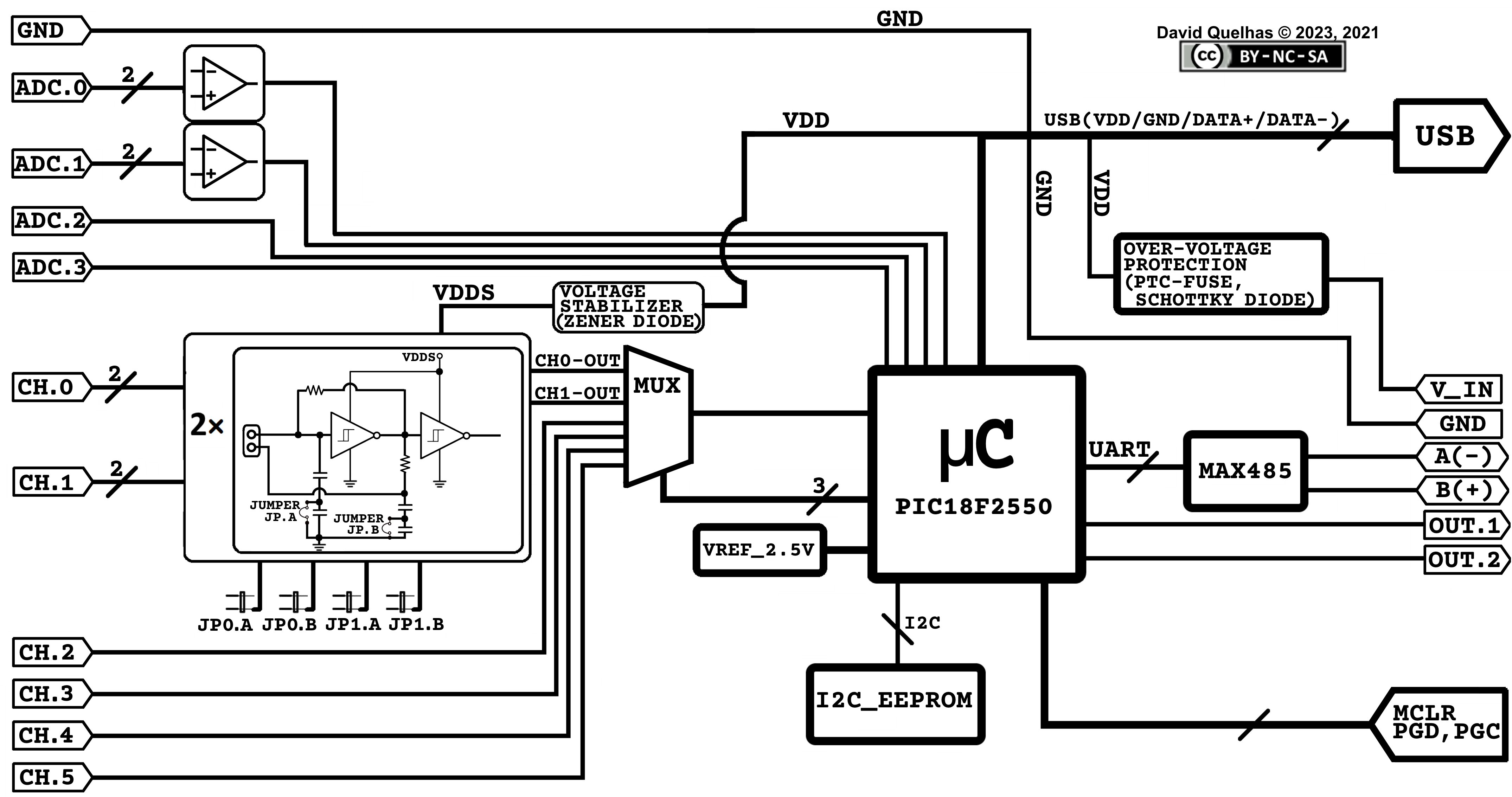}
\caption{Diagram of the Multiple Sensor Interface device.\label{FIG:diagramMultipleSensorInterface}}
\end{figure*}

\subsection{Prior art review}
The topic and devices commonly described in literature as 'Multiple Sensor Interface' and also as 'Universal Sensor Interface', commonly fall under 3 distinct categories:   a)  Device that has a more versatile interface or signal conditioning circuit capable of interfacing various sensor types;   b) Device that includes various specialized interfaces or signal conditioning circuits for each sensor type, typically built using various PCB boards for the sensors, to connect or stack into a PCB board with a microcontroller that will register and/or transmit the measurements, or alternatively have all these different circuits integrated inside a single integrated circuit (micro-chip);   c) Hardware and/or software systems that collect or register sensor data from various distinct sensing devices/circuits, that may apply some processing to the raw data for obtaining measurements, then to be transmitted to other systems or to a data storage, and so these hardware/software systems may also be called/named 'interface'.\par
The article review presented here is about a 'more versatile interface or signal conditioning circuit' which is the category most similar to this article. Types of versatile sensor interface found in prior art:\par
1- Interfacing resistive or capacitive sensors by measuring the charge-discharge time of an RC circuit, or measuring the frequency or PWM from an oscillator whose pace is controlled by the speed of a capacitor charge-discharge through a resistor; for example: \cite{InterfaceCircuitRCSensorsRelaxationOscillator}, \cite{WideDynamicRangeUniversalSensorReadoutPulsewidthModulation}, \cite{UniversalInterfaceForCapacitiveResistiveSensorElements} .\par
2-  Interfacing sensors based on the variation of impedance (includes sensor based on variation of resistance, capacitance or inductance) by an LCR meter, impedance meter, or potentiostat circuit; for example: \cite{ImpedanceToDigitalConverterForSensorArrayMicrosystems}, \cite{AutomatedComplexImpedanceMeasurementSystem}, \cite{HiokiWebsite}, \cite{UltrahighHumiditySensitivityOfGrapheneOxide} .\par
3- Interfacing a sensor as part of a bridge circuit (example:  resistive sensor on a resistance bridge, capacitive sensor on a capacitance bridge) \cite{ConfigEffInterfaceCircuitMultiSensorMicrosystems} .\newline \par

The Multiple Sensor Interface presented in this article has a working principle more similar to the circuits mentioned as type 1 (RC time or frequency or PWM of oscillator), however in comparison with the mentioned references/articles, the interface circuit of this article can interface more distinct sensor types, namely besides interfacing resistive and capacitive sensors it also interfaces inductive sensors and sensors by frequency measurement using exactly the same circuit and connector/port, also it is a simple circuit with a reduced number of components.\par
The Multiple Sensor Interface presented in this article in comparison to the circuits mentioned as type 2 (LCR or impedance meters), has the advantage of not requiring an AC voltage/signal generator for exciting the measurement circuit, and not requiring the complex hardware and/or complex post processing for digitizing voltage signal waveforms, that is typically required for the calculation of amplitude and phase difference of voltage signals, on the measurement of impedance by LCR or impedance meters.\par
The Multiple Sensor Interface presented in this article in comparison to the circuits mentioned as type 3 (measuring a sensor as part of a bridge), has the advantage of using the same circuit for all sensor types (resistive, capacitive, inductive, frequency), instead of requiring a different circuit (the bridge circuit) for each sensor type; thus the interface circuit of this article is a simple circuit with a reduced number of components probably much simpler than any circuitry required for obtaining a single output signal usable for measuring various sensor types on multiple bridge circuits.\par

A comparison regarding the accuracy or precision of this sensor interface and other interfaces/devices was not made, since the stated focus of the article is how to achieve a most versatile sensor interface, and also considering that such comparison may be easier when considering specific type(s) of sensor/application.\par
Also regarding a possible use case of the Multiple-Sensor interface presented in this article, where is possible/allowed that an end-user can replace/change a sensor component while maintaining the sensor interface, probably without a sensor calibration; this device is not meant as a measurement device, but as a sensor interface that can provide only qualitative measurements. Example of a qualitative scale (0 to 6 levels) for light intensity \cite{PerfAnalysisVisibleLightCommCMOSSensors} : 0- Total starlight (0.0001 lux), 1- Full moon (1 lux), 2- Hallways in office buildings (80 lux), 3- Office lighting (300-500 lux), 4- Overcast day (1000 lux), 5- Full daylight (10000-25000 lux), 6- Direct sunlight (32000-130000 lux) .

\section{Design, Materials and Methods}\label{SEC:DesignMaterialsMethods}

\subsection{Sensor Interface Device}
Here is presented the Multiple Sensor Interface (Fig.\ref{FIG:diagramMultipleSensorInterface}), the interface main components / sub-circuits are:
The connectors and sensor interface circuits (oscillators) for inductance, resistance, capacitance, and frequency (CH.0, CH.1); the connectors and over-voltage protection(zener diode) for frequency measurement (CH.2, CH.3, CH.4, CH.5); the connectors and interface circuit for voltage measurement (ADC.0, ADC.1, ADC.2, ADC.3); analog multiplexer for the sensor channels, the microcontroller (PIC18F2550); I2C EEPROM for storing calibration tables; USB connector; connector and circuit for RS-485 and UART; digital outputs connector (OUT.1, OUT.2). \par
The digital outputs have the value of a boolean function defined by the user, boolean functions with logic variables that are the result of a comparison ('bigger' or 'smaller' than), between the value/measurement of a sensor channel and a configurable threshold value. The connectors used for frequency measurement may be connected to external single sensor interface circuits (oscillators).

\subsection{The sensor interface circuit (oscillator)}
\begin{figure}[t!]
	\centering
		\includegraphics[width=\columnwidth]{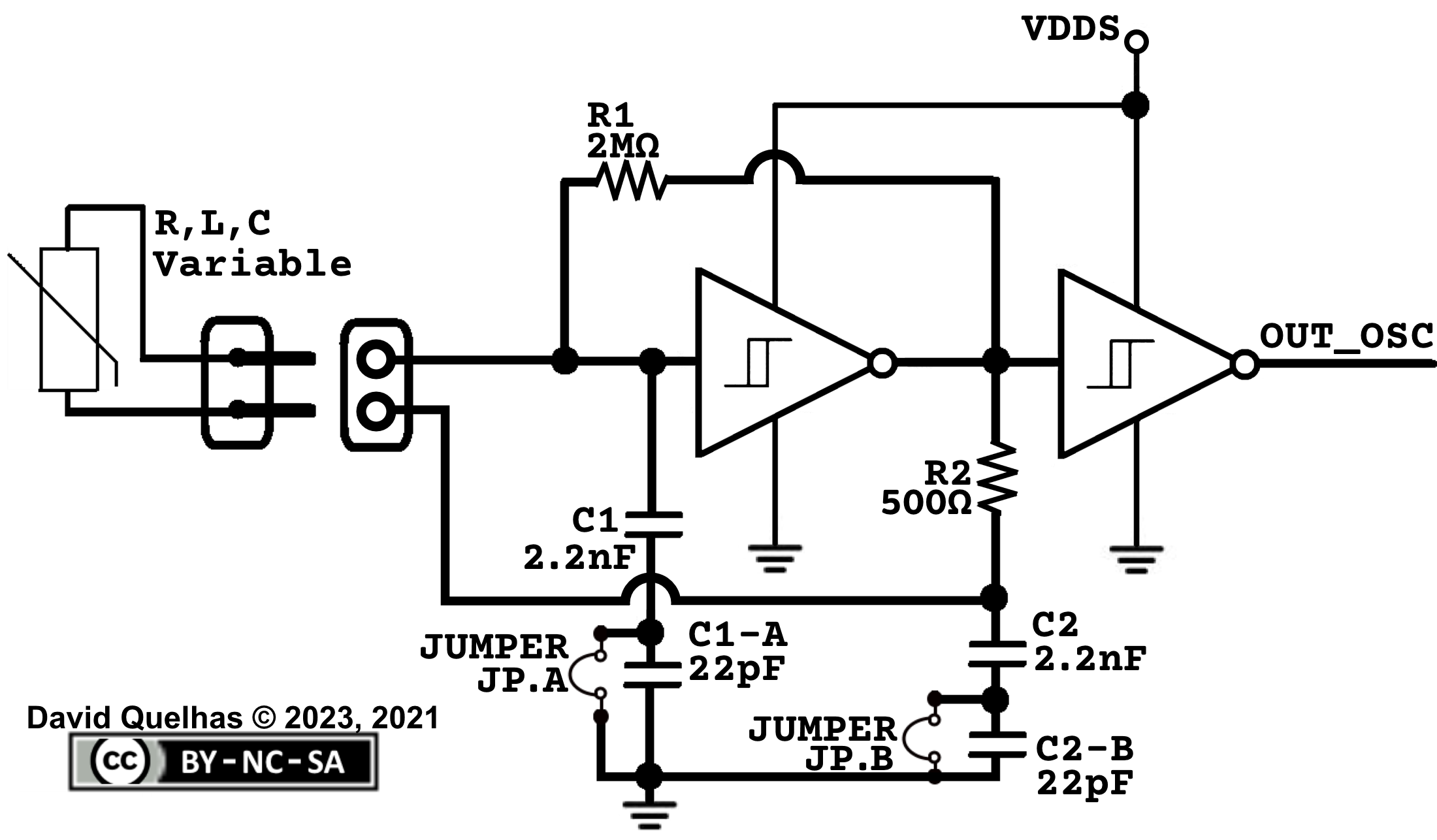}
	\caption{Schematic of the sensor interface circuit (oscillator).}\label{FIG:schSensorInterface}
\end{figure}

The sensor interface (Fig.\ref{FIG:schSensorInterface}) is an oscillator with a circuit design based on the Pierce oscillator with some modifications. The 1st difference is that there is no quartz crystal, and on the location of the crystal will be connected the sensor to be measured (variable inductance or resistance or capacitance), the 2nd difference is that instead of simple inverters ('NOT' gates) will be used Schmitt-trigger inverters(high-speed Si-gate CMOS, 74HC14), this is a very relevant difference that will allow the oscillator to work even with a resistive or capacitive sensor, in fact the interface circuit works with sensors mostly as Schmitt-trigger oscillator. Also the Schmitt-trigger inverters output a noise-free square-wave signal, as oscillator or as external signal converter.\par
The sensor interface circuit has 2 pairs of series capacitors (C1 2.2nF, C1-A 22pF and C2 2.2nF, C2-B 22pF) instead of just 2 capacitors(C1,C2) so the value of C1 and C2 can be adjusted just by placing/removing a jumper; placing a jumper removes C1-A or C2-B from the circuit making 2.2nF the value of C1 or C2; removing the jumper lets the capacitors in series making 21.78pF the total value of (C1, C1-A) or (C2, C2-B). So on the rest of the article, whenever is mentioned C1 or C2 is meant the resulting capacitor value that can be 2.2nF(jumper on) or 21.78pF(jumper off), accordingly with mentioned jumper configuration. \par
The sensors can be connected directly on the Multiple-Sensor Interface(on the screw terminals/connectors), or by using a cable; for a cable longer than 20cm is recommended the use of shielded twisted-pair(STP) cable to prevent cross-talk between sensor channels or external EMI.

\subsection{Measurement process}
The Multiple-Sensor device has a microcontroller (PIC18F2550) that is able to make frequency and voltage measurements, so the device makes frequency measurements for sensor channels CH.0 to CH.5 ; and makes voltage measurements for sensor channels ADC.0 to ADC.3; these frequency and voltage measurements made by the device are designated as the raw\_value of a sensor channel. To obtain the measurement of a sensor channel, the device uses a 2 column calibration table, that is a long list of points (raw\_value; measurement) relating the measurement value (obtained during calibration by an external reference device) to the corresponding raw\_value obtained on the Multiple-Sensor device, these calibration tables are stored on an I2C EEPROM memory on the Multiple-Sensor device.\par
\begin{figure}[ht!]
	\centering
		\includegraphics[width=0.4\columnwidth]{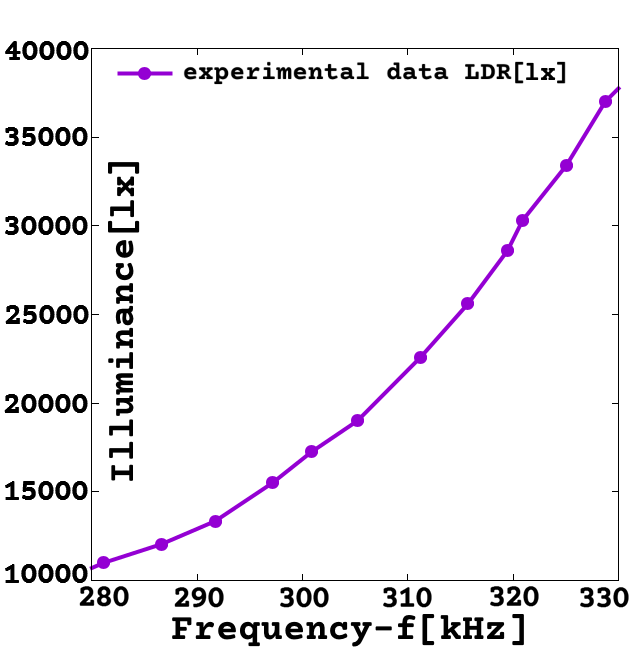}
	\caption{Plot experimental data with line, LDR light(brightness) sensor connected on Multiple-Sensor Interface; example of a calibration table exclusively from experimental data.}	\label{FIG:experimentalDataLineLDRSensor}
\end{figure}
The Multiple-Sensor device can work in two modes: single-channel or multiple-channel, the CH.0 to CH.5 raw\_value(frequency) are calculated through a counter/timer of the PIC18F2550 by periodically reading its value and calculating the frequency f=count/period ([Hz]=[cycles]/[s]). So in single-channel mode the frequency is always calculated on the selected/enabled sensor channel, in multiple-channel mode the frequency is calculated for each sensor channel sequentially (time-division multiplexing), since there are 6 channels to measure but only on counter/timer of the microcontroller for that job. Thus in multiple-channel mode a measurement takes 6x more time to be updated/refreshed than in single-channel mode.\par
For the sensor channels ADC.0 to ADC.3 the raw\_value is the voltage of those channels measured by using the ADC (Analog to Digital Converter) of the microcontroller and also reading a 2.5V voltage reference. \par
The sensor measurements are calculated by searching the raw\_value on the corresponding calibration table, and by using from the table 2 points (raw\_value,measurement) referenced here as points P and Q such that the measured raw\_value is bigger than raw\_value of P and is lower than raw\_value of Q; then is calculated a linear equation: $measurement=a{\cdot}(\text{raw\_value})+b$, defined by the points P and Q. So every-time the device calculates a sensor measurement, it will calculate the corresponding linear equation for the current raw\_value and use it to obtain the current measurement (Fig.\ref{FIG:experimentalDataLineLDRSensor}) .

\subsection{Device calibration for a sensor}
Device calibration is about obtaining calibration tables for each sensor channel, here are 2 ways to obtain it:\par
1- Do a full manual calibration using an external meter as reference where both the reference meter and the Multiple-Sensor device(with a sensor connected) are exposed to same stimulus/environment that is controllable by the user to produce all adequate variations/intensities necessary to record an extensive calibration table, with all experimental pairs of (raw\_value,measurement).\par
\begin{figure}[t!]
	\centering
		\includegraphics[width=\columnwidth]{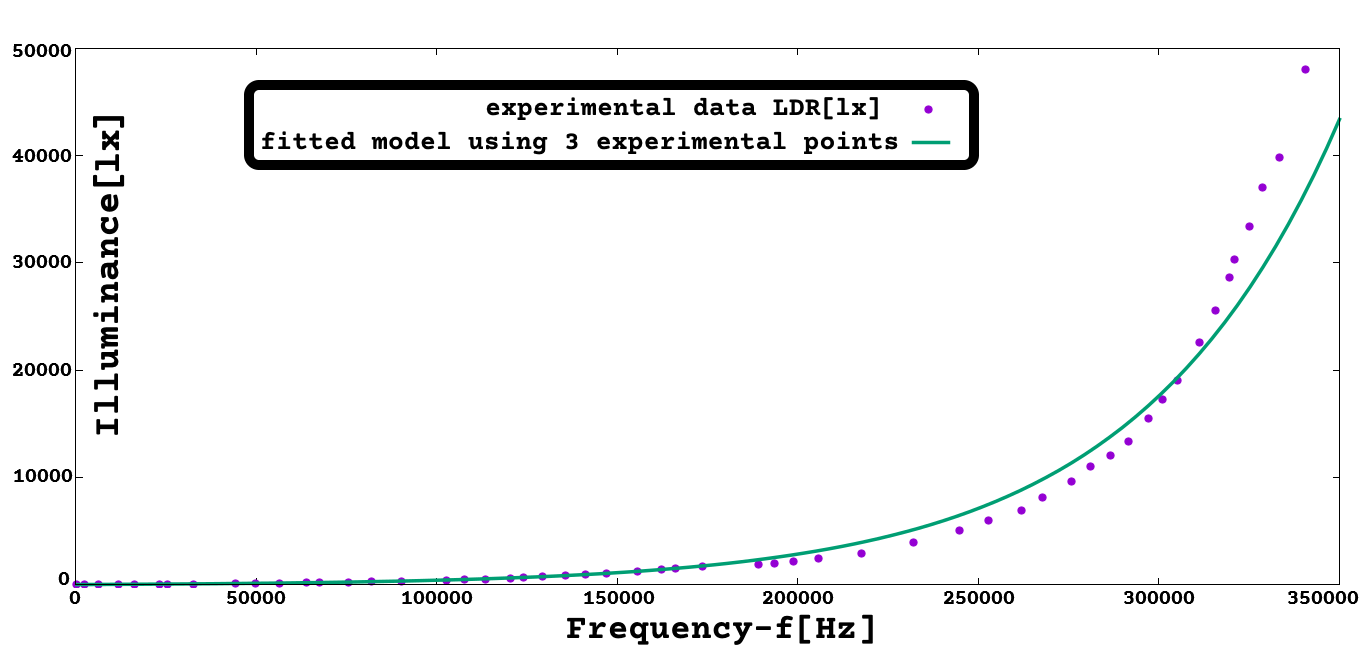}
	\caption{Plot experimental data and fitted model(by using 3 points), LDR light(brightness) sensor connected on Multiple-Sensor Interface.}\label{FIG:experimentalFittedModelLDRSensor}
\end{figure}

2- Using a known function that relates the measured phenomena to the obtained raw\_value on the Multiple-Sensor device (obtained by theoretical or experimental study), although a purely theoretical calibration could be used, probably is better or easier to obtain a calibration table by using a known function and have its constants/parameters calculated by a data fitting to some few experimental data points (raw\_value, measurement) obtained for the device calibration. So for example if the known function had 3 constants/parameters, it would require at least 3 different experimental measurements to obtain the function for that sensor channel, then having the function is just a question of calculating a longer list of pairs (raw\_value,measurement) on the desired measurement range. Fig.\ref{FIG:experimentalFittedModelLDRSensor} is the result of fitting the model function $Illuminance(f) = a+b{\cdot}({\emph{e}}^{c{\cdot}f})$, to the points: (244Hz, 0.01 [lx]), (25320Hz, 30 [lx]), (232041Hz, 3950 [lx]); obtaining the values: \mbox{$a$$=$$-$80.2359}; \mbox{$b$$=$79.8743}; \mbox{$c$$=$1.79972${\cdot}$$10^{-5}$}. The symbol $\emph{e}$ is the Euler-Napier constant. The point at 244Hz was changed from 0[lx] to 0.01[lx] as it may facilitate/improve the model function fit.

\subsection{Methods for error analysis of Multi-Sensor device}\label{SectionMethodsErrorAnalysisMultiSensorSchTriOsc}
A theoretical analysis of the measurement error of the Multiple-Sensor Interface can be made based on the theoretical results obtained on the following sections, for example by making a comparison between the expected theoretical values of ${R_s}(f)$, ${C_s}(f)$, ${L_s}(f)$ as Schmitt-trigger oscillator and the measured values (experimental data). Also it can be explored the theoretical predictions of how much the uncertainty of each electrical parameter (circuit constants) contributes to the uncertainty of the sensor measurement; that is, how the measurement error theoretical correlates with the error/uncertainty of the electrical parameters of circuit components.
A possible method for the theoretical error analysis can be based on the formula of the upper bound of the error propagation, similar to the General Formula for Error Propagation, but always valid regardless from the fact if the errors on independent variables (here the electrical parameters) are independent or random. The formula of the upper bound of the error propagation is: if $q=q(x,...,z)$ is any function of x,...,z, then ${\delta}q{\leqslant}{|\frac{{\partial}q}{{\partial}x}|{\delta}x}+...+{|\frac{{\partial}q}{{\partial}z}|{\delta}z}$  , \cite{IntroErrorAnalysisJohnTaylor}, \cite{ExperimentationDCBaird} .

Also, one typical term/concept for the error analysis of oscillators is the study of the 'frequency stability' (${{\Delta}f}/f$), that commonly is analysed and/or tested against external factors like power supply voltage and temperature. Since the oscillator studied here has the purpose of being a sensor interface, and not the typical purpose as generator of cyclic voltage signal, the analysis focus will be on the measurement error with specific sensor(s) and regarding the interface circuit will be based on the functions ${R_s}(f)$, ${C_s}(f)$, ${L_s}(f)$. Relevant to the error analysis done here is to recognize that variations in temperature and in supply-voltage ($V_{DDS}$) are interesting to be considered regarding the measurement error (just like variations in $R_1$, $R_2$, $C_1$, $C_2$). Also, variations in supply-voltage ($V_{DDS}$) will affect the theoretical measurement error through the H parameter that will be defined in the theoretical analysis as $\qquad H=ln \left ( \frac{({V_T^{-}}-{V_{DDS}}){V_T^{+}}}{({V_T^{+}}-{V_{DDS}}){V_T^{-}}} \right )$ .

Variations over time of the electrical parameters $R_1$, $R_2$, $C_1$, $C_2$, contribute to the measurement error; and in case of a calibration based purely on theoretical formulas the tolerance/uncertainty of $R_1$, $R_2$, $C_1$, $C_2$ will also contribute to the measurement error.
The tolerance of the components on the Multiple-Sensor device/PCB, that was used on the experimental work/tests is: $\pm$5$\%$ for resistors ($R_1$, $R_2$), and $\pm$10$\%$ for capacitors ($C_1$, $C_2$).
By tolerance is meant, the maximum deviation allowed from the specified nominal value of the electrical parameter of a component.
On the entire article, the values used for calculating the theoretical results where the nominal values of the components $R_1$, $R_2$, $C_1$, $C_2$.\par

The author considers that isn't convenient and/or interesting to include the theoretical (and numeric) prediction of the error (for example using the General Formula for Error Propagation) based on the theoretical analysis (equations of $R_s$(f), $C_s$(f), $L_s$(f)) in the scope of this article, because: 1- Is a calculation intensive process that produces an output that requires a long written content (tables and text) to be presented; 2- the proximity/conformity between experimental data and the theoretical models is already depicted on various graphs available on the article that show both theoretical and experimental data; 3- The presented theoretical analysis is already stated to include some mathematical approximation and so it should be expected some deviation/displacement between the theoretical models and the experimental data.\par
So a numeric characterization of the measurement error may be more interesting if applied to specific combination of sensor plus the Multiple-Sensor Interface. The characterization of the measurement error that is available on the Appendix sections, is a comparison/difference between large set of experimental data measurements against the values obtained from theoretical models that were fitted with a small subset of experimental data; so these fitted models may be a convenient way to obtain device calibration where is meaningful to analyse the measurement error, and where the known/nominal values of circuit parameters ($R_1$, $R_2$, $C_1$, $C_2$) and sensor parameters(sensor specific) are used as the initial value of the parameters that are calculated by the model fitting.

On the Appendixes \ref{SEC:LDRSensorCalibAndError} and \ref{SEC:FluidLevelSensorCalibAndError}, are graphs of fitted models (for example a theoretical model fitted with 8 samples of experimental data) that where obtained by using the 'fit' command of the GNUPlot software (it applies the "nonlinear least-squares Marquardt-Levenberg algorithm" \cite{GNUplot}); also is included/listed the experimental data samples used on the model fitting, the obtained model parameters, and error analysis graphs of the models.\par

\subsubsection{Temperature variation and measurement error}
The variation of temperature may influence indirectly the measurement error, as for example, typical room temperature variations are expected to cause small variations of the values/parameters of the resistors and capacitors used on the multiple sensor interface ($R_1$, $R_2$, $C_1$, $C_2$), and these small temperature variations if not taken into account on the measurement process, as is the case with basic/inexpensive sensor interfaces like this one, will cause a small increase of the measurement error. More relevant is to check the effect of temperature on the measurement error related with semiconductor devices, like the integrated circuit 74HC14 (Schmitt-trigger inverters) that was used on the sensor interface; since by the physics of semiconductors the relation between resistivity and temperature is given by an exponential function. The only electrical parameters used on the theoretical models of this article related to the Schmitt-trigger inverter are: ${V_T^{-}}$, ${V_T^{+}}$; and the relation of these parameters with temperature can be consulted on the datasheet of the IC used (Nexperia/NXP 74HC14) \cite{Nexperia74HC14datasheet} and on similar hex Schmitt-trigger gates that do have more information/detail on how ${V_T^{-}}$, ${V_T^{+}}$ change versus temperature.
The datasheet of 74HC14 by Nexperia/NXP \cite{Nexperia74HC14datasheet}, and also datasheets of similar devices like: MC74HC14A by Onsemi \cite{ONsemiMC74HC14datasheet}; CD40106B \cite{TexasInstrumentsCD40106Bdatasheet} and SN74LS14 \cite{TexasInstrumentsSN74LS14datasheet}
 by Texas Instruments; basically indicate that ${V_T^{-}}$, ${V_T^{+}}$ are nearly constant over the entire operational temperature range, since in parameters tables, they indicate a single value of ${V_T^{-}}$ and of ${V_T^{+}}$ for the entire temperature range (-40 ºC to +125 ºC ). 
The datasheet of SN74LS14A by Texas Instruments \cite{TexasInstrumentsSN74LS14datasheet} provides on page 8 a graph showing that the variation of ${V_T^{+}}$ versus temperature and ${V_T^{-}}$ versus temperature is minimal (almost constant).\par
So the overall effect of typical room temperature variations on the Multiple-Sensor interface is of small relevance to the type of low-end (not high precision) applications intended for this device.

\subsubsection{\texorpdfstring{$V_{DDS}$}{VDDS} variation and measurement error}
The variation of the power supply voltage ($V_{DDS}$) may influence indirectly the measurement error, since a variation of $V_{DDS}$ is expected to cause a variation of the H parameter that is used on the following theoretical analysis. Also is relevant to notice that in accordance with the tables on the datasheet of the IC used (Nexperia/NXP 74HC14) \cite{Nexperia74HC14datasheet} the values of ${V_T^{-}}$, ${V_T^{+}}$ have significant variation versus the power supply voltage (indicated as $V_{CC}$).
The datasheet of MC74HC14A by Onsemi \cite{ONsemiMC74HC14datasheet} provides on page 5 a graph showing a linear relation between ${V_T^{+}}$ versus $V_{CC}$ and ${V_T^{-}}$ versus $V_{CC}$.
So from the information on the datasheet for the purpose of the measurement error analysis, is appropriate and interesting, to replace the ${V_T^{+}}$ and ${V_T^{-}}$ by a function that has $V_{DDS}$ as independent variable, and that would be the most accurate way to study the influence of $V_{DDS}$ on the measurement error. 
For a measurement error analysis where variations/uncertainty of $V_{DDS}$ are considered, the 'H' parameter will be the only model parameter that is influenced by $V_{DDS}$, accordingly with the expression $\qquad H=ln ( ({({V_T^{-}}-{V_{DDS}}){V_T^{+}}})/({({V_T^{+}}-{V_{DDS}}){V_T^{-}}}) )$, that will be introduced on the theoretical analysis. Also more information on how the 'H' parameter is influenced by power supply voltage is available on page 9 and 10 of the datasheet of the IC used (Nexperia/NXP 74HC14) \cite{Nexperia74HC14datasheet}, where is mentioned the expression $f={1/T}\approx{1/({KRC})}$, where the 'K' parameter of the datasheet corresponds to the 'H' parameter of this article, and the RC product of constants corresponds to the ${\tau=-1/\lambda}$ parameter of this article. On the mentioned datasheet \cite{Nexperia74HC14datasheet} there is plot of $K({V_{CC}})$ that is consistent/similar with the $H({V_{DDS}})$ expression that was defined in this article.\par

However, for the sake of simplicity in this article, the values of ${V_T^{+}}$ and ${V_T^{-}}$ will be considered as constants on the theoretical analysis, and the error analysis available on the appendix sections doesn't includes/considers the variations/uncertainty of power supply voltage ($V_{DDS}$).

\subsubsection{Suggestions for minimizing the measurement error}
A sensor connected on the sensor interface may intentionally be matched with some corresponding values of ${C_1}$, ${C_2}$ (or also ${R_1}$, ${R_2}$), for example by a changeable jumper connection, or alternatively adding/connecting a fixed value resistor/capacitor/inductor in parallel with the sensor, to better align the functions ${R_{s}}(f)$, or ${C_{s}}(f)$, or ${L_{s}}(f)$ that define the sensor interface with the values of $R_{s}$, or $C_{s}$, or $L_{s}$ expected for the sensor output when measuring whatever dimension/phenomena, maximizing the use of the sections of ${R_{s}}(f)$, or ${C_{s}}(f)$, or ${L_{s}}(f)$ that have the lowest measurement error.  Also having components ${C_1}$, ${C_2}$, ${R_1}$, ${R_2}$ with lower/better tolerance can give a small contribution for reducing the error between theoretical results and experimental data.
Also, it may be possible to obtain fitted functions (of sensor measurement, or of ${R_{s, Fit.}}(f)$, or ${C_{s, Fit.}}(f)$, or ${L_{s, Fit.}}(f)$) with smaller measurement error, by selecting a different set of experimental data to use on the model fitting, or by limiting/reducing the range of values where the calibration is valid/used (and so using for model fitting a narrower set of experimental data that actually is the most relevant for the measurement range of the specific sensor).

\section{Results and Analysis}\label{SEC:ResultsAnalysis}

\subsection{Device testing}
The author developed and built prototypes of the described device, made of the components described in the previous section and in Fig.\ref{FIG:diagramMultipleSensorInterface} diagram. Then the device was experimented with various different sensors; including various common sensor components, namely: LDR (Light Dependent Resistor or also designated as photo-resistor), RTD(Resistance Temperature Detector), FSR(Force Sensitive Resistor), Relative-Humidity sensor (RH to impedance); also a water level / soil moisture sensor made with 2-layer PCB, and as well some handmade sensors done by the author for the purpose of exploring the device usability, namely: a proximity sensor (based on the variation of inductance of a flat coil, caused by the vicinity of a metallic object, or the vicinity of a non-metallic object covered with aluminium or copper adhesive tape), a force sensor (based on resistance variation when pressed) made using 'carbon impregnated foam' (also known as ESD/antistatic foam), aluminium foil and adhesive tape.\par
The tests done with the device connected on the various mentioned sensors, were made with the purpose of verifying that the device is indeed usable with various types of sensors, but those tests are not the most appropriate for studying how the device works, or for characterizing the device itself by determining its usability range, or for gathering quantitative data about the device to be used along with the data from a sensor datasheet for determining its compatibility.\par
\begin{figure}[t!]
	\centering
		\includegraphics[width=\columnwidth,keepaspectratio]{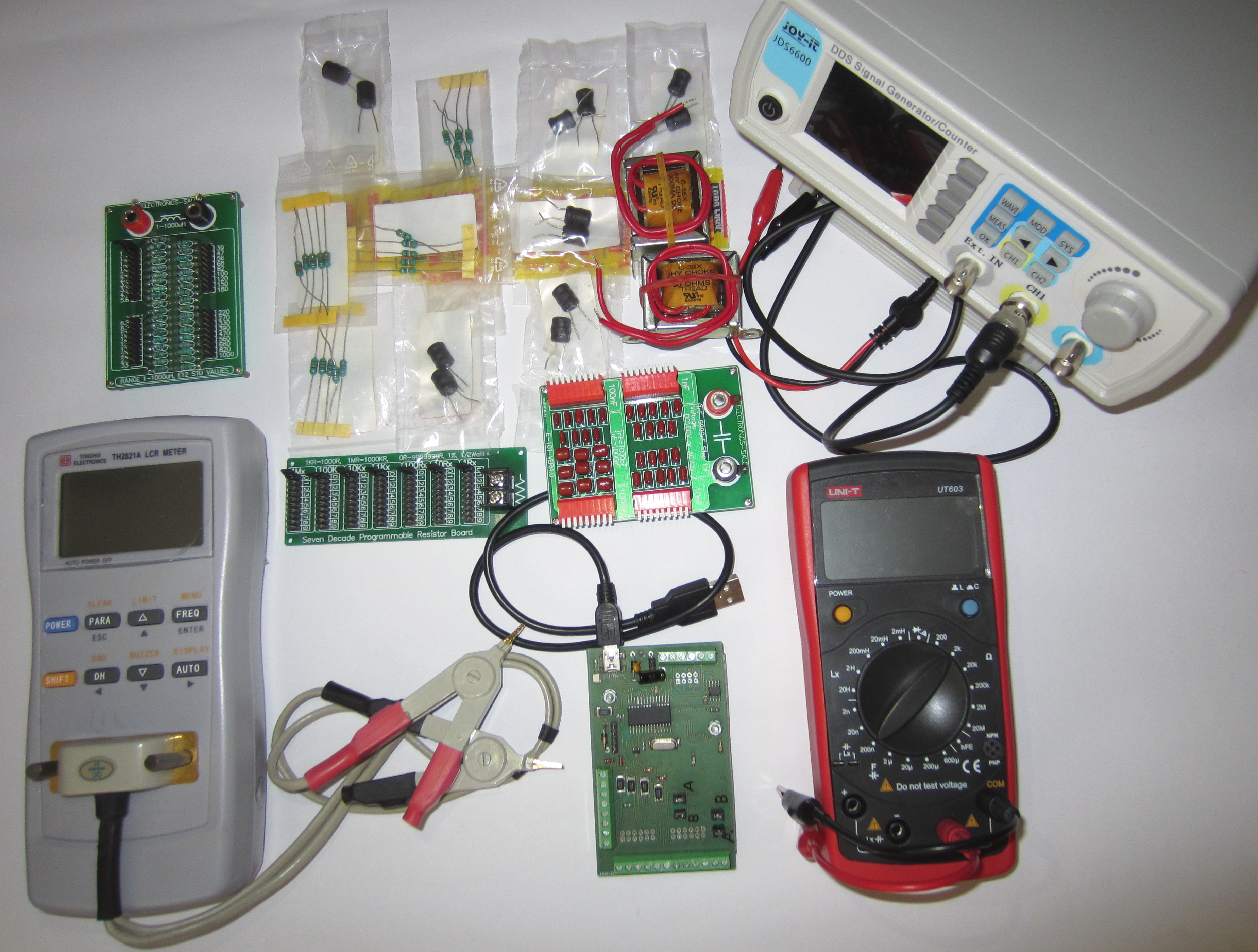}
	\caption{Equipment used for testing; R, C, L test components (top); and the Multiple-Sensor Interface (center bottom).}\label{FIG:showTestEquipmentMultipleSensor}
\end{figure}

So the tests chosen for characterizing the device were records (in 2 column tables) of the measured values of inductance, resistance, capacitance paired with measured frequency on the Multiple Sensor Interface device. For these tests (Fig.\ref{FIG:showTestEquipmentMultipleSensor}) were used arrays(PCBs) of inductors, resistors, capacitors that allow to obtain various different values just by changing a jumper/switch, also were used single components (including in series or parallel association); these fixed value components were connected as the sensor on the device. \quad
The various figures with plots/graphs in this article will show both the experimental data obtained from the mentioned tests, as well the theoretical graphs obtained from circuit analysis of the device, for comparison purposes.\par
The experimental data of the mentioned tests is on:\newline \mbox{Appendix \ref{SEC:AppxExpDatasets}} ( Tab. \ref{TAB:tabLsvsFreq_abc}, Tab. \ref{TAB:tabRsvsFreq_c},  Tab.  \ref{TAB:tabCsvsFreq_c} ) is the ${C_{sensor}}(f)$, ${L_{sensor}}(f)$, ${R_{sensor}}(f)$ tables with $C_1$, $C_2$ as 2.2nF(JP on) or 21.8pF(JP off); \mbox{Appendix \ref{SEC:AppxAdditionalExpDatasets}} ( Tab. \ref{TAB:tabRsvsFreq_f} , Tab. \ref{TAB:tabLsvsFreq_def} , Tab. \ref{TAB:tabCsvsFreq_f} ) is the ${R_{sensor}}(f)$, ${L_{sensor}}(f)$, ${C_{sensor}}(f)$ tables with $C_1$, $C_2$ as 93nF(JP on) or 21.8pF(JP off); \mbox{Appendix \ref{SEC:LDRSensorDataset}} ( Tab. \ref{TAB:tabIlluminancevsFreq_LDR_c} ) is the ${Illum}(f)$ table with $C_1$=$C_2$=2.2nF(JP on) of a LDR sensor;  \mbox{Appendix \ref{SEC:FluidLevelSensorDataset}} ( Tab. \ref{TAB:tabWaterLevelvsFreq_WaterLevelSensor_c} ) is the ${h_{water}}(f)$ table with $C_1$=$C_2$=2.2nF(JP on) of a water level sensor.

\subsection{Sensor Interface Circuit Analysis}

\subsubsection{Multiple-Sensor Interface for inductive sensors}
\begin{figure}[ht!]
	\centering
		\includegraphics[width=0.75\columnwidth]{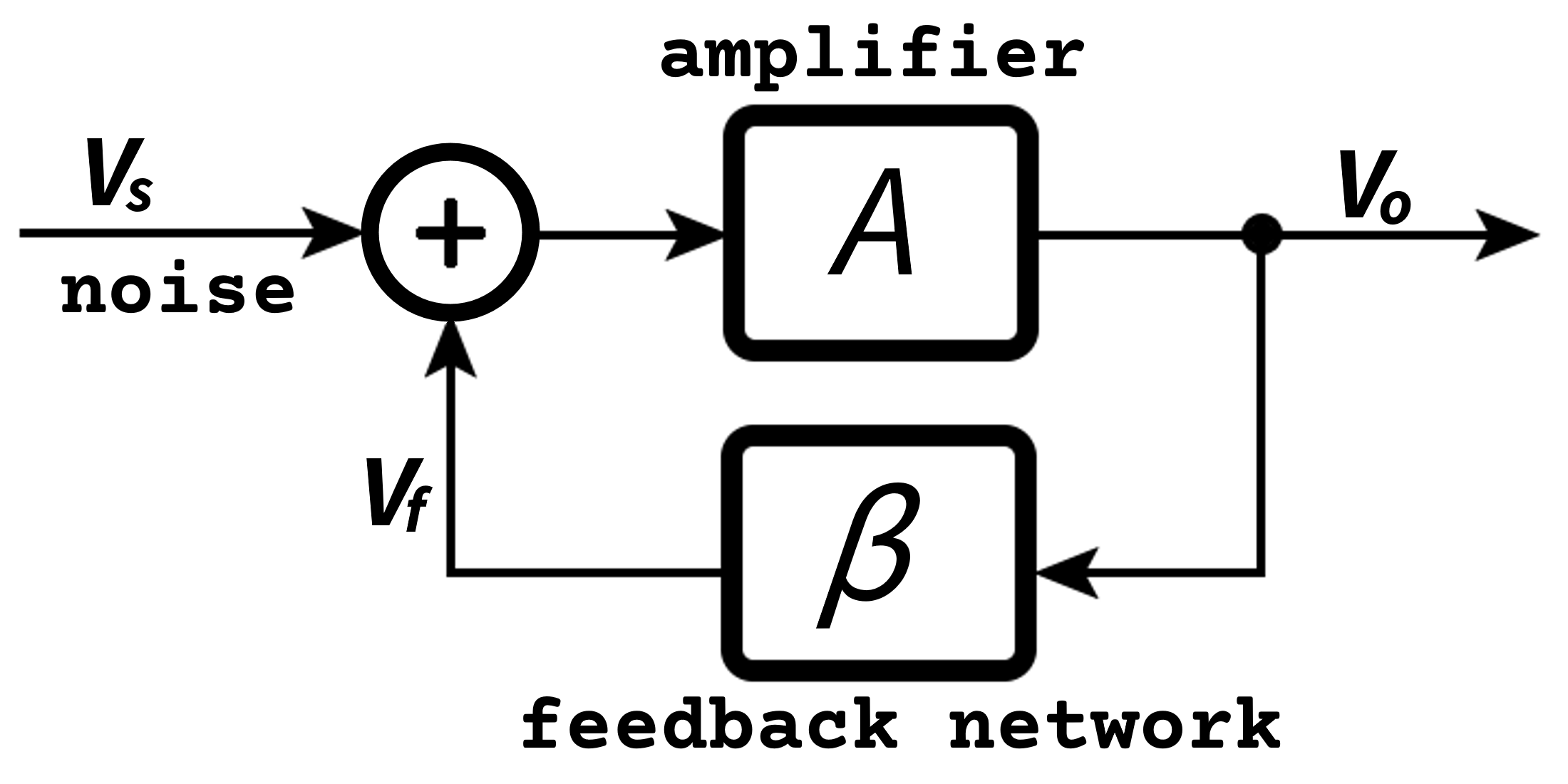}
	\caption{Diagram of model for the oscillator with an inductive sensor working as Pierce oscillator (model of feedback linear oscillator).}\label{FIG:modelFeedbackLinearOscillator}
\end{figure}

When is connected an inductor or inductive sensor the Multiple-Sensor Interface (Fig.\ref{FIG:schSensorInterface} ) may work as a Pierce oscillator(where the sensor is connected instead of a quartz crystal). The theoretical analysis for this type of oscillator, can be based on a model of 2 circuit blocks named 'A' and '$\beta$' connected for feedback by connecting the output of one to the input other. The 'A' is an electronic amplifier providing voltage gain, the '$\beta$' is an electronic filter providing frequency selection (resonance), so whatever voltage signal amplified by 'A' is frequency selected by '$\beta$' and feed back to the input of A for further amplification. As known, this oscillator is start-up by whatever noise ($v_{s}$) available at the input of 'A', Fig.\ref{FIG:modelFeedbackLinearOscillator} is a diagram depicting this concept. \par
With this type of oscillator, for determining the frequency of oscillation, may be used the Barkhausen stability criterion that says $A\beta=1$ to be possible to occur sustained oscillations (oscillations on steady state analysis); where $A$ and $\beta$ represent the transfer function of the correspondingly named block.  

The circuit analysis of the Pierce oscillator is available on various text book (and also class notes); references about Pierce oscillator circuit analysis that are the author preference for understating how the Multiple-Sensor Int. may work as Pierce oscillator, are: \quad  "Crystal Oscillators" of "Digital Electronics" class notes by Peter McLean \cite{analysisPierceOscillatorPMcLean} ; \quad "Microelectronic Circuit Design (4th ed.)" by R.C. Jaeger, T.N. Blalock \cite{hParametersAndPierceOscJaegerBlalock} ; \quad "Microelectronic Circuits (8th i. ed.)" by A. S. Sedra, K. C. Smith, T. C. Carusone, V. Gaudet \cite{PierceOscSedraSmith} .

About the circuit analysis of the Pierce oscillator (and in generally with Colpitts type oscillators), a relevant conclusion is that the feedback network (block $\beta$) composed of 3 electrical components, that include the capacitors $C_1$ and $C_2$, must have an inductive part on the remaining (3rd) electrical component (for example: the piezoelectric crystal of a typical Pierce oscillator, or simply an inductor, or an inductive sensor, etc...), to be able to satisfy $A\beta=1$ and so have oscillations on steady state analysis.

For the Pierce oscillator the equation that relates the oscillation frequency with the inductance and capacitance is a typical equation of LC oscillator circuits, using the definition of a "load capacitance" $C_L$ as $(1/{C_L})=(1/{C_1})+(1/{C_2})$.\par

The frequency of oscillation on the Pierce oscillator (using $C_1$, $C_2$, $L_s$) is:
\vspace{-3mm}
\begin{equation}\label{EQ:frequencyOscillationPierceUsingC1C2L3}{\omega}=\frac{1}{\sqrt{L_sC_L}} \hspace{5mm} \Leftrightarrow \hspace{5mm} f=\frac{1}{2\pi\sqrt{L_sC_L}}\end{equation}
\vspace{-1mm}
So the expression (theoretical) of a value for inductance ($L_{s}$) as a function of frequency(f) is:
\vspace{-1mm}
\begin{equation}\label{EQ:LSensorVersusFrequencyOscillationPierce}L_{sensor} = L_{s} = \frac{1}{4{\pi^2}{C_L}{f^2}} = \frac{C_1+C_2}{4{\pi^2}{C_1}{C_2}{f^2}}\end{equation}
\vspace{-3mm}

\begin{figure}[b!]
	\centering
		\includegraphics[width=\columnwidth,keepaspectratio]{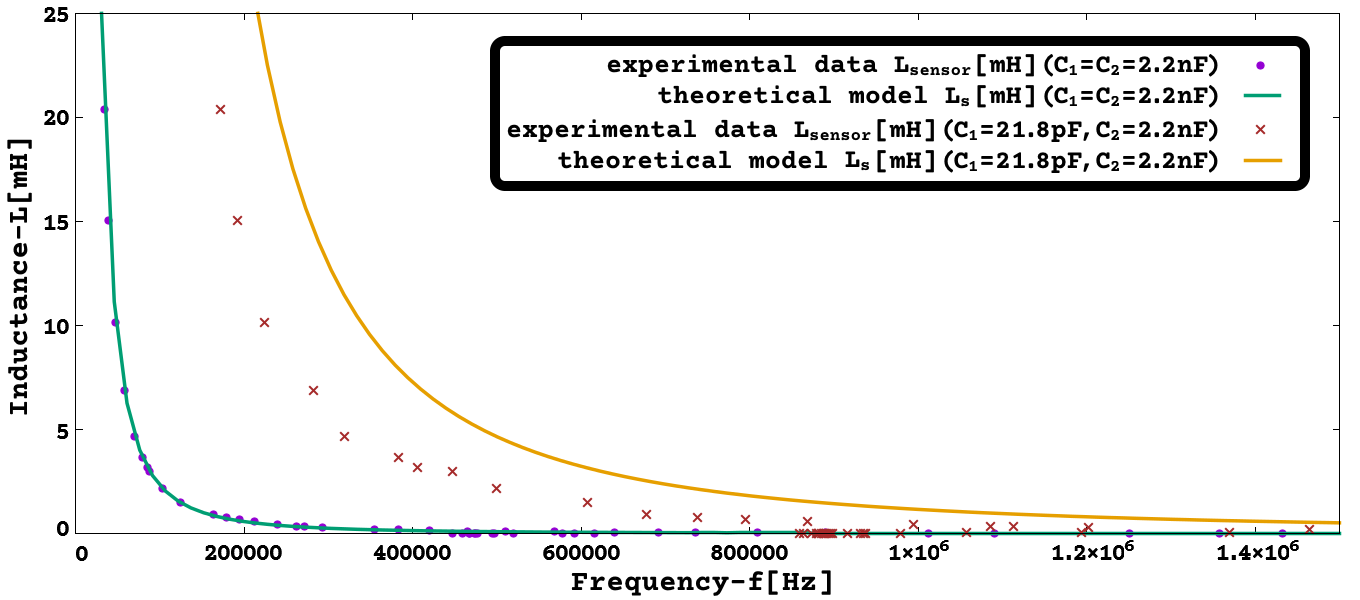}
	\caption{${L_s}(f)$[mH]([Hz]) with $C_1$=$C_2$=2.2nF and $C_1$=21.78pF, $C_2$=2.2nF (Pierce osc., Multi-Sensor Int. with inductive sensor)}\label{FIG:LsVersusFrequencyOscillationPierce}

	\centering
		\includegraphics[width=\columnwidth,keepaspectratio]{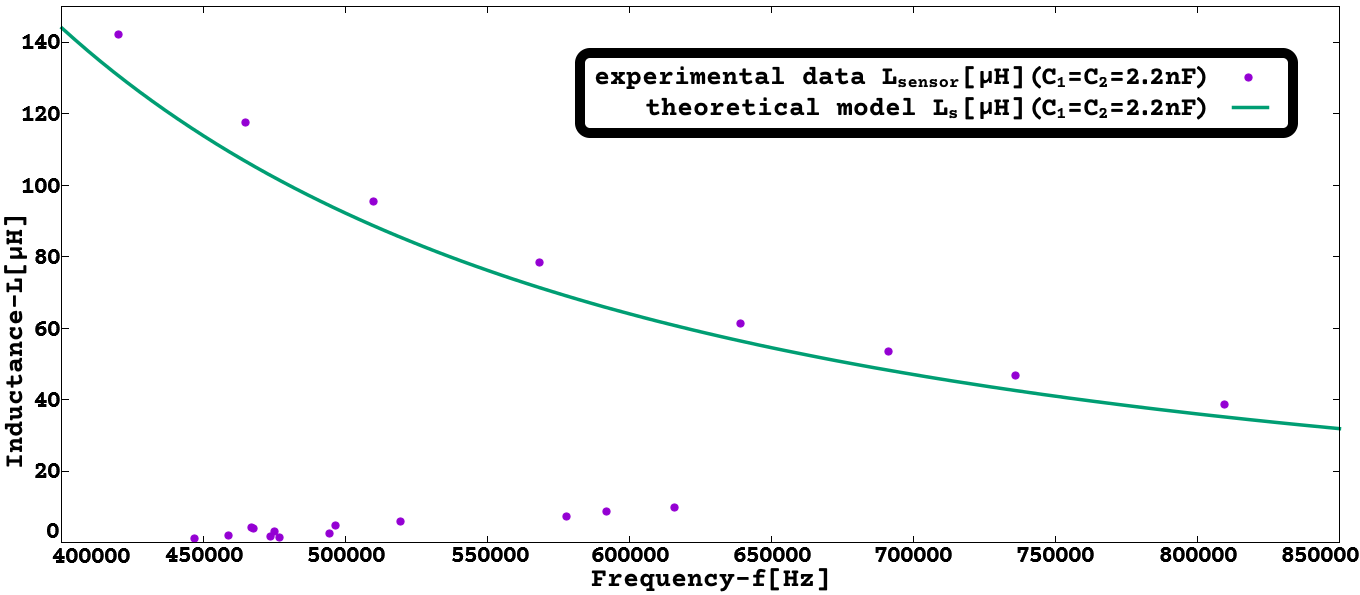}
	\caption{Frequency jump of ${L_s}(f)$[$\mu$H]([Hz]) with $C_1$$=$$C_2$$=$2.2nF (Pierce osc., Multi-Sensor Int. with inductive sensor).}\label{FIG:LsVersusFrequencyFreqJumpC1-C2-2200}

\end{figure}

On experimental tests done was observed that when using $C_1$=$C_2$=2.2nF (JP.A and JP.B closed) or when using $C_1$=21.78pF (JP.A open), $C_2$=2.2nF (JP.B closed), with decreasing values of $L_{s}$ connected, the oscillation frequency exhibited a sudden change, at some small value of L(around $10 \mu H$ for $C_1$=$C_2$=2.2nF), not coherent with theoretical model of Pierce oscillator. This may be related to the fact that the same circuit also implements a Schmitt-trigger oscillator(next section), that oscillates under different criteria, so the author opinion is when $L_s$ approaches some small value it may change from Pierce oscillator to Schmitt-trigger oscillator. Fig.\ref{FIG:LsVersusFrequencyOscillationPierce} and Fig.\ref{FIG:LsVersusFrequencyFreqJumpC1-C2-2200} shows the experimental data for various inductance values connected as the sensor and the plot $L_{s}(f)$ using (\ref{EQ:LSensorVersusFrequencyOscillationPierce}) with $C_1$=$C_2$=2.2nF and $C_1$=21.78pF, $C_2$=2.2nF.

\begin{figure}[t!]
	\centering
		\includegraphics[width=\columnwidth,keepaspectratio]{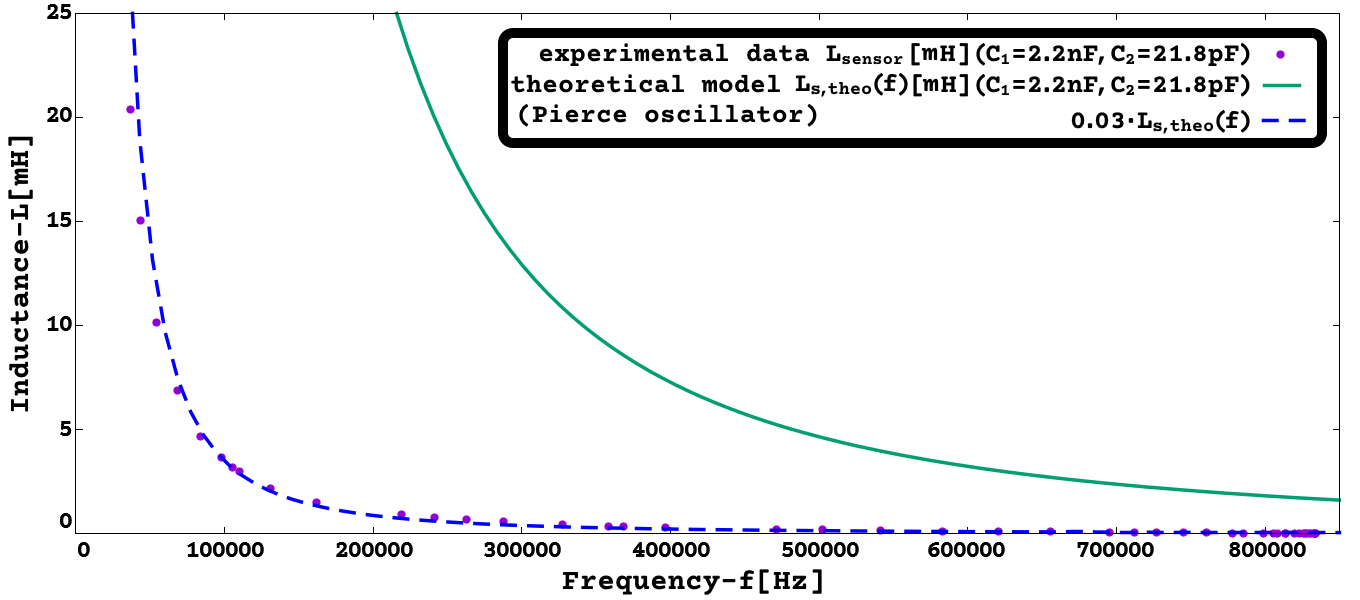}
	\caption{${L_s}(f)$[mH]([Hz]) with $C_1$=2.2nF, $C_2$=21.78pF (theoretical ${L_s}(f)$ as a Pierce osc.).}\label{FIG:LsVersusFreq--C1-2200--C2-22}

	\centering
		\includegraphics[width=\columnwidth,keepaspectratio]{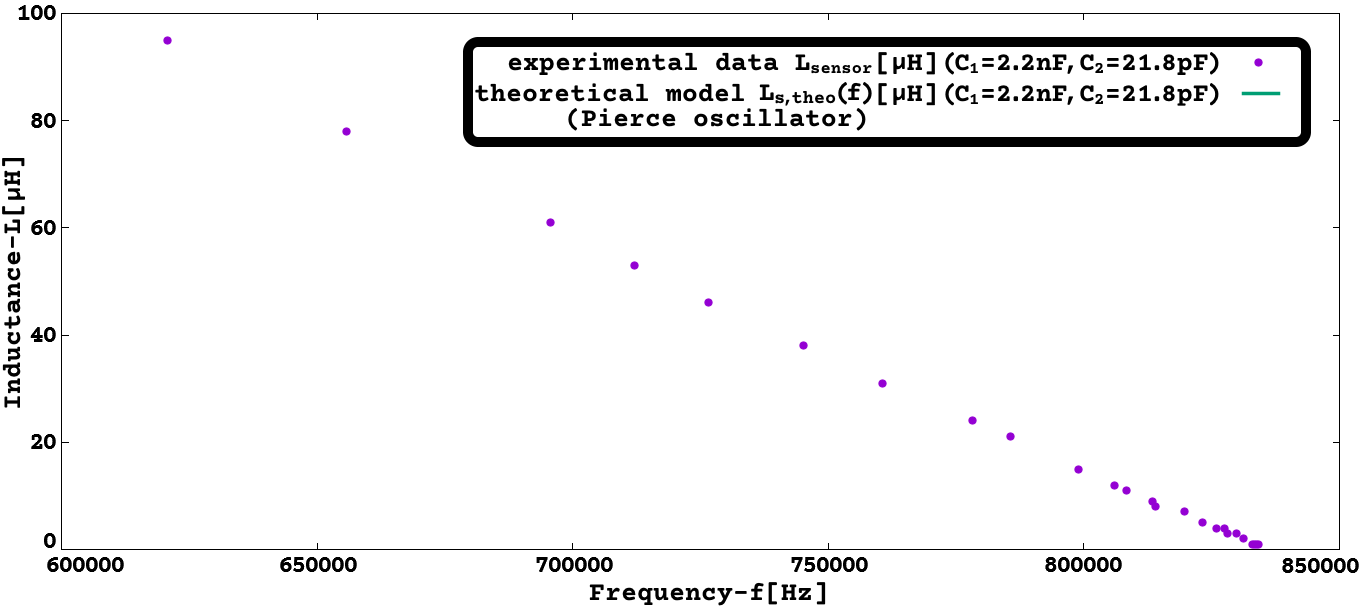}
	\caption{Experimental data of $L_{sensor}$ in $[0 \mu H; 100 \mu H]$, with $C_1$=2.2nF, $C_2$=21.78pF (theoretical ${L_s}(f)$ as a Pierce osc. is out of plot range).}\label{FIG:LsVersusFreqExprData100uH--C1-2200--C2-22}
\end{figure}

Since the mentioned sudden change of oscillator mode and frequency is not adequate on a $L_{s}(f)$ function usable for sensor interfacing; then on experimental tests with jumper configuration: JP.A on and JP.B off ($C_1$=$2.2nF$, $C_2$=21.78pF), it was observed a continuous and progressive $f(L_{sensor})$ function. With $C_1$=2.2nF, $C_2$=21.78pF, the experimental $L_{sensor}$ values followed a straight line for $L_{sensor}$ in $[0 \mu H; 100 \mu H]$ (Fig.\ref{FIG:LsVersusFreqExprData100uH--C1-2200--C2-22}), then for $L_{sensor}$$>$100$\mu H$ the experimental $L_{sensor}$ has a shape with some visual similarity to theoretical (as Pierce oscillator), but with very different $L_{sensor}$ values. Also apparently for larger values of $L_{sensor}$ the theoretical (Pierce oscillator, JP.A on, JP.B off) $L_{s, theo}(f)$ could be approximated to the experimental data by a constant multiplicative factor ($L_{sensor} \approx 0.03 \cdot L_{s, theo}(f)$, for $L_{sensor}$$>$1mH), as visible in Fig.\ref{FIG:LsVersusFreq--C1-2200--C2-22}.\par
For modeling(data fitting) purposes, the author knows that a function $L_{s}(f)\text{=}(a+(b/(c+d{\cdot}f))){\cdot}(n{\cdot}f+m)$, where 'a,b,c,d,m,n' are constants to fit, can be fitted to experimental data on both low and high values of $L_s$.

On following section \ref{SectionAltApproxCircuitAnalysis} was used an approximated model (for Schmitt-trigger oscillator) applied to Multiple-Sensor Interface with inductive sensor (JP.A on, JP.B off), that exhibited a theoretical ${L_s}(f)$ plot much closer to the experimental data, corroborating the hypothesis that with jumper configuration JP.A on, JP.B off ($C_1$=2.2nF, $C_2$=21.78pF), it operates as a Schmitt-trigger oscillator, where $L_s$ acts as an impedance influencing $C_1$ charge and discharge speed.\par
On following section \ref{SectionInterfForInductiveSensorsSchTriOsc}, also about inductive sensor (JP.A on, JP.B off), was used a theoretical model(for Schmitt-trigger oscillator) with some rude/obscene approximations; that model was successful in predicting the correct shape of $L_{s}(f)$ although with a somewhat considerable displacement to the experimental data.\par

\subsubsection{Multiple-Sensor Interface for resistive sensors}
In case of connecting a resistive sensor (or capacitive) to the Multiple-Sensor Interface it will not be able to satisfy the conditions for oscillation consequent of the Barkhausen criterion applied to the circuit as Pierce oscillator; so the conclusion is when a resistive sensor (or capacitive) is connected, it no longer is a Pierce oscillator. The Multiple-Sensor Interface is made using Schmitt-trigger inverters(high-speed Si-gate CMOS, 74HC14), and the hysteresis of the Schmitt-trigger can be used to implement another type of oscillator, the relaxation oscillator. So in the case of a resistive sensor the circuit to analyze is a Schmitt-trigger inverter connected to a network of resistors and capacitors.\par
To analyze this circuit the Schmitt-trigger inverter was replaced by a theoretical switch that changes the voltage of node $v_o$ to $V_{DDS}$ (power supply stabilized voltage for the sensor interface) when voltage $v_i$ is lower than $V_T^{-}$, and changes $v_o$ to GND when voltage $v_i$ is higher than $V_T^{+}$.\par
\begin{figure}[b!]
	\centering
		\includegraphics[width=\columnwidth,keepaspectratio]{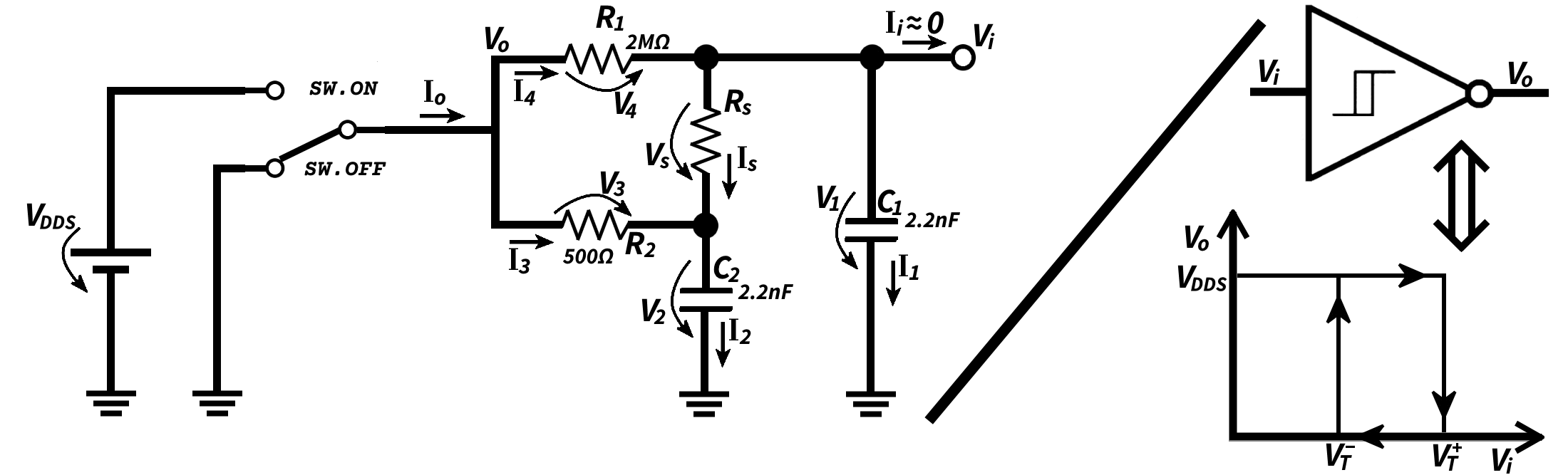}
	\caption{Multi-Sensor Int., resistive sensor(Schmitt-trigger osc.)}\label{FIG:schTheoryOscillatorResistiveSensor}
\end{figure} \par
So the circuit of Fig.\ref{FIG:schTheoryOscillatorResistiveSensor} is here analyzed to obtain $f(R_s)$, and then its inverse function ${R_s}(f){\approx}R_{sensor}$ that may be useful for using/configuring the Multiple-Sensor Interface. Notice that $i_{i}$$\approx$0 since $v_i$ is the input of a Schmitt-trigger inverter (high-speed Si-gate CMOS) that has a very high input impedance and so $i_{i}$$\approx$0 is an appropriate approximation simplifying the circuit. So from the circuit are obtained the equations: \qquad
Nodes and loops: \quad $i_4=i_s+i_1$,\newline $i_3+i_s=i_2$, \hspace{3mm} $i_o=i_3+i_4$, \hspace{3mm} $i_1+i_2=i_o$, \hspace{3mm} $i_1+i_2=i_3+i_4$,\newline $v_1-v_2-v_s=0$, \hspace{3mm} $v_4+v_s+v_2-v_o=0$, \hspace{3mm} $v_4+v_s-v_3=0$, \hspace{3mm} $v_3=v_o-v_2$, \hspace{3mm} $v_4=v_o-v_i$, \hspace{3mm} $v_2=v_i-v_s$ \quad .\par
Components: $i_1\text{=}{C_1}(dv_1/dt)$, \hspace{2mm} $i_2\text{=}{C_2}(dv_2/dt)$, \hspace{2mm} $v_3\text{=}R_2i_3$, \hspace{2mm} $v_4\text{=}R_1i_4$, \hspace{2mm} $v_s\text{=}R_si_s$ \quad .\par
Solving:
\begin{equation}\label{EQ:nodeI4IsI1OscillatorRsensor_1}\frac{v_o-v_i}{R_1}=\frac{v_s}{R_s}+{C_1}\frac{dv_i}{dt} \end{equation}
\begin{equation}\label{EQ:nodeI3IsI2OscillatorRsensor_1}\frac{v_o-v_i+v_s}{R_2}+\frac{v_s}{R_s}={C_2}\frac{dv_2}{dt} \end{equation}

Solving: $v_2=v_i-v_s \Rightarrow dv_2/dt=d(v_i-v_s)/dt \Rightarrow dv_2/dt=(dv_i/dt)-(dv_s/dt) $ \par
So using the previous result the (\ref{EQ:nodeI3IsI2OscillatorRsensor_1}) can be changed to:
\begin{equation}\label{EQ:nodeI3IsI2OscillatorRsensor_2}\frac{v_o-v_i}{R_2}+\left(\frac{1}{R_2}+\frac{1}{R_s}\right)v_s={C_2}\left(\frac{dv_i}{dt}-\frac{dv_s}{dt}\right) \end{equation}
Solving (\ref{EQ:nodeI4IsI1OscillatorRsensor_1}) for $v_s$ is obtained:
\begin{equation}\label{EQ:nodeI4IsI1OscillatorRsensor_2}v_s=\frac{R_s(v_o-v_i)}{R_1}-{R_sC_1}\frac{dv_i}{dt} \end{equation}
Calculating the derivative on both sides of (\ref{EQ:nodeI4IsI1OscillatorRsensor_2}) is obtained (remember $v_o$ is a constant equal to $V_{DDS}$ or GND depending on the position of the switch 'SW'):
\begin{equation}\label{EQ:nodeI4IsI1OscillatorRsensor_3}\frac{dv_s}{dt}=\frac{-R_s}{R_1}\frac{dv_i}{dt}-{R_sC_1}\frac{d^2 v_i}{dt^2} \end{equation}
Now using (\ref{EQ:nodeI4IsI1OscillatorRsensor_2}) and (\ref{EQ:nodeI4IsI1OscillatorRsensor_3}) to remove the variables $v_s$ and $dv_s/dt$ from equation (\ref{EQ:nodeI3IsI2OscillatorRsensor_2}) is obtained an equation solvable for determining $v_{i}(t)$:
\begin{multline}\label{EQ:diffEqOscillatorRsensor}\left( \frac{1}{R_2} + \frac{R_s}{{R_2}{R_1}} + \frac{1}{R_1} \right) ( v_o - v_i ) = \\ \hspace{5mm} \left( C_1 \left( 1 + \frac{R_s}{R_2} \right) + C_2 \left( 1 + \frac{R_s}{R_1} \right) \right)\frac{dv_i}{dt} + {R_s}{C_1}{C_2}\frac{d^2 v_i}{dt^2}\end{multline}

The equation (\ref{EQ:diffEqOscillatorRsensor}) is of the type: $c(v_o-v_i)=b(dv_i/dt)+a(d^2v_i/dt^2)$ that has the general solution: $v_i(t)=v_o+k_1\emph{e}^{{\lambda_1}t}+k_2\emph{e}^{{\lambda_2}t}$, where $k_1$,$k_2$ are integration constants to be defined by 'initial conditions' and  $\lambda_1$, $\lambda_2$ are defined by: $a{\lambda^2}+b\lambda+c=0 \Leftrightarrow \lambda=\frac{-b\pm\sqrt{{b^2}-4ac}}{2a}$ and $\emph{e}$ is the Euler-Napier constant $\emph{e} = \sum_{n=0}^{\infty} (1/(n!))$.\par
So defining here $a_r$, $b_r$, $c_r$ as the values of a, b, c for obtaining ${v_i}(t)$ when using a resistive sensor, as:\par
${a_r}={R_s}{C_1}{C_2}$, \par
${b_r}=({C_1}(1+({R_s}/{R_2})))+({C_2}(1+({R_s}/{R_1})))$, \par
${c_r}=(1/{R_2})+({R_s}/({R_2}{R_1}))+(1/{R_1})$ .\par

In order to obtain ${v_i}(t)$ for this circuit is required to calculate $k_1$ and $k_2$, that are constants to be defined by 'initial conditions', the value of these constants is related to the voltage (or electrical charge) on capacitors $C_1$ and $C_2$ at the moment the inverter gate changes its output voltage (high to low, or low to high), on the model used for analyzing the circuit that is when the 'theoretical switch' $v_o$ changes state.
Also, is only known the value of $v_i$ (to be $V_T^{-}$ or $V_T^{+}$) when the inverter gate $v_o$ changes value, so is very difficult to calculate both $k_1$ and $k_2$ by algebraic manipulation.
Admitting that $R_1 \gg R_2$ and that $C_1 \geqslant C_2$  (that is the case of the circuit that was studied and tested where $R_{1}\text{=}2M\Omega$ and $R_{2}\text{=}500\Omega$), then is known that the capacitor $C_2$ will charge faster than $C_1$ for all values of $R_s$, and in case $R_s$ has an impedance comparable to $R_2$ then $C_2$ will charge much faster than $C_1$. So the voltage (and electrical charge) of $C_2$ follows closely the values of $v_o$ and so will be of small relevance to the initial conditions of the circuit, thus $v_i(t)=v_o+k_1\emph{e}^{{\lambda_1}t}+k_2\emph{e}^{{\lambda_2}t}$ can be approximated as single exponential function $v_i(t){\approx}v_o+k\emph{e}^{{\lambda}t}$, this approximation is implied on the following calculations.\par

Is selected the solution $\lambda_{2} \text{=} \frac{-b+\sqrt{{b^2}-4ac}}{2a}$ by setting $k_1\text{=}0$, because is the one that provides an adequate value for $v_i(t)$, $f(R_s)$, consistent with experimental data, however for obtaining the (approximate) function ${R_s}(f)$ any may be used.\par
For convenience of a $v_i(t)$ more similar to typical RC circuits is defined ${\tau=-1/\lambda}$, and so ${v_i(t)=v_o+k_2\emph{e}^{-t/{\tau_2}} }$ (or more appropriately: ${v_i(t){\approx}v_o+k_2\emph{e}^{-t/{\tau_2}} }$).\par
\vspace{2mm}
Charging time of $C_1$: \quad $v_o={V_{DDS}}$ \newline
\noindent $v_i(t\text{=}0)\text{=}{V_T^{-}} \rightarrow {V_T^{-}}={V_{DDS}}+k_2\emph{e}^{0} \rightarrow k_2={V_T^{-}}-{V_{DDS}}$ \newline
\noindent $v_i(t\text{=}T_C)\text{=}{V_T^{+}} \rightarrow V_T^{+}={V_{DDS}}+k_2\emph{e}^{{-T_C}/{\tau_2}} $   $\quad \rightarrow \newline 
\noindent \phantom{'} \hfill \rightarrow \quad T_C=-\tau_2 ln((V_T^{+}-{V_{DDS}})/({V_T^{-}}-{V_{DDS}}))$ \par
\vspace{2mm}
Discharging time of $C_1$: \quad $v_o=0$ \newline
\noindent $v_i(t\text{=}0)\text{=}{V_T^{+}} \rightarrow {V_T^{+}}=0+k_2\emph{e}^{0} \quad \rightarrow \quad k_2=V_T^{+}$ \newline
\noindent $v_i(t\text{=}T_D)\text{=}{V_T^{-}} \rightarrow {V_T^{-}}=0+k_2\emph{e}^{{-T_D}/{\tau_2}} \rightarrow$\newline 
\noindent \phantom{'} \hfill $ \quad \rightarrow \quad T_D=-\tau_2 ln({V_T^{-}}/{V_T^{+}})$\par
\vspace{2mm}
The time for a complete cycle of charge and discharge of $C_1$ is: $T=T_C+T_D$; the frequency of $v_i(t)$ is $f=1/T$.\par
Solving: $T=-\tau_2 \left ( ln \left ( \frac{V_T^{+}-{V_{DDS}}}{{V_T^{-}}-{V_{DDS}}} \right ) + ln \left ( \frac{V_T^{-}}{V_T^{+}} \right ) \right )$   $\qquad \Leftrightarrow \qquad T=\tau_2 ln \left ( \frac{({V_T^{-}}-{V_{DDS}}){V_T^{+}}}{({V_T^{+}}-{V_{DDS}}){V_T^{-}}} \right ) $ \par

For convenience defining the constant 'H' by: \newline
\noindent \phantom{'} \hfill $\qquad H = ln \left ( \frac{({V_T^{-}}-{V_{DDS}}){V_T^{+}}}{({V_T^{+}}-{V_{DDS}}){V_T^{-}}} \right )$, \qquad \newline
\noindent then $f=1/T \Leftrightarrow f = 1/({\tau_2} H) \Leftrightarrow f = -{\lambda_2} / H$ . \par
So the expression (theoretical) of an approximate value for resistance ($R_{s}$) as a function of frequency(f) is:\newline
\begin{equation}\label{EQ:RSensorVersusFrequencyOscillationSchmittTrigger}R_{sensor} {\approx} R_s = \frac{(C_1+C_2){R_2}{R_1}Hf-R_2-R_1}{({C_2}{R_2}Hf-1)({C_1}{R_1}Hf-1)}\end{equation}

\begin{figure}[H]
	\centering
		\includegraphics[width=\columnwidth,keepaspectratio]{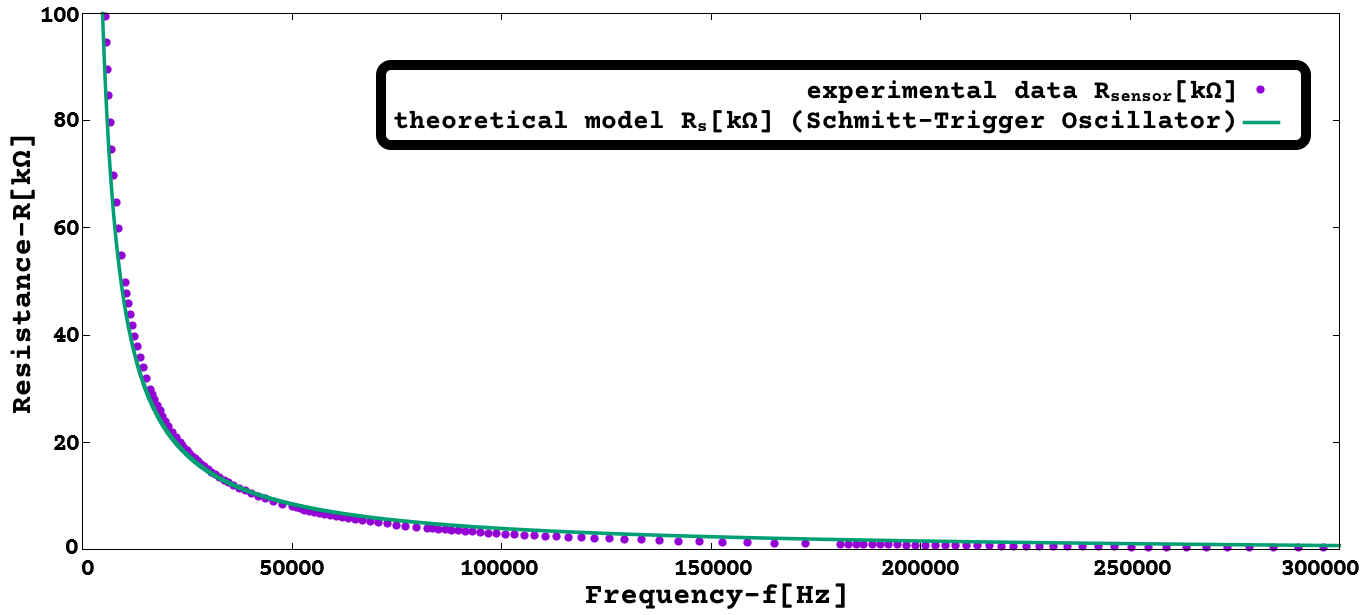}
	\caption{${R_s}(f)$ [k$\Omega$], frequency in [0Hz, 300kHz] (Schmitt-trigger oscillator, Multiple-Sensor Interface with resistive sensor).}\label{FIG:RSensorVersusFrequencyOscillationSchmittTrigger2}
\end{figure}

Using the values $C_1\text{=}C_2\text{=}2.2nF$, $R_2\text{=}500\Omega$, $R_1\text{=}2M\Omega$, ${V_T^{-}}\text{=}1.2V$, ${V_T^{+}}\text{=}2.2V$, ${V_{DDS}}\text{=}4.18V$, is obtained $H\text{=}1.01496$, Fig.\ref{FIG:RSensorVersusFrequencyOscillationSchmittTrigger2} shows experimental data for Multiple-Sensor Interface with various resistance values connected as the sensor and also shows $R_{s}(f)$ using (\ref{EQ:RSensorVersusFrequencyOscillationSchmittTrigger}) with the mentioned values of $C_1$, $C_2$, $R_2$, $R_1$, $H$.

\subsubsection{Multiple-Sensor Interface for capacitive sensors}
In case of connecting a capacitive sensor (or resistive) to the Multiple-Sensor Interface it will not be able to satisfy the conditions for oscillation consequent of the Barkhausen criterion applied to the circuit as Pierce oscillator; and so it is again a Schmitt-trigger relaxation oscillator. To analyze this circuit the Schmitt-trigger inverter was replaced by a theoretical switch (Schmitt-trigger), just like previously with resistive sensors.\par
\begin{figure}[ht!]
	\centering
		\includegraphics[width=\columnwidth,keepaspectratio]{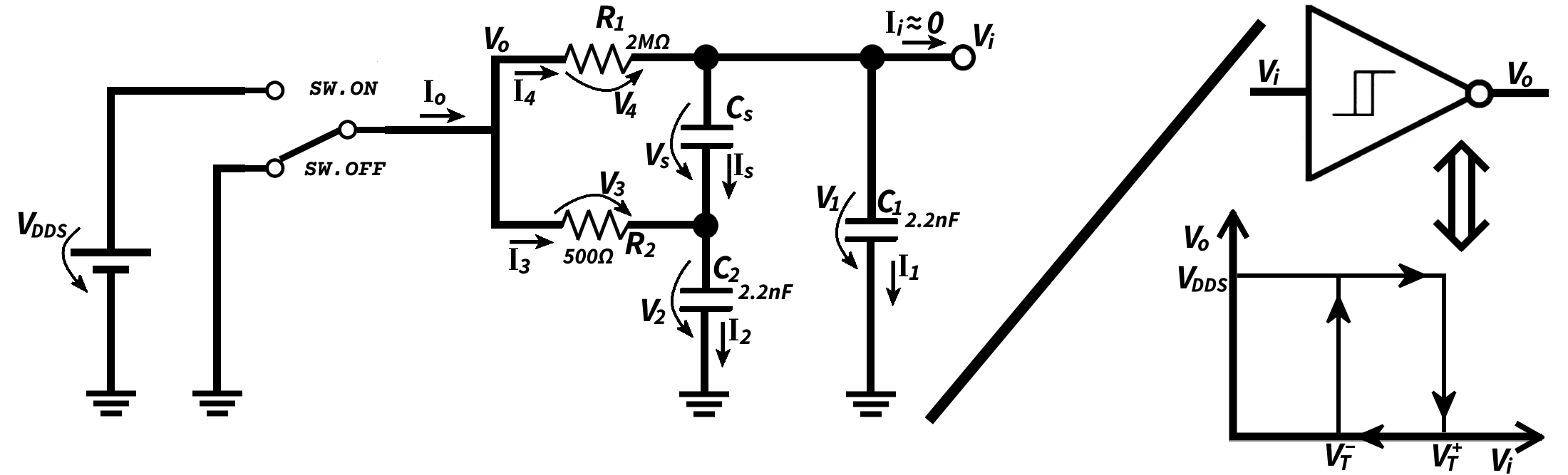}
	\caption{Multi-Sensor Int., capacitive sensor (Schmitt-trigger osc.)}\label{FIG:schTheoryOscillatorCapacitiveSensor}
\end{figure} \par

So the circuit of Fig.\ref{FIG:schTheoryOscillatorCapacitiveSensor} is here analyzed to obtain $f(C_s)$, and then its inverse function ${C_s}(f){\approx}C_{sensor}$ that is useful for using/configuring the Multiple-Sensor Interface. Notice that $i_{i}$$\approx$0 since $v_i$ is the input of the Schmitt-trigger inverter(high-speed Si-gate CMOS) that has a very high input impedance and so $i_{i}$$\approx$0 is an appropriate approximation simplifying the circuit. So from the circuit are obtained the equations: \par
Nodes and loops: \quad $i_4=i_s+i_1$, \hspace{3mm} $i_3+i_s=i_2$, \newline $i_o=i_3+i_4$, \hspace{3mm} $i_1+i_2=i_o$, \hspace{3mm} $i_1+i_2=i_3+i_4$,\newline $v_1-v_2-v_s=0$, \hspace{3mm} $v_4+v_s+v_2-v_o=0$, \hspace{3mm} $v_4+v_s-v_3=0$, \hspace{3mm} $v_3=v_o-v_2$, \hspace{3mm} $v_4=v_o-v_i$, \hspace{3mm} $v_2=v_i-v_s$ \quad .\par
Components: $i_1\text{=}{C_1}(dv_1/dt)$, \hspace{2mm} $i_2\text{=}{C_2}(dv_2/dt)$, \hspace{2mm} $v_3\text{=}R_2i_3$, \hspace{2mm} $v_4\text{=}R_1i_4$, \hspace{2mm} $i_s\text{=}{C_s}(dv_s/dt)$ .\par
Solving:
\begin{equation}\label{EQ:nodeI4IsI1OscillatorCsensor_1}\frac{v_o-v_i}{R_1}={C_s}\frac{dv_s}{dt}+{C_1}\frac{dv_i}{dt} \end{equation}
\begin{equation}\label{EQ:nodeI3IsI2OscillatorCsensor_1}\frac{v_o-v_2}{R_2}+{C_s}\frac{dv_s}{dt}={C_2}\frac{dv_2}{dt} \end{equation}
\begin{equation}\label{EQ:nodeI1I2I3I4OscillatorCsensor_1}\frac{v_o-v_2}{R_2}+\frac{v_o-v_i}{R_1}={C_1}\frac{dv_i}{dt}+{C_2}\frac{dv_2}{dt} \end{equation}

Since $v_s=v_i-v_2$ then $dv_s/dt=(dv_i/dt)-(dv_2/dt)$, and so using it on equation (\ref{EQ:nodeI4IsI1OscillatorCsensor_1}), is obtained:
\begin{equation}\label{EQ:nodeI4IsI1OscillatorCsensor_2}\frac{dv_2}{dt}=\left ( 1+\frac{C_1}{C_s} \right )\frac{dv_i}{dt}-\frac{v_o-v_i}{C_sR_1}\end{equation}

Using $dv_s/dt=(dv_i/dt)-(dv_2/dt)$ and (\ref{EQ:nodeI4IsI1OscillatorCsensor_2}) on equation (\ref{EQ:nodeI3IsI2OscillatorCsensor_1}), is obtained:
\begin{equation}\label{EQ:nodeI3IsI2OscillatorCsensor_2}\resizebox{\hsize}{!}{$v_2=v_o+\frac{R_2(C_2+C_s)(v_o-v_i)}{C_sR_1} + R_2C_s \left ( 1 - \left (1+\frac{C_2}{C_s} \right ) \left (1+\frac{C_1}{C_s} \right ) \right )\frac{dv_i}{dt}$}\end{equation}

Calculating the derivative of (\ref{EQ:nodeI3IsI2OscillatorCsensor_2}) is obtained:
\begin{equation}\label{EQ:nodeI3IsI2OscillatorCsensor_3}\resizebox{\hsize}{!}{$\frac{d(v_2)}{dt}=-\frac{R_2(C_2+C_s)}{C_sR_1}\frac{dv_i}{dt}+ R_2C_s \left ( 1 - \left (1+\frac{C_2}{C_s} \right ) \left (1+\frac{C_1}{C_s} \right ) \right )\frac{d^2v_i}{dt^2}$}\end{equation}

Now using (\ref{EQ:nodeI3IsI2OscillatorCsensor_2}) and (\ref{EQ:nodeI3IsI2OscillatorCsensor_3}) to remove the variables $v_2$ and $dv_2/dt$ from the equation (\ref{EQ:nodeI1I2I3I4OscillatorCsensor_1}), is obtained an equation solvable for determining $v_i(t)$:
\begin{equation}\label{EQ:diffEqOscillatorCsensor}\resizebox{\hsize}{!}{$v_o - v_i = (R_2(C_2+C_s)+R_1(C_1+C_s))\frac{dv_i}{dt} + {R_2}{R_1}({C_1}{C_2}+C_s(C_1+C_2))\frac{d^2 v_i}{dt^2}$}\end{equation}

The equation (\ref{EQ:diffEqOscillatorCsensor}) is of the type: $c(v_o-v_i)=b(dv_i/dt)+a(d^2v_i/dt^2)$, that has the general solution: \newline
\noindent $v_i(t)=v_o+k_1\emph{e}^{{\lambda_1}t}+k_2\emph{e}^{{\lambda_2}t}$, where $k_1$,$k_2$ are integration constants to be defined by 'initial conditions' and  $\lambda_1$, $\lambda_2$ are defined by: $a{\lambda^2}+b\lambda+c=0 \Leftrightarrow \lambda=\frac{-b\pm\sqrt{{b^2}-4ac}}{2a}$ and $\emph{e}$ is the Euler-Napier constant $\emph{e} = \sum_{n=0}^{\infty} (1/(n!))$ . \par
So defining here $a_c$, $b_c$, $c_c$ as the values of a, b, c for obtaining ${v_i}(t)$ when using a capacitive sensor, as: ${c_c}=1$, \par
${b_c}=({R_2}(C_2+C_s)+{R_1}(C_1+C_s))$, \par
${a_c}={R_2}{R_1}({C_1}{C_2}+{C_s}(C_1+C_2))$ .\par

In order to obtain ${v_i}(t)$ for this circuit is required to calculate $k_1$ and $k_2$, to be defined by 'initial conditions'; notice that this circuit has 3 capacitors that store electrical charge defining 'initial conditions', but the location of $C_s$ connected between $C_1$ and $C_2$ implies that the voltage of $C_s$ (or its electrical charge) is completely defined/known by the voltages (or electrical charges) on $C_1$ and $C_2$. \par
Here are made the same approximation/simplification of $v_i(t){\approx}v_o+k\emph{e}^{{\lambda}t}$, as explained and done on the case of resistive sensor.

Is selected the solution $\lambda_{2} \text{=} \frac{-b+\sqrt{{b^2}-4ac}}{2a}$ by setting $k_1\text{=}0$, because is the one that provides an adequate value for $v_i(t)$, $f(C_s)$, consistent with experimental data, however for obtaining the function ${C_s}(f)$ any may be used.\par
For convenience of a $v_i(t)$ more similar to typical RC circuits is defined ${\tau=-1/\lambda}$, and so ${v_i(t)=v_o+k_2\emph{e}^{-t/{\tau_2}} }$ (or more appropriately: ${v_i(t){\approx}v_o+k_2\emph{e}^{-t/{\tau_2}} }$).\par
So when is connected a capacitive sensor ($C_s$) the differential equation and solution $v_i(t)$ are the same as when is connected a resistive sensor ($R_s$), the only differences are in the values of $a$, $b$, $c$; and as such the equations of $f$(frequency) and $T$(period) are also the same and are reused from the previous section.\par
The constant 'H' defined by:\newline
\noindent \phantom{'} \hfill $\qquad H=ln \left ( \frac{({V_T^{-}}-{V_{DDS}}){V_T^{+}}}{({V_T^{+}}-{V_{DDS}}){V_T^{-}}} \right )$, \qquad \qquad \newline
\noindent and $f=1/T \Leftrightarrow f = 1/({\tau_2} H) \Leftrightarrow f = -{\lambda_2} / H$ . \par

So the expression (theoretical) of an approximate value for capacitance ($C_{s}$) as a function of frequency(f) is:
\begin{equation}\label{EQ:CSensorVersusFrequencyOscillationSchmittTrigger}C_{sensor} {\approx} C_s = \frac{({C_1}{R_1}+{C_2}{R_2})Hf-1-{C_1}{C_2}{R_1}{R_2}{H^2}{f^2}}{Hf((C_1+C_2){R_1}{R_2}Hf-R_1-R_2)}\end{equation} \par
\begin{figure}[ht!]
	\centering
		\includegraphics[width=\columnwidth,keepaspectratio]{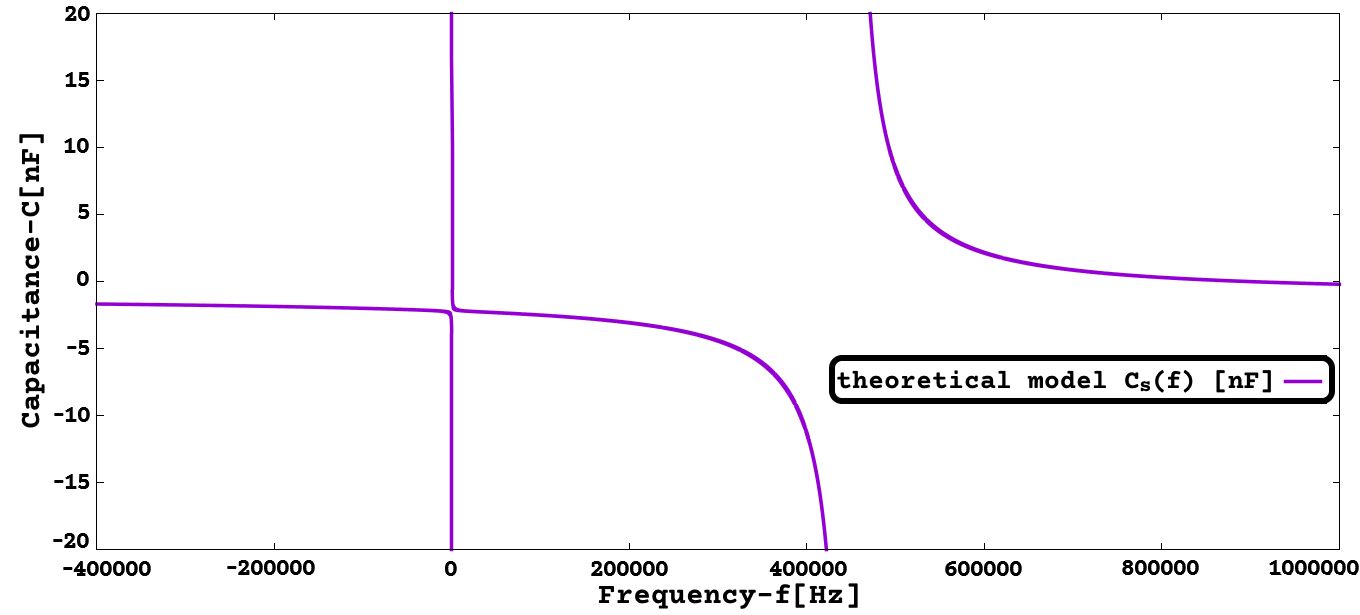}
	\caption{${C_s}(f)$ [nF], frequency in [-400kHz, 1MHz] (Schmitt-trigger oscillator, Multiple-Sensor Interface with capacitive sensor).}\label{FIG:CSensorVersusFrequencyOscillationSchmittTriggerTheoryEq}
\end{figure}

Using the values $C_1\text{=}C_2\text{=}2.2nF$, $R_2\text{=}500\Omega$, $R_1\text{=}2M\Omega$, ${V_T^{-}}\text{=}1.2V$, ${V_T^{+}}\text{=}2.2V$, ${V_{DDS}}\text{=}4.18V$, is obtained $H\text{=}1.01496$, Fig.\ref{FIG:CSensorVersusFrequencyOscillationSchmittTriggerTheoryEq} shows the plot of $C_{s}(f)$ using (\ref{EQ:CSensorVersusFrequencyOscillationSchmittTrigger}) with the mentioned values of $C_1$, $C_2$, $R_2$, $R_1$, $H$.

Analyzing the plot in Fig.\ref{FIG:CSensorVersusFrequencyOscillationSchmittTriggerTheoryEq} by firstly looking at plot regions with $C_s$$>$0, its visible that the ${C_s}(f)$ plot is close to being over the vertical axis (f=0) and this would mean that for all values of $C_s$ the frequency would be close to zero ($f{\approx}0$), but visible to the right is another curve that is also placed on the area of $C_s$$>$0 ( for frequency [447957Hz, 895689Hz] ) and at a first view this curve would seem appropriate. But strangely on $C_s$$>$0, f$>$0 for each value of $C_s$ are 2 values of frequency, while for $C_s$$<$0, f$>$0 each value of $C_s$ has only one possible value of frequency (f$<$0 is considered meaningless/ignored).\par
\begin{figure}[ht!]
	\centering
		\includegraphics[width=\columnwidth,keepaspectratio]{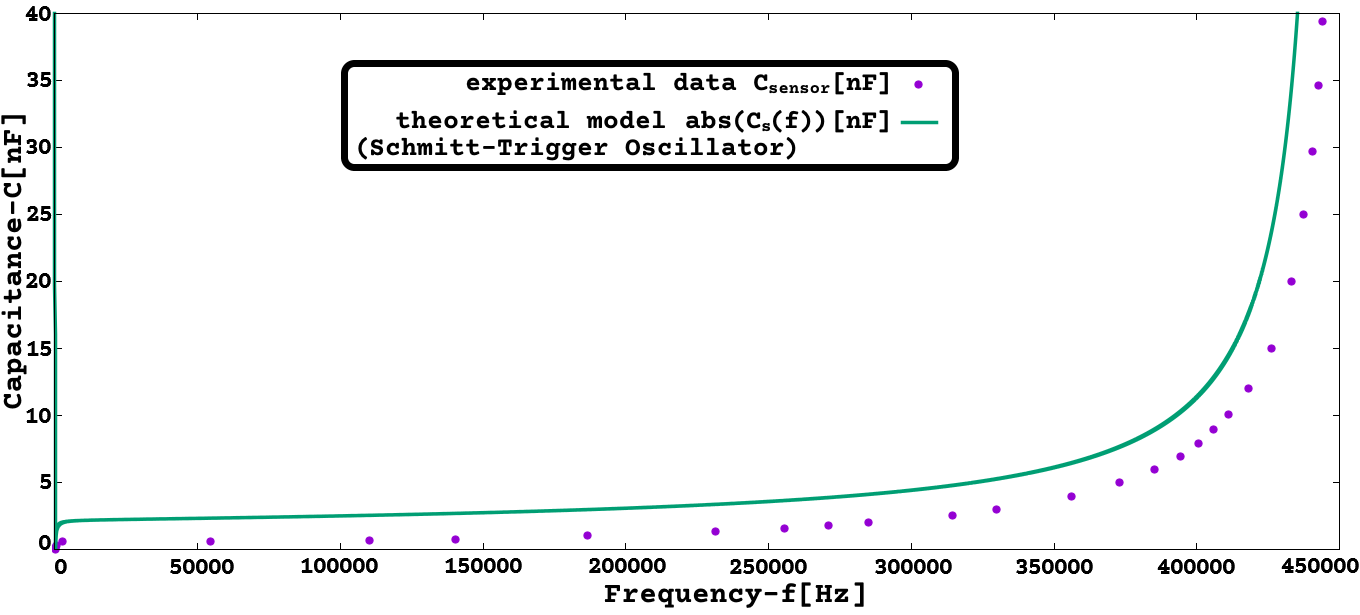}
	\caption{$|{C_s}(f)|$ [nF], frequency in [0Hz, 450kHz] (Schmitt-trigger osc., Multi-Sensor with capacitive sensor)}\label{FIG:CSensorVersusFrequencyOscillationSchmittTrigger1}

	\vspace{2mm}

	\centering
		\includegraphics[width=\columnwidth,keepaspectratio]{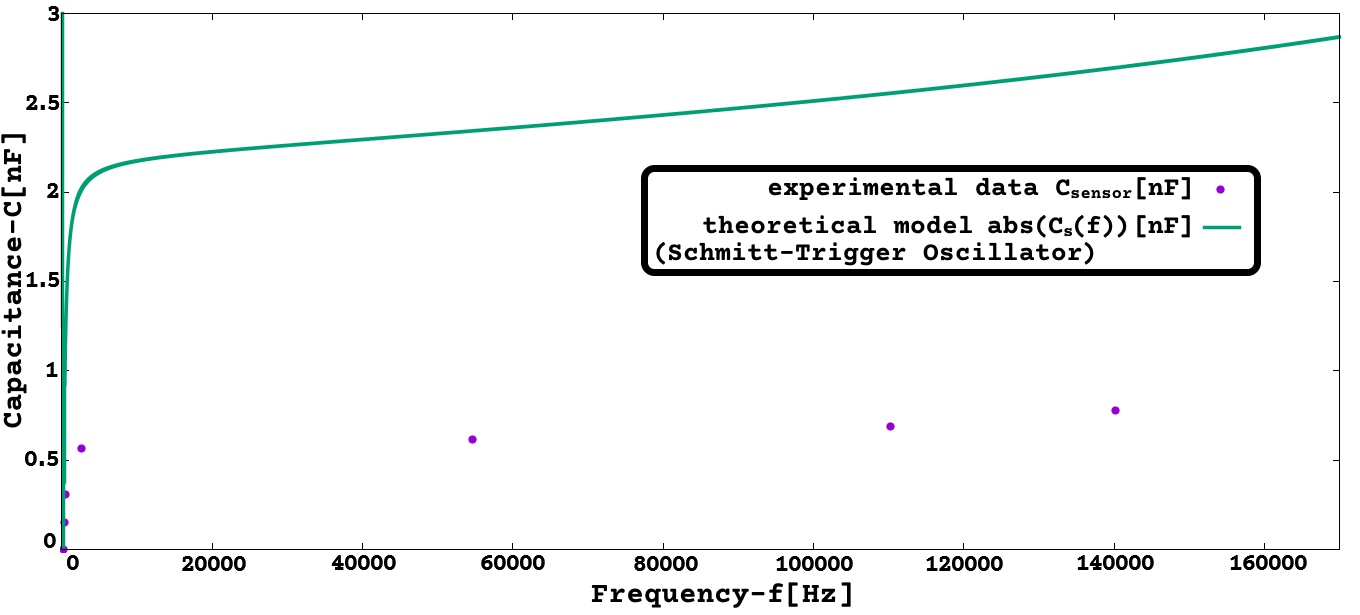}
	\caption{$|{C_s}(f)|$ [nF], frequency in [0Hz, 170kHz] (Schmitt-trigger osc., Multi-Sensor with capacitive sensor)}\label{FIG:CSensorVersusFrequencyOscillationSchmittTrigger2}
\end{figure}

The experimental data shows that the way the oscillator works using a capacitive sensor is different from what some would expect on a first view of the plot ${C_s}(f)$, in order to compare the experimental data with the theoretical model is shown in Fig.\ref{FIG:CSensorVersusFrequencyOscillationSchmittTrigger1} and Fig.\ref{FIG:CSensorVersusFrequencyOscillationSchmittTrigger2} the experimental data for Multiple-Sensor Interface with various capacitance values connected as the sensor and also the plot of $abs( {C_s}(f) )$ ($=|{C_s}(f)|$) using (\ref{EQ:CSensorVersusFrequencyOscillationSchmittTrigger}) with the mentioned values of $C_1$, $C_2$, $R_1$, $R_2$, $H$. \quad So it seems that the obtained function of ${C_s}(f)$ although strangely indicates negative values for the sensor capacitance it can provide a theoretical curve/plot similar to what was obtained on the experimental data for $C_{sensor}$. On the following sections is given a better insight on why ${C_s}(f)$ has a negative value.

\subsubsection{Multiple-Sensor Int. for measuring frequency}
For measuring frequency of an external voltage signal (between 0V and $V_{DDS}$, so preferentially a digital signal or in case of analog signal it should be limited/trimmed before) is possible to use the mentioned Multiple-Sensor Interface and so using the same port/connector of the device. For this the user should remove/open the jumpers "JP.A", "JP.B" making the capacitors C1-A, C2-B active on the circuit, this will make $C_1=C_2=21.8pF$ that is a quite low capacitance that will have an insignificant effect on the external voltage signal. The external voltage signal should be connected to the 1st pin of the sensor channel that is the one connected directly to the input of the Schmitt-trigger inverter, so that the inverter is directly driven by the external voltage signal, then the Multiple-Sensor Interface is just a converter of the voltage signal to a square wave signal where its frequency will be measured through the counter/timer of the PIC18F2550.\par
The external voltage signal would preferentially be from a sensor with a square wave output, and the sensor have its power supplied by one of the $V_{DDS}$,GND ports/connectors of the sensor interface device or by an external connection to the same power supply used to power the device.

\subsection{Alternative Approximate Circuit Analysis}\label{SectionAltApproxCircuitAnalysis}

\subsubsection{Sensor interface circuit simplified}

The Multiple-Sensor Interface circuit when working as Schmitt-trigger oscillator (using $R_s$, $C_s$, or $L_s$ with specific $C_1$,$C_2$ values) can be studied and understood in a more intuitive way by making some simplification/approximation that may be inaccurate for quantitative purposes but still captures its essence, with the benefit of exposing how it works and resulting in much simpler differential equations. So the interface circuit is a more complex Schmitt-trigger oscillator, but its essence is the same, it is just some capacitors being charged by currents that pass through some resistors, and the voltage on a capacitor($v_i$) will trigger(at $V_T^{-}$ or $V_T^{+}$) a switch(electronic inverter) to change the voltage($v_o$) \cite{howSchmittTriggerOscWork} . \par

\begin{figure}[ht!]
	\centering
		\includegraphics[width=0.5\columnwidth,keepaspectratio]{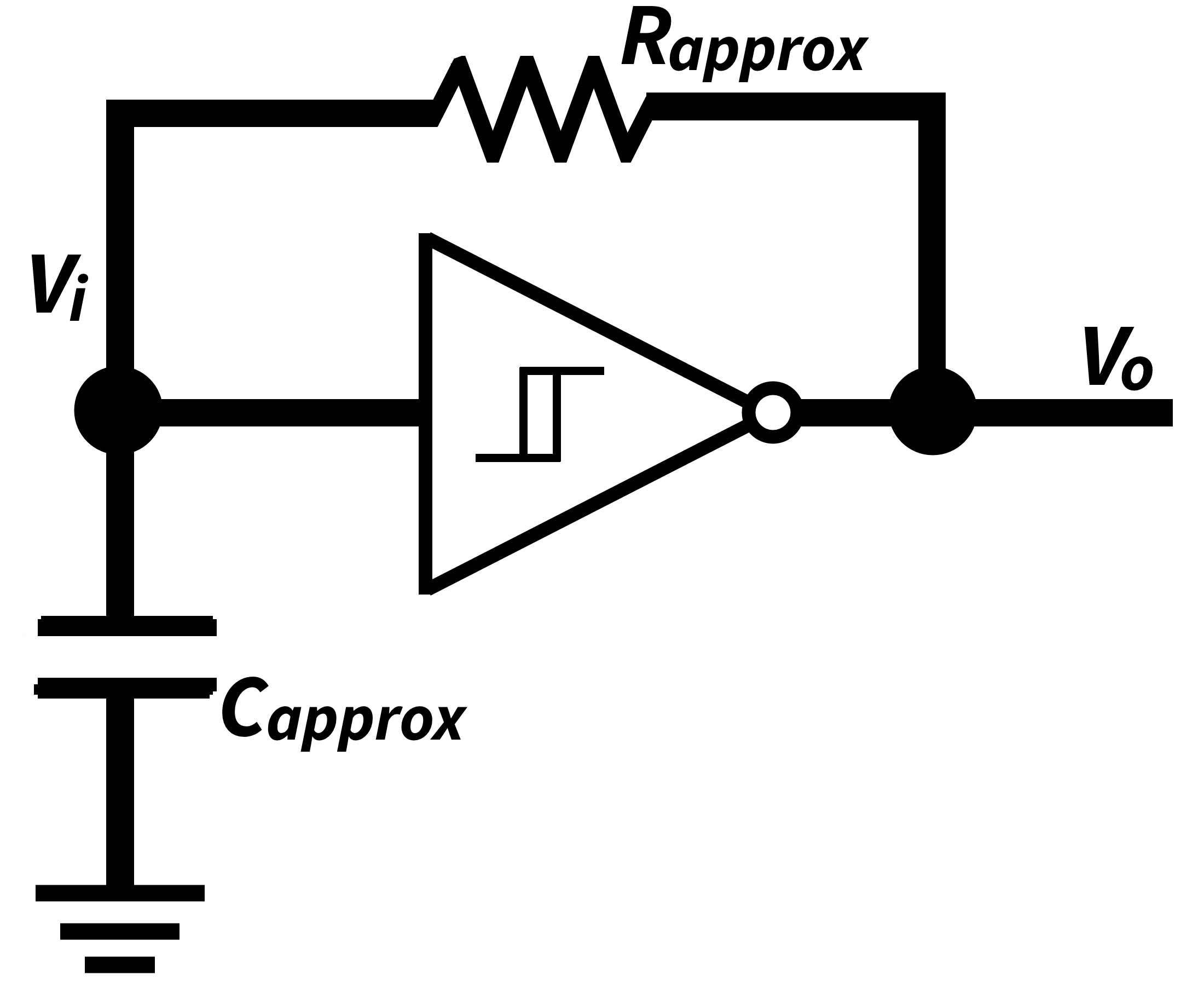}
	\caption{Schematic of a basic Schmitt-trigger oscillator to be used as an approximation of the circuit of Multiple-Sensor Interface}\label{FIG:BasicApproxOscillationSchmittTrigger}
\end{figure}

So the sensor and interface circuit can be described approximately as a basic Schmitt-trigger oscillator that only has one capacitor and one resistor (that determine the frequency of oscillation), and so was used the simplified circuit in Fig.\ref{FIG:BasicApproxOscillationSchmittTrigger} where $C_{approx}$ is a capacitor and $R_{approx}$ is a resistor that approximate in overall the capacitance and resistance of the sensor interface oscillator. 

 \par

\begin{figure}[ht!]
	\centering
		\includegraphics[width=\columnwidth,keepaspectratio]{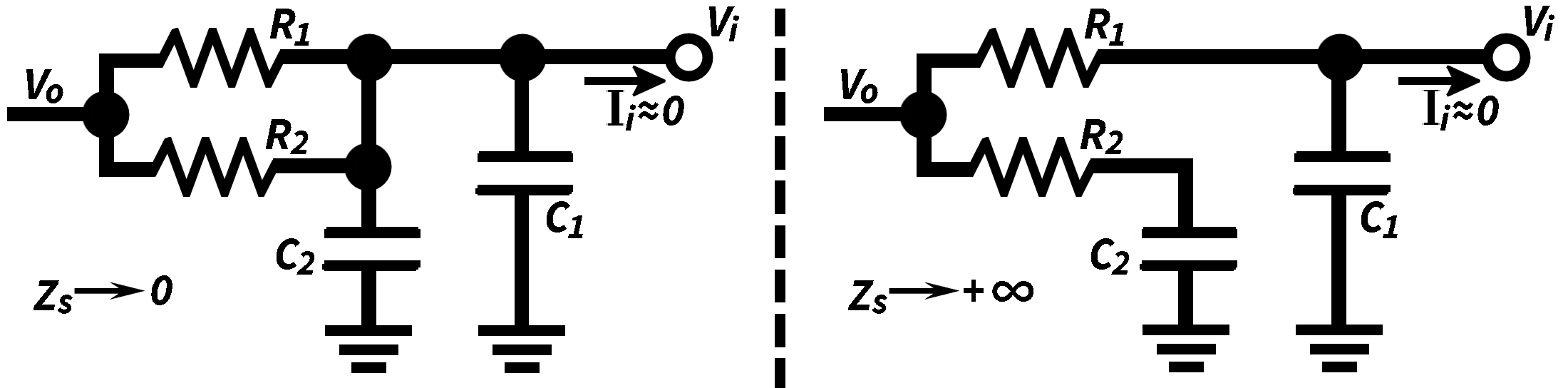}
	\caption{Schematic of the RC network of the Schmitt-trigger oscillator for the 2 extreme cases of sensor impedance ($Z_s$).}\label{FIG:extreme_cases_Zs_RC_OscillatorSchmittTrigger}
\end{figure}

To build expressions of $C_{approx}$ and $R_{approx}$ that include $R_1$, $R_2$, $C_1$, $C_2$ are considered initially 2 extreme cases of the sensor impedance($Z_s$): 1st $|Z_s|\text{=}0$ the sensor can be replaced by a wire, and 2nd $|Z_s|\text{=}\infty$ the sensor can be removed(open circuit), these 2 extreme cases possible for the sensor impedance are represented in Fig.\ref{FIG:extreme_cases_Zs_RC_OscillatorSchmittTrigger} . 

Now the sensor can be described as an electric connection that can be weakened or intensified depending on the sensor impedance, so when $|Z_s|$ changes progressively from 0 to $+\infty$ the circuit behavior changes progressively and smoothly from the behavior  of the left circuit to the behavior of right circuit of Fig.\ref{FIG:extreme_cases_Zs_RC_OscillatorSchmittTrigger}. So to obtain equations for $R_{approx}$ and $C_{approx}$ was selected an expression that allows to change smoothly the resistance and capacitance of the left side circuit to the resistance and capacitance of the right side circuit of Fig.\ref{FIG:extreme_cases_Zs_RC_OscillatorSchmittTrigger}.\par

So as in Fig.\ref{FIG:extreme_cases_Zs_RC_OscillatorSchmittTrigger}, here are the values of $C_{approx}$ and $R_{approx}$ for the 2 extreme values of $|Z_s|\text{=}0$ and $|Z_s|\text{=}\text{+}\infty$:  \newline
\noindent $C_{approx}(Z_s\text{=}0)\text{=}C_1\text{+}C_2$; \hfill $C_{approx}(Z_s\text{=}\infty)\text{=}C_1$; \hspace{2mm} \newline
$R_{approx}(Z_s\text{=}0)\text{=}({R_1}{R_2})/({R_1}\text{+}{R_2})$; \hfill $R_{approx}(Z_s\text{=}\infty)\text{=}R_1$;

\subsubsection{\texorpdfstring{$R_{approx}$ and $C_{approx}$ for a resisitve sensor ($R_s$)}{Rapprox and Capprox for a resisitve sensor Rs} }
Here are functions modeled to describe $C_{approx}$ and $R_{approx}$ (with resistive sensor) with a smooth transition from its values at $|Z_s|\text{=}0$ and $|Z_s|\text{=}+\infty$, where $|Z_s|\text{=}R_s$:
\vspace{-2mm}
\begin{equation}\label{EQ:CapproxOscillationSchmittTriggerWithRs}C_{approx} = (C_1+C_2)\frac{R_1}{|Z_s|+R_1}+C_1\frac{|Z_s|}{|Z_s|+R_1}\end{equation}
\vspace{-2mm}
\begin{equation}\label{EQ:RapproxOscillationSchmittTriggerWithRs}R_{approx} = \frac{{R_1}{R_2}}{(R_1+R_2)}\frac{2R_1}{(|Z_s|+2R_1)}+R_1\frac{|Z_s|}{|Z_s|+2R_1}\end{equation}
\phantom{.}\par
Using the equations: $f \text{=} 1/T \Leftrightarrow f \text{=} 1/(\tau H) \Leftrightarrow f \text{=} \text{-} \lambda / H$, and using $\tau \text{=} R_{approx}C_{approx}$, where $|Z_s|$ was removed using $|Z_s| \text{=} R_{s}$, it can be obtained $R_{s}(f)$.\par
Using values ${V_T^{-}}\text{=}1.2V$, ${V_T^{+}}\text{=}2.2V$, ${V_{DDS}}\text{=}4.18V$, is obtained $H\text{=}1.01496$ (valid for any type of sensor).\par
\begin{figure}[H]
	\centering
		\includegraphics[width=\columnwidth,keepaspectratio]{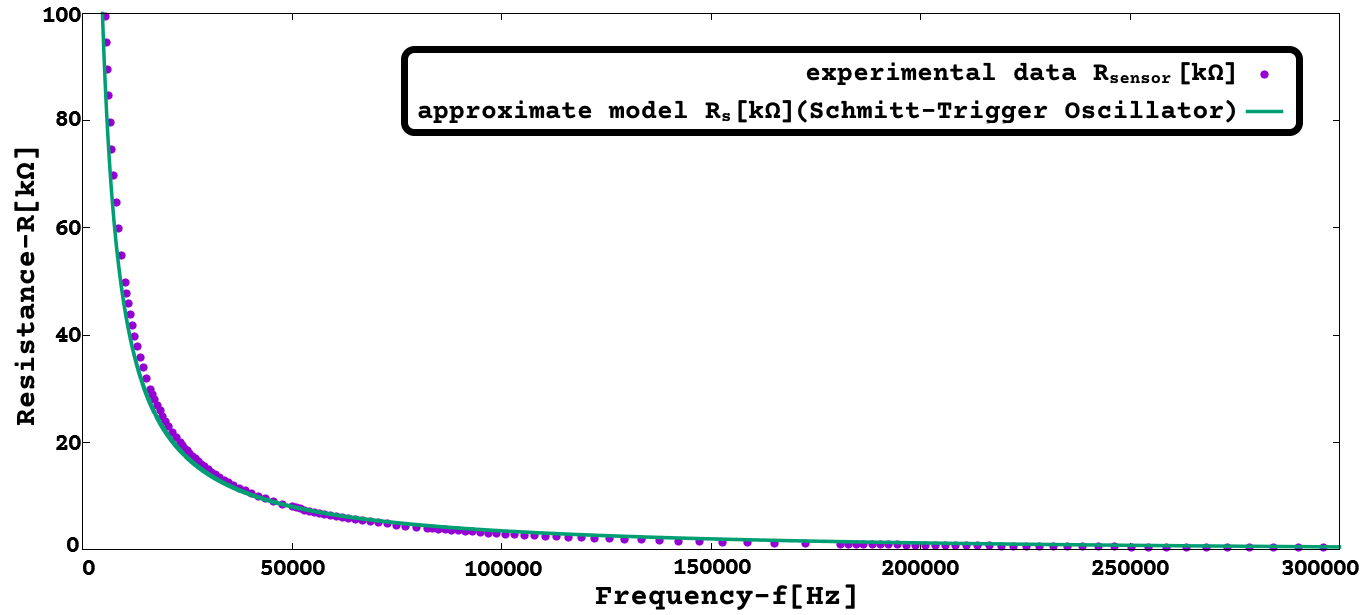}
	\caption{${R_s}(f)$ [k$\Omega$] by approximate model ($C_{approx}$, $R_{approx}$; Multi-Sensor Int. with resistive sensor).}\label{FIG:RsVersusFrequencyApproxModelSchmittTrigger2}
\end{figure}

Using the values $C_1\text{=}C_2\text{=}2.2nF$, $R_2\text{=}500\Omega$, $R_1\text{=}2M\Omega$, $H\text{=}1.01496$ with the approximate model ($C_{approx}$, $R_{approx}$), is obtained the plot of ${R_s}(f)$ in Fig.\ref{FIG:RsVersusFrequencyApproxModelSchmittTrigger2}.

\subsubsection{\texorpdfstring{$R_{approx}$ and $C_{approx}$ for an inductive sensor ($L_s$)}{Rapprox and Capprox for an inductive sensor Ls} }
This approximate model for the interface circuit with inductive sensor is only valid for jumper configuration(capacitor values) that make it work as Schmitt-trigger oscillator, as is expected for JP.A closed, JP.B open ($C_1$=2.2nF, $C_2$=21.78pF).
Here are functions modeled to describe $C_{approx}$ and  $R_{approx}$ (with inductive sensor) with a smooth transition from its values at $|Z_s|\text{=}0$ and $|Z_s|\text{=}+\infty$, where $|Z_s|\text{=}2 \pi f L_s$:
\vspace{-2mm}
\begin{equation}\label{EQ:CapproxOscillationSchmittTriggerWithLs}C_{approx} = (C_1+C_2)\frac{R_1}{|Z_s|+R_1}+C_1\frac{|Z_s|}{|Z_s|+R_1}\end{equation}
\vspace{-2mm}
\begin{equation}\label{EQ:RapproxOscillationSchmittTriggerWithLs}R_{approx} = \frac{{R_1}{R_2}}{(R_1+R_2)}\frac{R_1}{(|Z_s|+R_1)}+R_1\frac{|Z_s|}{|Z_s|+R_1}\end{equation}
\phantom{.}\par
\begin{figure}[H]
	\centering
		\includegraphics[width=\columnwidth,keepaspectratio]{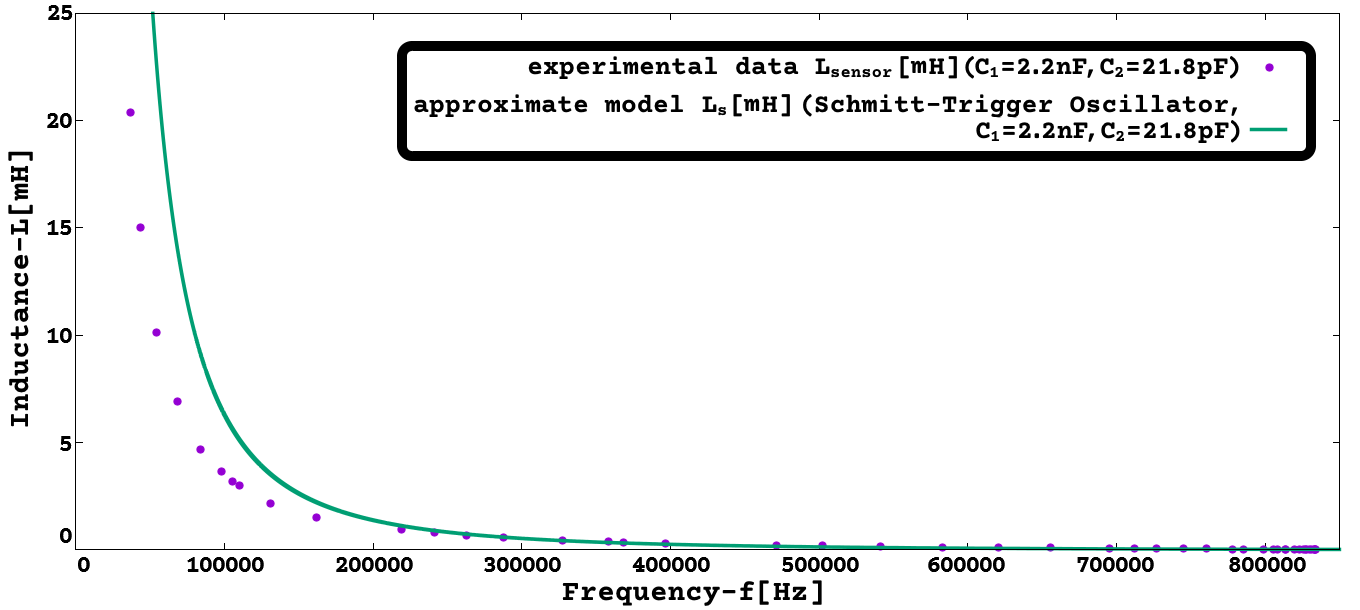}
	\caption{${L_s}(f)$ [mH] by approximate model, frequency in [0Hz, 850kHz] ($C_{approx}$, $R_{approx}$; Multi-Sensor Int. with inductive sensor)}\label{FIG:LsVersusFreqApproxModelSchmittTriggerJPAonJPBoff1}

	\vspace{1mm}

	\centering
		\includegraphics[width=\columnwidth,keepaspectratio]{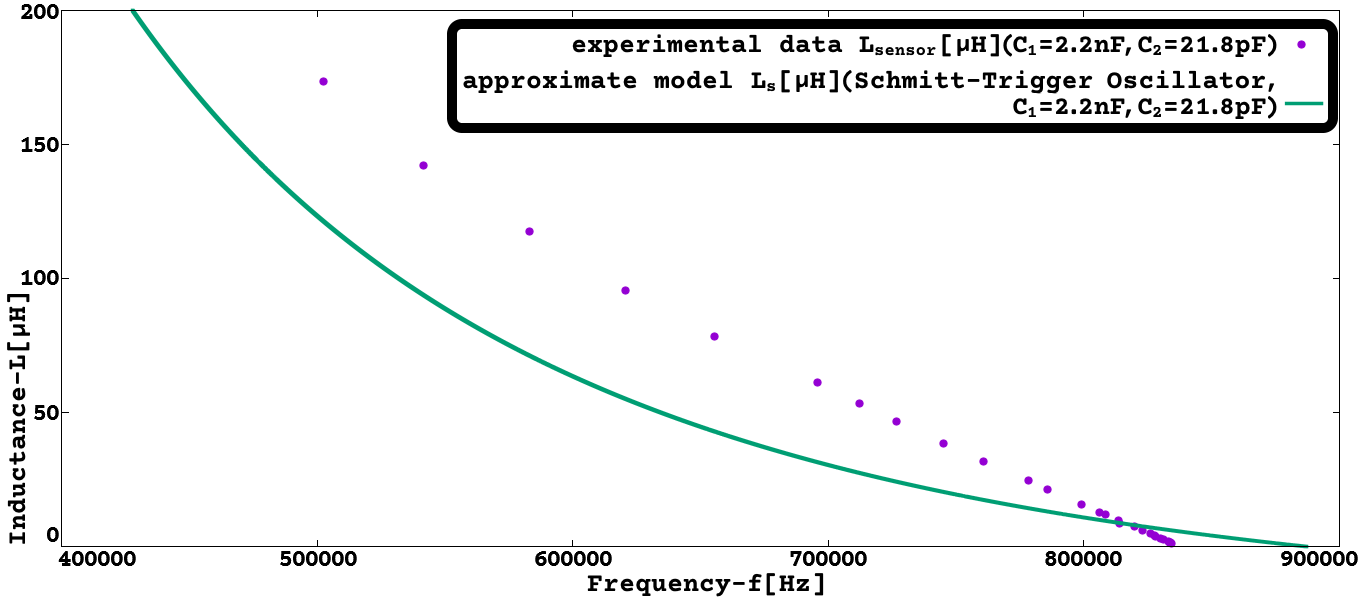}
	\caption{${L_s}(f)$ [$\mu$H] by approximate model, frequency in [400kHz, 900kHz] ($C_{approx}$, $R_{approx}$; Multi-Sensor Int. with inductive sensor)}\label{FIG:LsVersusFreqApproxModelSchmittTriggerJPAonJPBoff2}
\end{figure}

Using the values $C_1$=2.2nF, $C_2$=21.78pF, $R_2\text{=}500\Omega$, $R_1\text{=}2M\Omega$, $H\text{=}1.01496$ with the approximate model ($C_{approx}$, $R_{approx}$), is obtained the plot of ${L_s}(f)$ in Fig.\ref{FIG:LsVersusFreqApproxModelSchmittTriggerJPAonJPBoff1} and Fig.\ref{FIG:LsVersusFreqApproxModelSchmittTriggerJPAonJPBoff2}.

\subsubsection{\texorpdfstring{$R_{approx}$ and $C_{approx}$ for a capacitive sensor ($C_s$)}{Rapprox and Capprox for a capacitive sensor Cs} }
Here are functions modeled to describe $C_{approx}$ and  $R_{approx}$ (with capacitive sensor) with a smooth transition from its values at $|Z_s|\text{=}0$ and $|Z_s|\text{=}+\infty$, where $|Z_s|\text{=}1/(2 \pi f C_s)$:
\vspace{-2mm}
\begin{equation}\label{EQ:CapproxOscillationSchmittTriggerWithCs}C_{approx} = (C_1+C_2)\frac{R_1}{|Z_s|+R_1}+C_1\frac{|Z_s|}{|Z_s|+R_1}\end{equation}
\vspace{-2mm}
\begin{equation}\label{EQ:RapproxOscillationSchmittTriggerWithCs}R_{approx} = \frac{{R_1}{R_2}}{(R_1+R_2)}\frac{R_1}{(|Z_s|+R_1)}+R_1\frac{|Z_s|}{|Z_s|+R_1}\end{equation}
\phantom{.}\par
\begin{figure}[H]
	\centering
		\includegraphics[width=\columnwidth,keepaspectratio]{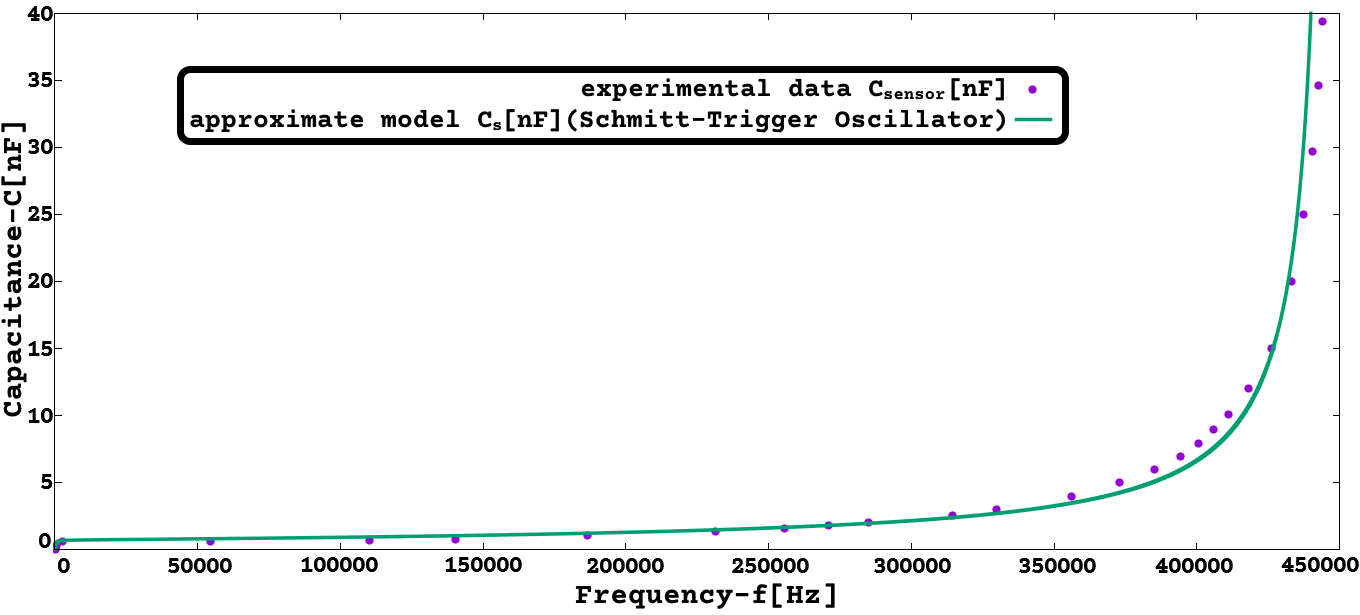}
	\caption{${C_s}(f)$ [nF] by approximate model, frequency in [0Hz, 450kHz] ($C_{approx}$, $R_{approx}$; Multi-Sensor Int. with capacitive sensor)}\label{FIG:CsVersusFrequencyApproxModelSchmittTrigger1}

	\vspace{1mm}

	\centering
		\includegraphics[width=\columnwidth,keepaspectratio]{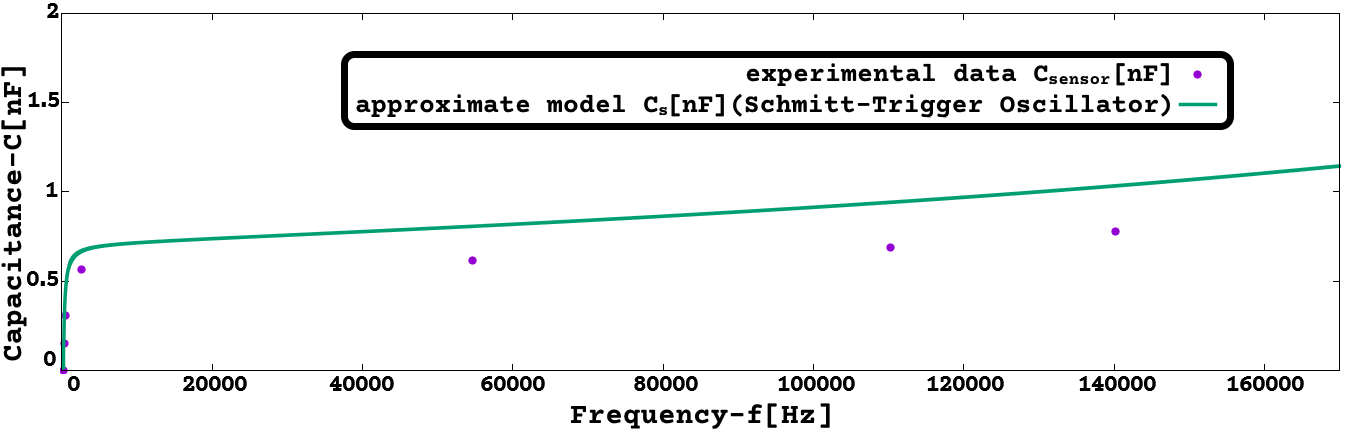}
	\caption{${C_s}(f)$ [nF] by approximate model, frequency in [0Hz, 170kHz] ($C_{approx}$, $R_{approx}$; Multi-Sensor Int. with capacitive sensor)}\label{FIG:CsVersusFrequencyApproxModelSchmittTrigger2}
\end{figure}

Using the values $C_1\text{=}C_2\text{=}2.2nF$, $R_2\text{=}500\Omega$, $R_1\text{=}2M\Omega$, $H\text{=}1.01496$, with the approximate model ($C_{approx}$, $R_{approx}$), is obtained the plot ${C_s}(f)$ in Fig.\ref{FIG:CsVersusFrequencyApproxModelSchmittTrigger1} and Fig.\ref{FIG:CsVersusFrequencyApproxModelSchmittTrigger2}.

\subsection{Multiple-Sensor Interface for inductive sensors as Schmitt-trigger oscillator}\label{SectionInterfForInductiveSensorsSchTriOsc}
Also is possible a theoretical analysis of the Multiple-Sensor Interface with an inductive sensor working continuously as a Schmitt-trigger oscillator observed when jumper configuration is JP.A on, JP.B off ($C_1$=2.2nF, $C_2$=21.78pF), although some approximations were required to be able to obtain a ${L_s}(f)$ function.\par

\begin{figure}[ht!]
	\centering
		\includegraphics[width=\columnwidth,keepaspectratio]{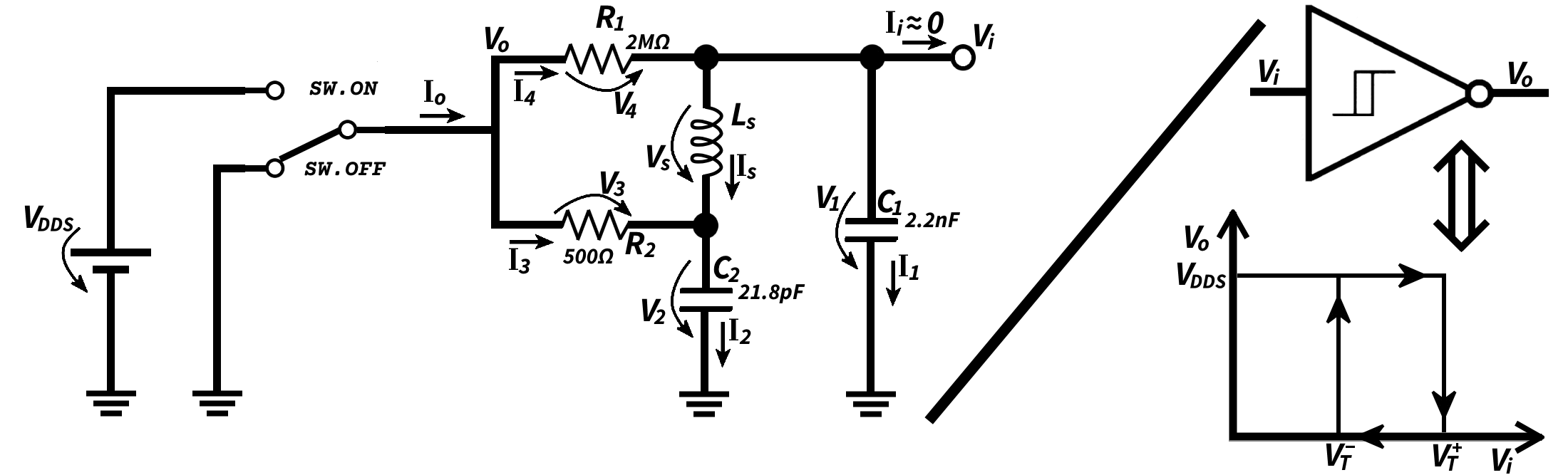}
	\caption{Multi-Sensor Int., inductive sensor(Schmitt-trigger osc.)}\label{FIG:schTheoryOscillatorInductiveSensor}
\end{figure}

So the circuit of Fig.\ref{FIG:schTheoryOscillatorInductiveSensor} is here analyzed to obtain an approximation of $f(L_s)$, and of its inverse function ${L_s}(f){\approx}L_{sensor}$ that is useful for using/configuring the Multiple-Sensor Interface. Notice that $i_{i}$$\approx$0 since $v_i$ is the input of the Schmitt-trigger inverter(high-speed Si-gate CMOS) that has a very high input impedance and so $i_{i}$$\approx$0 is an appropriate approximation simplifying the circuit. So from the circuit are obtained the equations: \par
Nodes and loops: \quad $i_4=i_s+i_1$, \hspace{3mm} $i_3+i_s=i_2$, \newline $i_o=i_3+i_4$, \hspace{3mm} $i_1+i_2=i_o$, \hspace{3mm} $i_1+i_2=i_3+i_4$,\newline $v_1-v_2-v_s=0$, \hspace{3mm} $v_4+v_s+v_2-v_o=0$, \hspace{3mm} $v_4+v_s-v_3=0$, \hspace{3mm} $v_3=v_o-v_2$, \hspace{3mm} $v_4=v_o-v_i$, \hspace{3mm} $v_2=v_i-v_s$ \quad .\par
Components: $i_1\text{=}{C_1}(dv_1/dt)$, \hspace{2mm} $i_2\text{=}{C_2}(dv_2/dt)$, \hspace{2mm} $v_3\text{=}R_2i_3$, \hspace{2mm} $v_4\text{=}R_1i_4$, \hspace{2mm} $v_s\text{=}{L_s}(di_s/dt)$ .\par
Solving:
\vspace{-4mm}
\begin{equation}\label{EQ:nodeI1I2I3I4OscillatorLsensor_1}\frac{v_o-v_2}{R_2}+\frac{v_o-v_i}{R_1}={C_1}\frac{dv_i}{dt}+{C_2}\frac{dv_2}{dt}\end{equation}
\vspace{-2mm}
\begin{equation}\label{EQ:loopV4VsV3OscillatorLsensor_1}{v_2}={v_i}-{L_s}\frac{di_s}{dt}\end{equation}
\vspace{-2mm}
\begin{equation}\label{EQ:nodeI4IsI1OscillatorLsensor_1}\frac{v_o-v_i}{R_1}={i_s}+{C_1}\frac{dv_i}{dt}\end{equation}
\vspace{-3mm}

From (\ref{EQ:loopV4VsV3OscillatorLsensor_1}) is obtained $dv_2/dt=(dv_i/dt)-{L_s}(d^2{i_s}/dt^2)$, and so using it on equation (\ref{EQ:nodeI1I2I3I4OscillatorLsensor_1}), is obtained:\newline
\vspace{-2mm}
\begin{equation}\label{EQ:nodeI1I2I3I4OscillatorLsensor_2}({v_o}-{v_i}){\left ( \frac{1}{R_1}+\frac{1}{R_2} \right )}+{\frac{L_s}{R_2}}{\frac{di_s}{dt}}=({C_1}+{C_2})\frac{dv_i}{dt}-{L_s}{C_2}\frac{d^2 i_s}{dt^2}\end{equation}
From equation (\ref{EQ:nodeI4IsI1OscillatorLsensor_1}) is obtained:
\vspace{-3mm}
\begin{multline}\label{EQ:nodeI4IsI1OscillatorLsensor_2} {i_s}=\frac{{v_o}-{v_i}}{R_1}-{C_1}\frac{dv_i}{dt} \hspace{3mm} \Rightarrow \hspace{3mm} \frac{di_s}{dt}=-{\frac{1}{R_1}}{\frac{dv_i}{dt}}-{C_1}\frac{d^2{v_i}}{dt^2} \\ \Rightarrow \hspace{3mm} \frac{d^2{i_s}}{dt^2}=-{\frac{1}{R_1}}{\frac{d^2{v_i}}{dt^2}}-{C_1}\frac{d^3{v_i}}{dt^3} \end{multline}
Using (\ref{EQ:nodeI4IsI1OscillatorLsensor_2}), $i_s$ can be eliminated from eq. (\ref{EQ:nodeI1I2I3I4OscillatorLsensor_2}), obtaining:
\vspace{-3mm}
\begin{multline}\label{EQ:diffEqOscillatorLsensor} {\left ( \frac{1}{R_1}+\frac{1}{R_2} \right )}({v_o}-{v_i}) = \\ {\left ({C_1}+{C_2}+{\frac{L_s}{{R_1}{R_2}}} \right)}{\frac{dv_i}{dt}}+{L_s}\left ( \frac{C_1}{R_2} + \frac{C_2}{R_1}  \right ) \frac{d^2 v_i}{dt^2} + {L_s}{C_1}{C_2}{\frac{d^3 v_i}{dt^3}}\end{multline}
\vspace{-2mm}

The equation (\ref{EQ:diffEqOscillatorLsensor}) is of the type: $d(v_o-v_i)=c(dv_i/dt)+b(d^2v_i/dt^2)+a(d^3v_i/dt^3)$ that has the general solution: $v_i(t)=v_o+k_1\emph{e}^{{\lambda_1}t}+k_2\emph{e}^{{\lambda_2}t}+k_3\emph{e}^{{\lambda_3}t}$, where $k_1$,$k_2$, $k_3$ are integration constants to be defined by 'initial conditions' and  $\lambda_1$, $\lambda_2$, $\lambda_3$ are defined by: $a{\lambda^3}+b{\lambda^2}+c\lambda+d=0$ and $\emph{e}$ is the Euler-Napier constant $\emph{e} = \sum_{n=0}^{\infty} (1/(n!))$.\par
So defining here $a_l$, $b_l$, $c_l$, $d_l$ as the values of a, b, c, d for obtaining ${v_i}(t)$ when using an inductive sensor, as: \par
${a_l}={L_s}{C_1}{C_2}$ , \quad 
${b_l}={L_s}(({C_1}/{R_2})+({C_2}/{R_1}))$ , \par
${c_l}=C_1+C_2+({L_s}/({R_1}{R_2}))$ , \quad
${d_l}=(1/{R_1})+(1/{R_2})$ . \par

In order to obtain ${v_i}(t)$ for this circuit is required to calculate $k_1$, $k_2$ and $k_3$, that are constants to be defined by 'initial conditions', the value of these constants is related to the voltage (or electrical charge) on capacitors $C_1$ and $C_2$, and also to the electric current on the inductive sensor, at the moment the inverter gate changes its output voltage (high to low, or low to high), on the model used for analyzing the circuit that is when the 'theoretical switch' $v_o$ changes state.
Also, is only known the value of $v_i$ (to be $V_T^{-}$ or $V_T^{+}$) when the inverter gate $v_o$ changes value, so is very difficult to calculate $k_1$, $k_2$, $k_3$ by algebraic manipulation.
Admitting that $R_1 \gg R_2$ and that $C_1 \geqslant C_2$  (that is the case of the circuit that was studied and tested where $R_{1}\text{=}2M\Omega$ and $R_{2}\text{=}500\Omega$), then is known that the capacitor $C_2$ will charge faster than $C_1$ for all values of $Z_s$ (determined by the value $L_s$), and in case $Z_s$ has an impedance comparable to $R_2$ then $C_2$ will charge much faster than $C_1$. So the voltage (and electrical charge) of $C_2$ follows closely the values of $v_o$ and so will be of small relevance to the initial conditions of the circuit.\par
The inclusion of an inductor (the inductive sensor) makes the behavior of this circuit more complex and so the appearance of a 3rd degree differential equation; so in order to obtain an expression for ${L_s}(f)$ in a similar way as previously for resistive and capacitive sensors, are made the following approximations:
1 - The determination of ${L_s}(f)$ will be made in two domains, one expression of ${L_s}(f)$ valid for high frequency and other expression of ${L_s}(f)$ for low and middle frequency of the operation of the Schmitt-trigger oscillator. The author observed that high frequency operation is dominated by the 1st-root of $a{\lambda^3}+b{\lambda^2}+c\lambda+d=0$ that is a real number; and observed that low and middle frequency operation is dominated by the 2nd-root and 3rd-root of $a{\lambda^3}+b{\lambda^2}+c\lambda+d=0$ that are imaginary numbers.\par
2 - Regarding the low and middle frequency operation that is dominated by the 2nd-root and 3rd-root of the 3rd degree equation, will be made a 'rude' approximation of $v_i(t){\approx}v_o+{k_{23}}\emph{e}^{{\lambda_{23}}t}$, where $k_{23}={k_2}+{k_3}$ and $\lambda_{23}=({\lambda_2}+{\lambda_3})/2$. Also is relevant to note that by adding $\lambda_2$ with $\lambda_3$ will nullify the imaginary part resulting in a real number, and also that $\lambda_{23}=({\lambda_2}+{\lambda_3})/2=Re[{\lambda_2}]=Re[{\lambda_3}]$.\par
3 - Regarding the high frequency operation that is dominated by the 1st-root of the 3rd degree equation, will be made the approximation of $v_i(t){\approx}v_o+{k_{1}}\emph{e}^{{\lambda_{1}}t}$.\par
The mentioned approximations resulting in $v_i(t){\approx}v_o+{k}\emph{e}^{{\lambda}t}$, allows the theoretical analysis already used for resistive and capacitive sensor to be reused again here, for obtaining an approximation of ${L_s}(f)$.\par

So for an inductive sensor, just like with resistive or capacitive sensor, is used the function ${{v_i}(t) {\approx} {v_o}+k\emph{e}^{-t/{\tau}} }$, where ${\tau=-1/\lambda}$, and the constant 'H' defined by:\newline
\noindent \phantom{'} \hfill $\qquad H=ln \left ( \frac{({V_T^{-}}-{V_{DDS}}){V_T^{+}}}{({V_T^{+}}-{V_{DDS}}){V_T^{-}}} \right )$, \qquad \qquad \newline
\noindent and $f=1/T \Leftrightarrow f = 1/(\tau H) \Leftrightarrow f = -\lambda / H$ . \par
So applying the previously mentioned approximations is obtained: ${f_{high}} \approx - {\lambda_1} / H$ ; \quad ${f_{low, middle}} \approx - {\lambda_{23}} / H$ .

So the expression (theoretical) of an approximate value for inductance ($L_{s, HF}$) as a function of frequency(f) for high frequency is:
\vspace{-3mm}
\begin{equation}\label{EQ:LSensorVersusFrequencyOscillationSchmittTriggerForHighFreq}L_{sensor, HF} {\approx} L_{s, HF} = \frac{{R_1}+{R_2}-({C_1}+{C_2}){R_1}{R_2}Hf}{Hf({C_1}{R_1}Hf-1)({C_2}{R_2}Hf-1)}\end{equation} \par
\vspace{-2mm}

So the expression (theoretical) of an approximate value for inductance ($L_{s, LMF}$) as a function of frequency(f) for low and middle frequency is:
\vspace{-3mm}
\begin{multline}\label{EQ:LSensorVersusFrequencyOscillationSchmittTriggerForLowMiddleFreq}L_{sensor, LMF} {\approx} L_{s, LMF} = ( {{C_1}^2}{{R_1}^2}{R_2} + {{C_2}^2}{{R_2}^2}{R_1} \\ - 2{{R_1}^2}{{R_2}^2}Hf( {{C_1}^2}{C_2} + {{C_2}^2}{C_1} ) ) / ( 8{{C_1}^2}{{C_2}^2}{{R_1}^2}{{R_2}^2}{H^3}{f^3} \\ - 8{C_1}{{C_2}^2}{R_1}{{R_2}^2}{H^2}{f^2} - 8{{C_1}^2}{C_2}{{R_1}^2}{R_2}{H^2}{f^2} + 2{{C_2}^2}{{R_2}^2}Hf \\ + 6{C_1}{C_2}{R_1}{R_2}Hf + 2{{C_1}^2}{{R_1}^2}Hf - {C_1}{R_1} - {C_2}{R_2} )\end{multline} \par
\begin{figure}[ht!]
	\centering
		\includegraphics[width=\columnwidth,keepaspectratio]{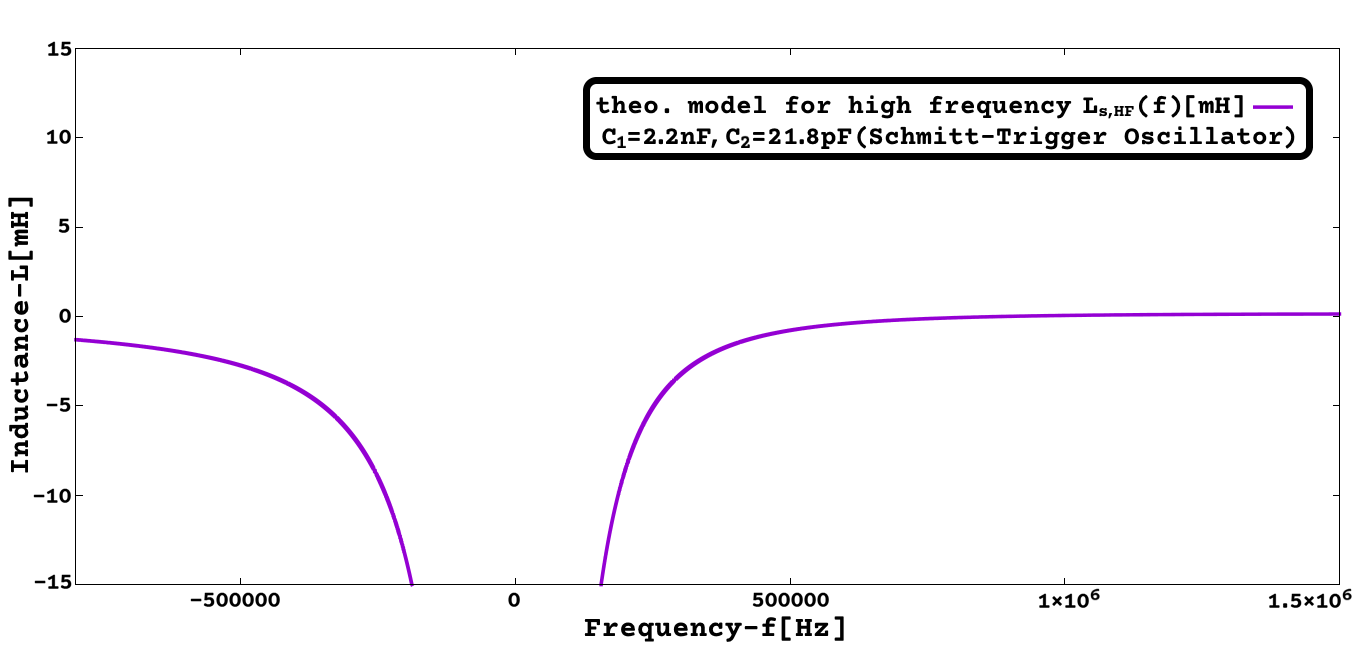}
	\caption{$L_{s,HF}(f)$ [mH], frequency in [-800kHz, 1.5MHz] (Sch-trig. osc., Multi-Sensor Int. with inductive sensor, HF approx. function).}\label{FIG:LSensorVersusFrequencyOscSchTrigHFApproxTheoryEq}
\end{figure}

\vspace{-2mm}
Using the values $C_1\text{=}2.2nF$, $C_2\text{=}21.78pF$, $R_2\text{=}500\Omega$, $R_1\text{=}2M\Omega$, ${V_T^{-}}\text{=}1.2V$, ${V_T^{+}}\text{=}2.2V$, ${V_{DDS}}\text{=}4.18V$, is obtained $H\text{=}1.01496$, Fig.\ref{FIG:LSensorVersusFrequencyOscSchTrigHFApproxTheoryEq} shows the plot of $L_{s, HF}(f)$ using (\ref{EQ:LSensorVersusFrequencyOscillationSchmittTriggerForHighFreq}) with the mentioned values of $C_1$, $C_2$, $R_2$, $R_1$, $H$.

 \par

\begin{figure}[t!]
	\centering
		\includegraphics[width=\columnwidth,keepaspectratio]{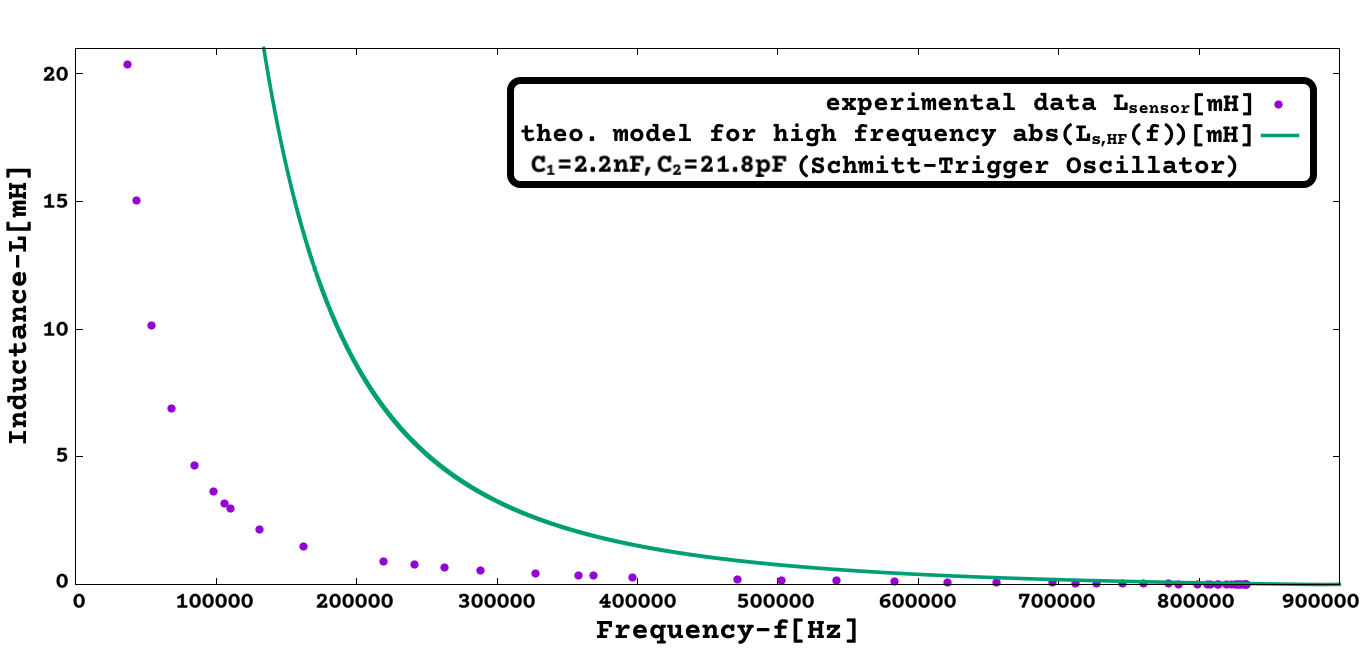}
	\caption{$abs(L_{s,HF}(f))$ [mH], frequency in [0Hz, 900kHz] (Sch-trig. osc., Multi-Sensor Int. with inductive sensor, HF approx. function).}\label{FIG:LSensorVersusFrequencyOscillationSchmittTriggerHF1}

	\vspace{0.5mm}

	\centering
		\includegraphics[width=\columnwidth,keepaspectratio]{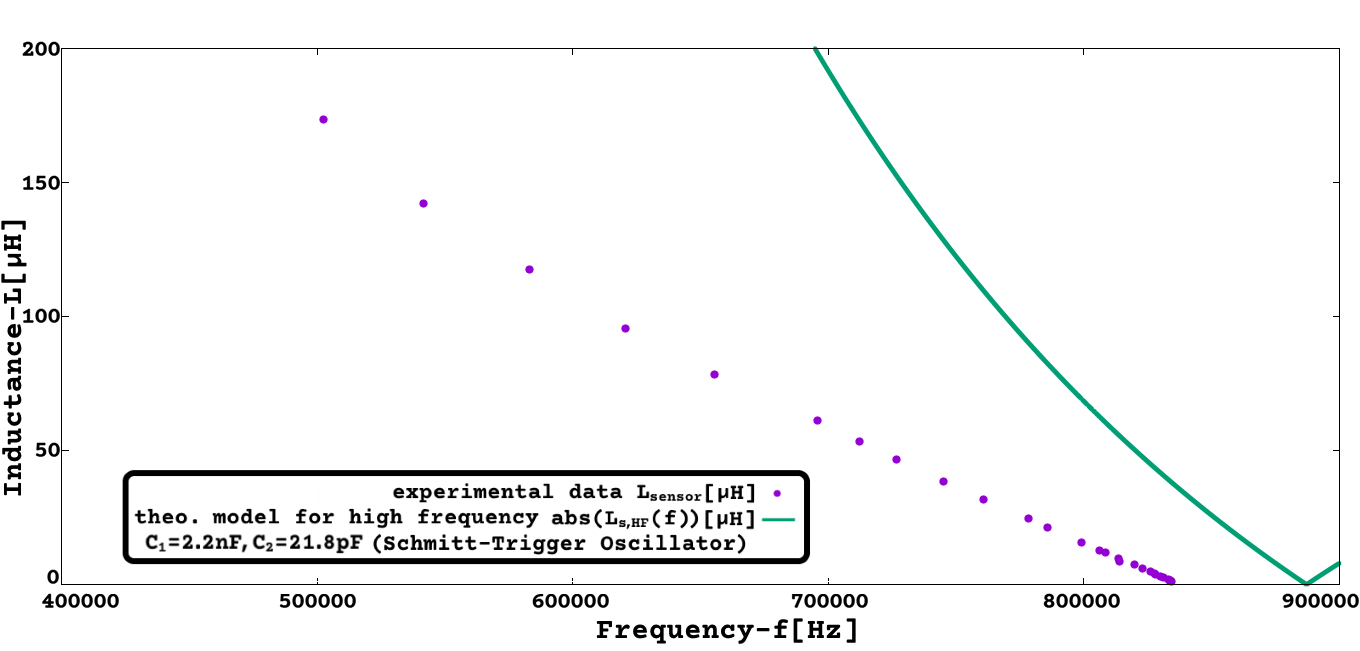}
	\caption{$abs(L_{s,HF}(f))$ [$\mu$H], frequency in [400kHz, 900kHz] (Sch-trig. osc., Multi-Sensor Int. with inductive sensor, HF approx. function).}\label{FIG:LSensorVersusFrequencyOscillationSchmittTriggerHF2}
\end{figure}

Observing Fig.\ref{FIG:LSensorVersusFrequencyOscSchTrigHFApproxTheoryEq} is noticeable that $L_{s,HF}(f)$ is almost always negative, except for frequency above around 900kHz. However drawing the plot of $abs( L_{s,HF}(f) )$ ($=|L_{s,HF}(f)|$) along with experimental data for Multiple-Sensor Interface with various inductance values connected as the sensor (in Fig.\ref{FIG:LSensorVersusFrequencyOscillationSchmittTriggerHF1} and Fig.\ref{FIG:LSensorVersusFrequencyOscillationSchmittTriggerHF2} ), like was done before for the plots of a capacitive sensor, is then visible that the shape of $abs( L_{s,HF}(f) )$ resembles the experimental data, however significantly displaced above the experimental data for low and middle values of frequency, but for high frequency the plot of $abs( L_{s,HF}(f) )$ is much closer to the experimental data and correctly predicts that a value of $L_{s}=0$ will result on an oscillation frequency somewhat above 800kHz but lower than 900kHz.

\begin{figure}[t!]
	\centering
		\includegraphics[width=\columnwidth,keepaspectratio]{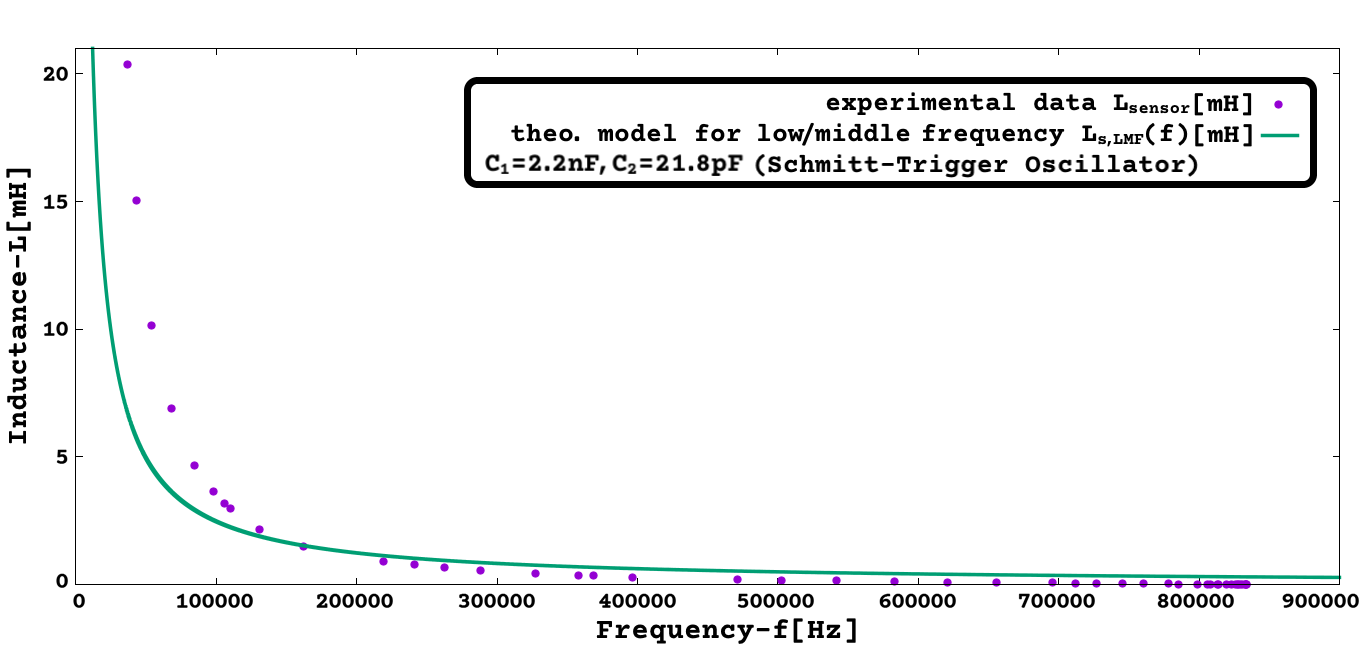}
	\caption{$L_{s,LMF}(f)$ [mH], frequency in [0Hz, 900kHz] (Sch-trig. osc., Multi-Sensor Int. with inductive sensor, LMF approx. function).}\label{FIG:LSensorVersusFrequencyOscillationSchmittTriggerLMF1}

	\vspace{0.5mm}

	\centering
		\includegraphics[width=\columnwidth,keepaspectratio]{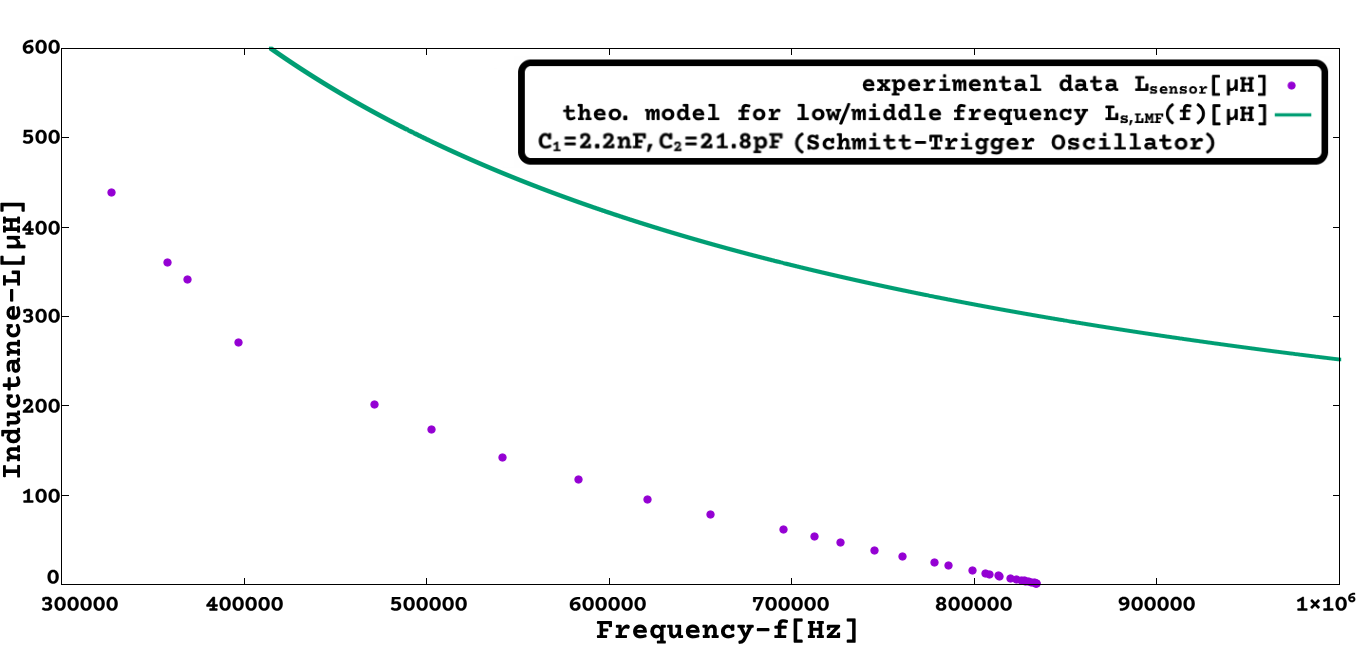}
	\caption{$L_{s,LMF}(f)$ [$\mu$H], frequency in [300kHz, 1MHz] (Sch-trig. osc., Multi-Sensor Int. with inductive sensor, LMF approx. function).}\label{FIG:LSensorVersusFrequencyOscillationSchmittTriggerLMF2}
\end{figure}

The Fig.\ref{FIG:LSensorVersusFrequencyOscillationSchmittTriggerLMF1} and Fig.\ref{FIG:LSensorVersusFrequencyOscillationSchmittTriggerLMF2}, shows experimental data for Multiple-Sensor Interface with various inductance values connected as the sensor, and the plot of $L_{s, LMF}(f)$ using (\ref{EQ:LSensorVersusFrequencyOscillationSchmittTriggerForLowMiddleFreq}) with the mentioned values of $C_1$, $C_2$, $R_2$, $R_1$, $H$. Observing Fig.\ref{FIG:LSensorVersusFrequencyOscillationSchmittTriggerLMF1} and Fig.\ref{FIG:LSensorVersusFrequencyOscillationSchmittTriggerLMF2} is noticeable that $L_{s,LMF}(f)$ follows the experimental data on low and middle frequency (however with some displacement), but for high frequency $L_{s,LMF}(f)$ has a large displacement to the experimental data and fails to predict the frequency at $L_{s}$=0.\par

A way for obtaining a theoretical function (by previously mentioned approximations/simplifications) valid for all values of frequency, that captures the behavior of the Multiple-Sensor Interface with an inductive sensor, is to combine both the expressions $abs( L_{s,HF}(f) )$ and $L_{s,LMF}(f)$ into a single function. That function can be obtained for example by multiplying $abs( L_{s,HF}(f) )$ with a moderating function that silences it on low and middle frequency and by multiplying $L_{s,LMF}(f)$ with a moderating function that silences it on high frequency, then sum these 2 parts to obtain $L_{s}(f)$.

A set of 2 good moderating functions, considering the used values of $C_1$, $C_2$, $R_2$, $R_1$, $H$, would be for example:\newline
\resizebox{\hsize}{!}{${m_{HF,ef}}(f)=(((-1+(\emph{e}^{(f/100000)-5}))/(1+(\emph{e}^{(f/100000)-5})))+1)/2$}\newline
\resizebox{\hsize}{!}{${m_{LMF,ef}}(f)=(((-1+(\emph{e}^{(-f/100000)+5}))/(1+(\emph{e}^{(-f/100000)+5})))+1)/2$}\par
These moderating functions are good because: ${m_{HF,ef}}(f)$+${m_{LMF,ef}}(f)$$\approx$1, $\forall f$, and also:\newline
${m_{HF,ef}}$($f$$=$0)$\approx$0; \hspace{2mm} ${m_{HF,ef}}$($f$$=$900kHz)$\approx$1;\newline
${m_{HF,ef}}$($f$$=$+$\infty$)$=$1; \hspace{2mm} ${m_{LMF,ef}}$($f$$=$+$\infty$)$=$0;\newline
${m_{LMF,ef}}$($f$$=$0)$\approx$1; \hspace{2mm} ${m_{LMF,ef}}$($f$$=$900kHz)$\approx$0;\newline
${m_{LMF,ef}}$($f$$=$500kHz)$\approx$${m_{HF,ef}}$($f$$=$500kHz)$\approx$0.5 .\par

However the moderating functions ${m_{HF,ef}}(f)$, ${m_{LMF,ef}}(f)$ are exponential functions that are undesirably complex and heavy to evaluate; so here is suggested another set of 2 moderating functions that are simpler and faster to evaluate, although the property ${m_{HF}}(f)$+${m_{LMF}}(f)$$\approx$1 will only be true on a small frequency range and that will cause some distortion (undesirable change) on the theoretical model. Since the theoretical model is not accurate because of the approximations that were used for obtaining a solution to the initial value problem (that had 3 unknowns $k_1$, $k_2$, $k_3$, but was solved as if had a single unknown), and also knowing that $abs(L_{s,HF}(f))$ is always above the experimental data and that $L_{s,LMF}(f)$ is above the experimental data on a portion of the plot (on this situation for f>250kHz), then can be selected moderating functions that have a 'distortion' that actually approximates(lowers the value of) $abs(L_{s,HF}(f))$ and $L_{s,LMF}(f)$ to the experimental data.

A set of 2 moderating functions that are more simple and fast/easy to evaluate, but 'distort' (lower) the value of $L_{s}(f)$:
\vspace{-2mm}
\begin{equation}\label{EQ:mHFf3}{m_{HF,f3}}(f)=({a_{HF}}{\cdot}{f^3})/(1+{a_{HF}}{\cdot}{f^3})\end{equation}
\vspace{-2mm}
\begin{equation}\label{EQ:mLMFf3}{m_{LMF,f3}}(f)=1/(1+{a_{LMF}}{\cdot}{f^3})\end{equation}
\vspace{-1mm}

Also let's have in consideration a frequency value $f_{0}$ defined as $abs(L_{s,HF}(f$=$f_{0}))$=$0$ (the frequency where the plot of $abs(L_{s,HF}(f))$ reaches the horizontal axis of $L_{s}=0$).\par
Then, $f_{0}$=$({R_1}+{R_2})/(({C_1}+{C_2})H{R_1}{R_2})$.\par

So by first determining the frequency $f_{0}$ for some selected values of the parameters $C_1$, $C_2$, $R_2$, $R_1$, $H$, then is possible to create/define $L_{s,LMF,HF}(f)$ as a 'theoretical' function (by the previously mentioned approximations/simplifications) through ${m_{HF,f3}}(f)$, ${m_{LMF,f3}}(f)$, to be valid for all values of frequency:
\vspace{-2mm}
\begin{equation}\label{EQ:LSensorVersusFrequencyOscillationSchmittTriggerLMFAndHFModel}\resizebox{\hsize}{!}{$L_{s,LMF,HF}(f) = (L_{s,LMF}(f))\cdot({m_{LMF,f3}}(f))+abs(L_{s,HF}(f))\cdot({m_{HF,f3}}(f))$}\end{equation}\par

The calculation of $a_{HF}$ and $a_{LMF}$ to create some moderating functions ${m_{HF,f3}}(f)$, ${m_{LMF,f3}}(f)$ may be done for example by defining target values of ${m_{HF,f3}}(f$=$2{\cdot}f_{0}/3)$, ${m_{LMF,f3}}(f$=${f_0})$, by this method is obtained:
\begin{equation}\label{EQ:aHFf3}{a_{HF}}={\frac{27{({C_1}+{C_2})^3}{H^3}{{R_1}^3}{{R_2}^3}{{m_{HF}}(f{\textnormal{=}}2{f_0}/3)}}{8{({R_1}+{R_2})^3}(1-{{m_{HF}}(f{\textnormal{=}}2{f_0}/3)})}}\end{equation}
\begin{equation}\label{EQ:aLMFf3}{a_{LMF}}={\frac{{({C_1}+{C_2})^3}{H^3}{{R_1}^3}{{R_2}^3}(1-{{m_{LMF}}(f{\textnormal{=}}{f_0})})}{{{({R_1}+{R_2})^3}{m_{LMF}}(f{\textnormal{=}}f_0)}}}\end{equation}

So the constant ${a_{HF}}$ may be selected/adjusted so that ${m_{HF,f3}}(f$=$2{\cdot}f_{0}/3)$=0.5, this is the same as to say that ${a_{HF}}$ will be selected/adjusted so that when frequency is at 2/3 of $f_{0}$ then $abs(L_{s,HF})\cdot({m_{HF,f3}})$ will be 50\% of $abs(L_{s,HF})$.\par
So the constant ${a_{LMF}}$ may be selected/adjusted so that ${m_{LMF,f3}}(f$=$f_{0})$=0.01, this is the same as to say that ${a_{LMF}}$ will be selected/adjusted so that when frequency is at $f_{0}$ then $(L_{s,LMF})\cdot({m_{LMF,f3}})$ will be 1\% of $(L_{s,LMF})$.\par

\begin{figure}[ht!]
	\centering
		\includegraphics[width=\columnwidth,keepaspectratio]{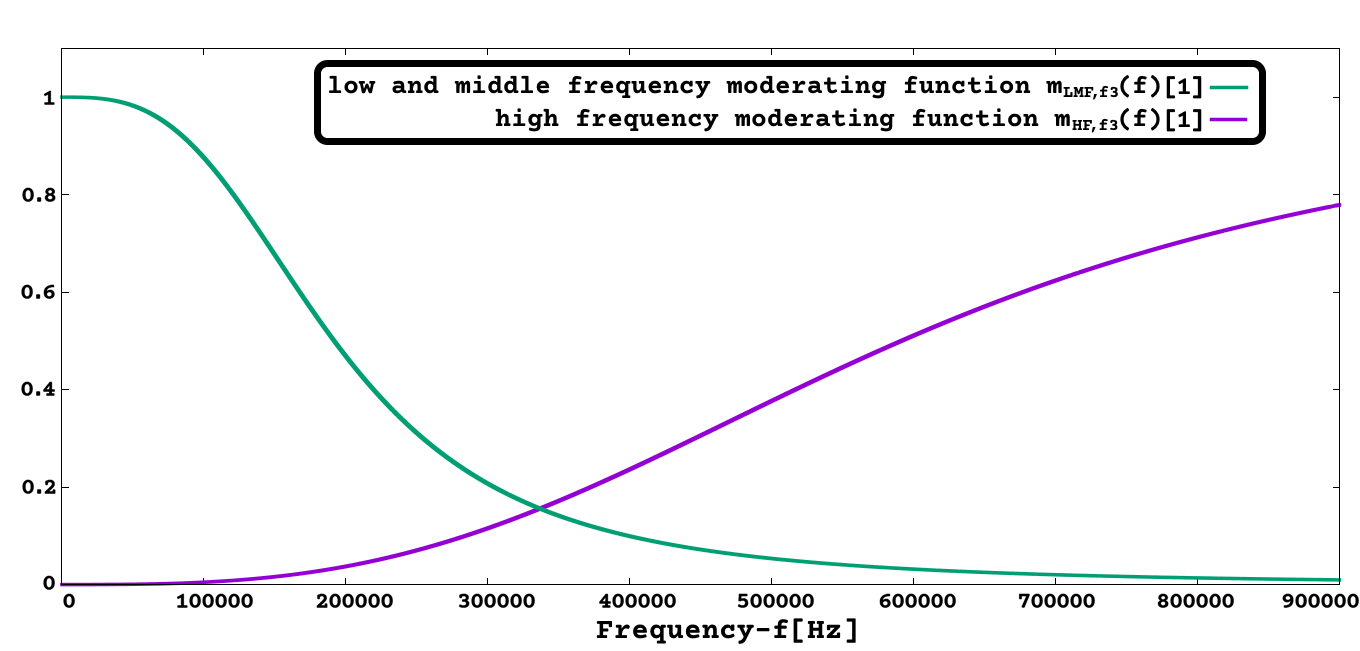}
	\caption{$m_{LMF,f3}(f)$ and $m_{HF,f3}(f)$ [1], frequency in [0Hz, 900kHz].}\label{FIG:ModeratingFunctionsWithFrequencyPower3ForLMFAndHF}
\end{figure}

So for this case ($R_1\text{=}2M\Omega$, $R_2\text{=}500\Omega$, $C_1$=2.2nF, $C_2$=21.78pF, $H\text{=}1.01496$), with ${m_{HF,f3}}(f$=$2{\cdot}f_{0}/3)$=0.5, ${m_{LMF,f3}}(f$=$f_{0})$=0.01, was selected/adjusted the values: ${a_{HF}}=4.834{\cdot}{10^{-18}}$ $s^3$, ${a_{LMF}}=1.418{\cdot}10^{-16}$ $s^3$ (Fig. \ref{FIG:ModeratingFunctionsWithFrequencyPower3ForLMFAndHF}).\par

\begin{figure}[ht!]
	\centering
		\includegraphics[width=\columnwidth,keepaspectratio]{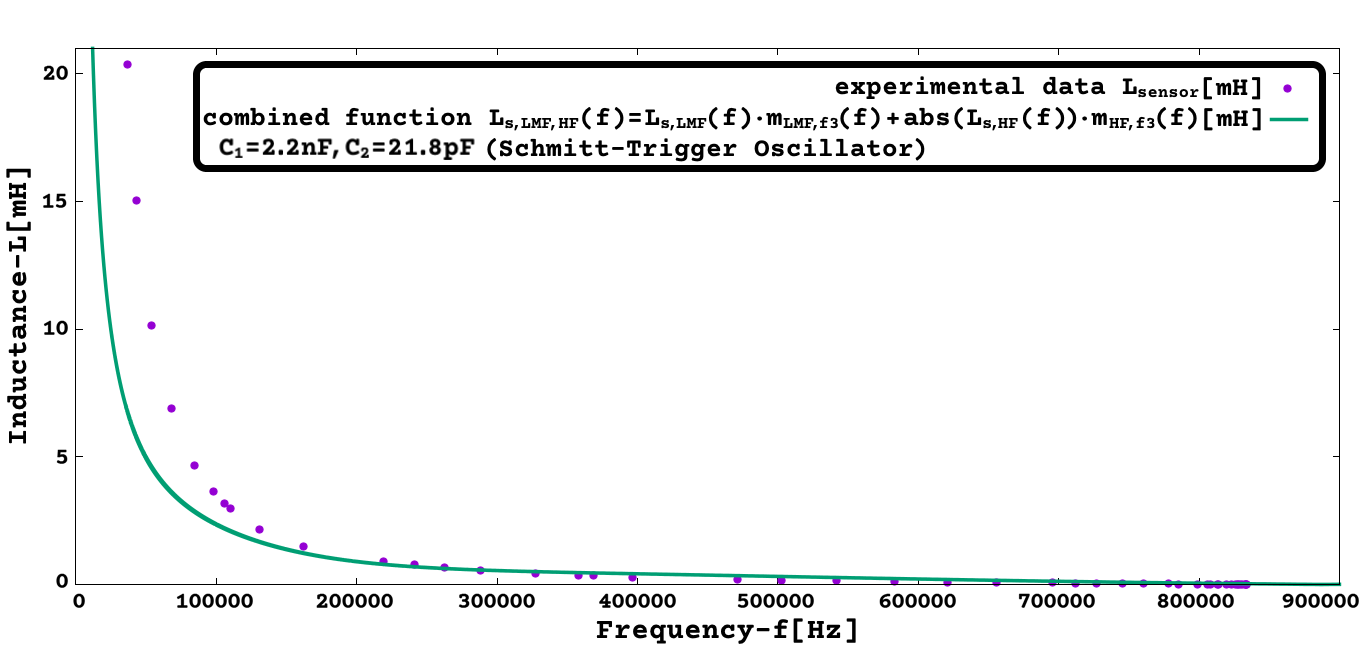}
	\caption{$L_{s,LMF,HF}(f)$ [mH], frequency in [0Hz, 900kHz] (Sch-trig. osc., Multi-Sensor Int. with inductive sensor, LMF and HF combined approx. function, with $C_1$=2.2nF, $C_2$=21.78pF).}\label{FIG:LSensorVersusFrequencyOscillationSchmittTriggerLMFAndHFCombined1}

	\vspace{0.5mm}

	\centering
		\includegraphics[width=\columnwidth,keepaspectratio]{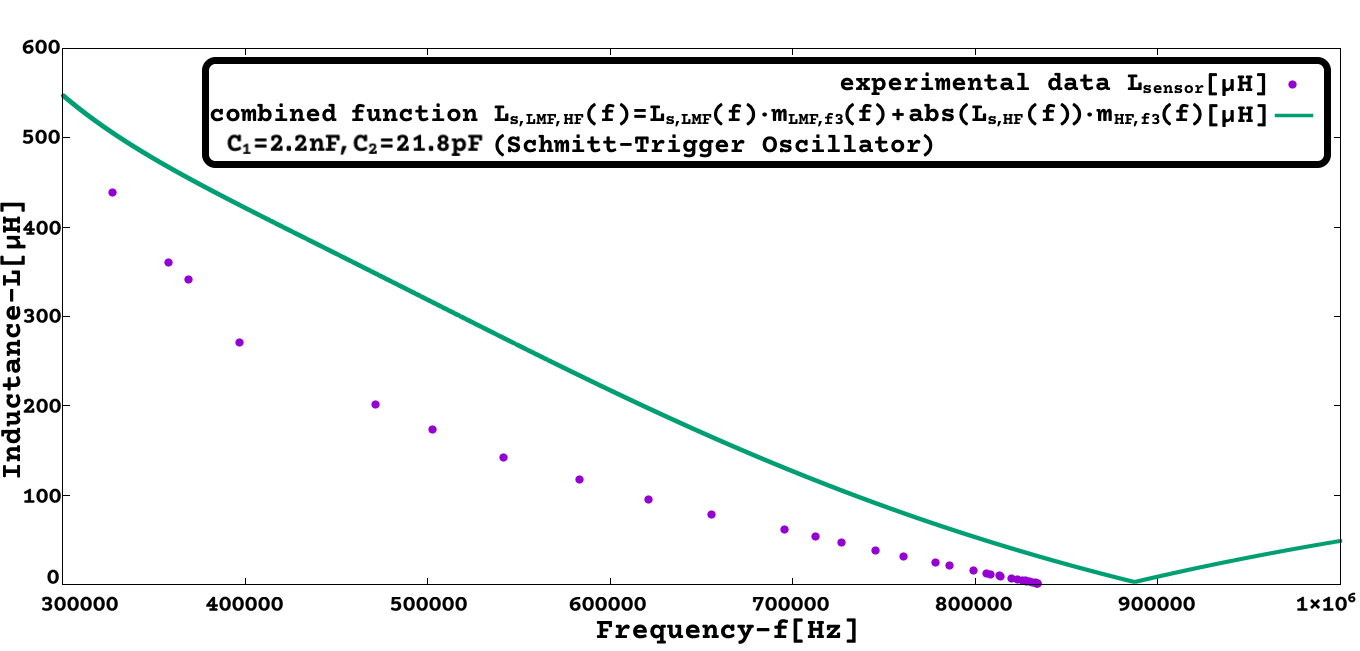}
	\caption{$L_{s,LMF,HF}(f)$ [$\mu$H], frequency in [300kHz, 1MHz] (Sch-trig. osc., Multi-Sensor Int. with inductive sensor, LMF and HF combined approx. function, with $C_1$=2.2nF, $C_2$=21.78pF).}\label{FIG:LSensorVersusFrequencyOscillationSchmittTriggerLMFAndHFCombined2}
\end{figure}

The Fig. \ref{FIG:LSensorVersusFrequencyOscillationSchmittTriggerLMFAndHFCombined1} and Fig. \ref{FIG:LSensorVersusFrequencyOscillationSchmittTriggerLMFAndHFCombined2}, shows experimental data for Multiple-Sensor Interface with various inductance values connected as the sensor, and the plot of $L_{s,LMF,HF}(f)$ using (\ref{EQ:LSensorVersusFrequencyOscillationSchmittTriggerLMFAndHFModel}) with $C_1$=2.2nF, $C_2$=21.78pF, $R_2\text{=}500\Omega$, $R_1\text{=}2M\Omega$, $H\text{=}1.01496$, ${a_{HF}}=4.834{\cdot}{10^{-18}}$ $s^3$, ${a_{LMF}}=1.418{\cdot}10^{-16}$ $s^3$.\par

\section{Discussion}\label{SEC:Discussion}

The experimental data obtained when testing the Multiple-Sensor Interface is in overall close to the values calculated using the formulas obtained from the circuit analysis, as it is visible on the various figures that show the plots of the experimental data along with graphs done using the mentioned formulas. In some graphs, there was a deviance or offset between the theoretical and experimental quantitative values, however the observable deviations are not reason for concern as they never affected the similitude between theoretical and experimental graphs (with exception of Fig.\ref{FIG:LsVersusFreq--C1-2200--C2-22} and Fig.\ref{FIG:LsVersusFreqExprData100uH--C1-2200--C2-22} that as explained, when using that specific configuration the device no longer behaves as Pierce oscillator, but instead as a Schmitt Trigger oscillator, as it was shown on later subsections). 
The information that is made available on the article, besides explaining how the device works, may also be useful for a user of the device/technology for determining if a specific sensor of his interest is compatible/usable when connected to the Multiple-Sensor device, namely by observing on the graphs (or tables), for what range of values (min. and max.) of the sensor electrical quantity ($R_s$ or $C_s$ or $L_s$) it is verified that a change on the quantity the sensor is measuring will produce/cause also a significant or measurable change of the signal frequency on the output of the oscillator used for sensor interfacing.

\subsection{Future Work}
Further work related to the sensor interface circuit may be developing a better understanding/prediction of why/when the Multiple-Sensor Interface with inductive sensor changes from working as Pierce oscillator to Schmitt-trigger oscillator depending on the values of $C_1$ and $C_2$ capacitors (whose values can be adjusted by jumpers JP.A and JP.B).\par
Other future work, outside the scope of this article, could be characterizing, testing, and comparing this sensor interface circuit for specific sensor types and/or applications, thus allowing a performance comparison with other technologies on specific use cases.

\subsection{Better accuracy on a commercial scenario}
A more commercial usage of Multiple-Sensor Interface possibly with better accuracy by using pre-calibrated sensors PCBs, may be:\par
1- Production of small single sensor PCBs that include: one sensing element (Ex: RTD, LDR, proximity inductive sensor, capacitive humidity sensor, etc ...), the oscillator circuit, and a voltage stabilizer (Ex: Zener diode). The voltage stabilizer is required to obtain a stable/fixed $V_{DDS}$ on the sensor PCB independent of the power supply, that is already set when the producer/seller does the sensor calibration. Also the sensor PCBs should have values of $C_1$ and $C_2$ selected/tuned for best measurement range or best accuracy/precision (and so jumpers JPA/JPB not required, or may be soldered/fixed), having as output the oscillator square wave signal.\par
2- Obtaining a calibration table for the single sensor PCB (pairing output oscillator frequency with measurements by reference calibration instruments), on an appropriate calibration environment done/performed by the producer/seller.\par
3- Distribution/sale of the sensor PCBs with their own calibration table included. Could be: a calibration table printed on paper (for example: to be typed by hand on the software/application of the device, or automatically by OCR or QR-code), or as digital file that can be downloaded from the distributor/seller by using a serial number associated with the single sensor PCB.\par
4- An end user would connect a single sensor PCB to the Multiple-Sensor Interface device (that contains the micro-controller, EEPROM, and USB, RS-485, GPIO interfaces) on one of the frequency measurement channels (and also connect the GND), preferably using the same power supply for both the sensor PCBs and the Multiple-Sensor device. The end-user would import the provided calibration table into the Multiple-Sensor device, using the provided software/application.\par

Also note that with an external single sensor PCB any oscillator circuit/design may be used as long its output is a square wave signal (that doesn't exceeds the power supply voltage); and also that a much longer distance/cable can be used between the sensor location and the Multiple-Sensor Interface device, since that any parasitic resistance/capacitance/inductance of the cable won't affect the oscillator frequency that will be measured by the device.\par
The production/sale and usage scenario here described is the author perspective of a modest compromise that could be made on the hardware configurability/versatility (of the single sensor PCB with square wave signal output), that allows a producer/seller to supply ready to use sensors without any changes on the proposed hardware design of the Multiple-Sensor Interface (since it already includes 4 channels dedicated to measuring frequency of an external signal) and so without restricting the end-user ability to use any sensors it wishes or even a custom made sensor for a very specific use/application.\par

\subsection{About \texorpdfstring{$C_s$(f)$<$0}{Cs(f) is negative} on Multi-Sensor with capacitive sensor}
About $C_s$(f)$<$0 have in mind the Multiple-Sensor with capacitive sensor is studied on transient behavior (relaxation oscillator), where 'frequency' is a measure of the speed of charge and discharge on $C_1$; and also of how fast the transient circuit analysis alternates between $v_o \text{=} V_{DDS}$ and $v_o \text{=} 0$. \par
To understand why a normal capacitor may behave as a negative capacitance when connected as the sensor of the Multiple-Sensor Interface (this is, a way to check that ${C_s}(f)\text{<}0$ is possible/expected), is important to highlight some things already explored on the previous sections:\par
1) $C_1\text{=}C_2$, $R_1 \gg R_2$.\par
2) The primary path (always available) to charge $C_1$ is through $R_1$, the primary path (always available) to charge $C_2$ is through $R_2$, since  $R_1 \gg R_2$ and $C_1\text{=}C_2$ this implies that capacitor $C_2$ will charge/discharge much faster(takes less time) than capacitor $C_1$.\par
3) The purpose of sensor $C_s$ on this circuit is to act as a variable impedance that can establish an alternative path on the circuit ($V_o$$\rightarrow$$R_2$$\rightarrow$$C_s$$\rightarrow$$C_1$) to charge/discharge capacitor $C_1$; so $C_s\nearrow \hspace{1mm} \Rightarrow \hspace{1mm} |Z_s|\searrow \hspace{1mm} \Rightarrow R_{approx}\searrow \Rightarrow \tau_2\searrow \Rightarrow C_1 \text{charges faster}$.\par
4) No matter how small $|Z_s|$ may be the capacitor $C_2$ will always charge/discharge faster than capacitor $C_1$, and on the limit where $|Z_s|\text{=}0$ the capacitors $C_1$ and $C_2$ will be charged/discharged simultaneously.\par
5) The capacitors $C_1$, $C_2$, $C_s$ have same working principles, but consequence of their position within the circuit they serve different functions; so $C_1$ and $C_2$ are working as storage of electrical charge (Ex: they start discharged and end charged), while $C_s$ is working as a connection with 'impedance' against charge flow (Ex: starts discharged and will end discharged, that is: ${{v_s}(t\text{=}0})\text{=}0$ and ${{v_s}(t\text{=}\text{+}\infty})\text{=}0$ ).\par
\begin{figure}[H]
	\centering
		\includegraphics[width=\columnwidth,keepaspectratio]{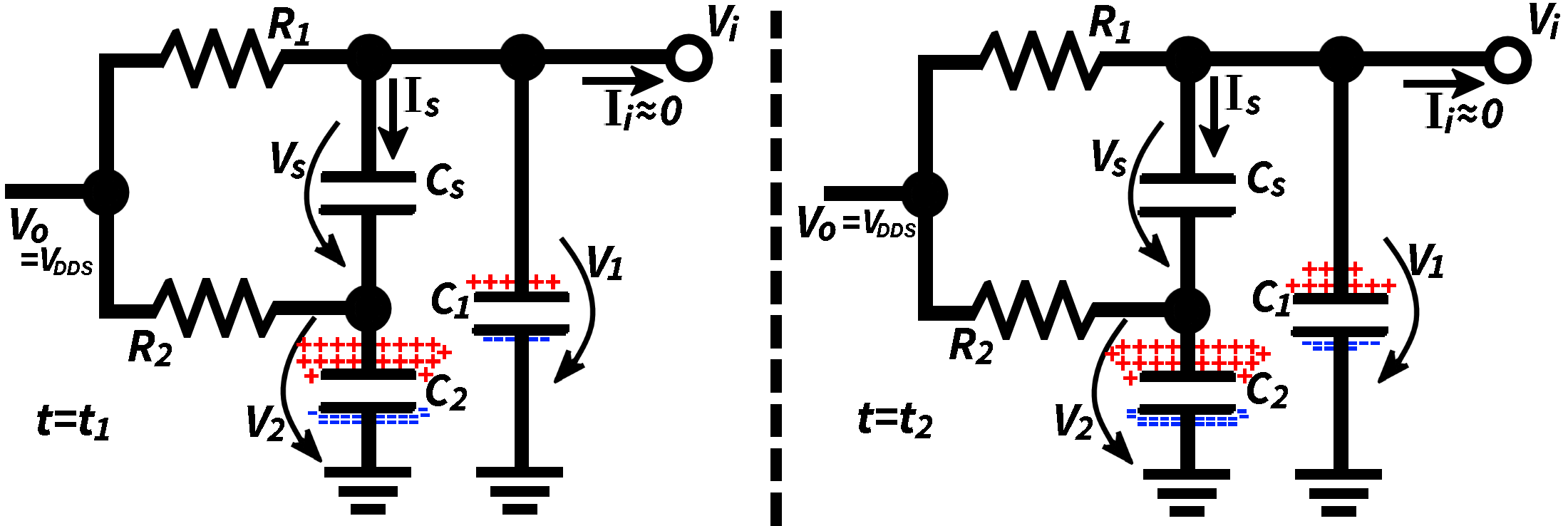}
	\caption{Schematic of RC network of the Schmitt-trigger osc. with a representation of electrical charge on $C_1$, $C_2$ on $t\text{=}t_1$ and  $t\text{=}t_2$.}\label{FIG:showNegativeCsRCOscillatorSchmittTrigger}
\end{figure}

For the following discussion was used as definition of capacitance the formula $C_s=i_s/(dv_s/dt)$, where the $|C_s|=|Q_s|/|v_s|$ ($C_s$: [F] farad; $Q_s$: [C] coulomb; $v_s$: [V] volt), and since the only purpose is to show how $C_s$ can be a negative number it was used the approximate expression $C_s \approx \overline{i_s} / (\Delta v_s / \Delta t)$ that provides exactly the same sign as the exact formula; the $\overline{i_s}$ is the average(mean) value of $i_s$ between $t\text{=}t_1$ and $t\text{=}t_2$ . To show is possible $C_s$$<$0 were considered qualitative relations of the circuit electrical parameters on the RC network of the oscillator, the relevant electrical parameters and their variation between $t\text{=}t_1$ and $t\text{=}t_2$ is represented in Fig.\ref{FIG:showNegativeCsRCOscillatorSchmittTrigger}.\par

It were assumed symbolic values for the voltages on the circuit, used as specimen values to determine how fast a voltage is changing between $t_1$ and $t_2$ time moments. So for representing a small amount of electrical charge are used the symbols: [+] for positive charge and [-] for negative charge, since already stated $C_1=C_2$ for each additional amount of [+] and [-] charge stored on each plate (of $C_1$ or $C_2$) will cause an increase of capacitor voltage that will be represented as $[+\underline{v}]$, where $C_1=C_2=[+]/[+\underline{v}]$. \par
As visible in Fig.\ref{FIG:showNegativeCsRCOscillatorSchmittTrigger}, $V_{o}\text{=}{V_{DDS}}$$\approx$+4.18V, and so ${V_{DDS}}$ will eventually be the voltage on $C_1$ and $C_2$ when t$\rightarrow$$\infty$. For making visual on the schematic the charging process, the charge accumulated in $C_1$, $C_2$ was divided in 20 sets, each represented by [+], [-]; and for each set of accumulated charge is associated a corresponding increase in voltage of $[+ \underline{v}]$, and so $[+ \underline{v}]={V_{DDS}}/20$. \par
Accordingly in Fig.\ref{FIG:showNegativeCsRCOscillatorSchmittTrigger} is represented that $C_2$ is charged to near the final value (${V_{DDS}}$) during the interval $[0;t_1]$ while $C_1$ charges much slower. During interval $[t_1;t_2]$ is visible that $C_2$ increased its charge only by 1[+] becoming charged to approximately(or practically) its final value($v_2 \approx {V_{DDS}}$), whether $C_1$ is still charging and $v_1$ is far from its final value(${V_{DDS}}$), but interestingly $v_1$ is now increasing faster than $v_2$, because $v_2$ already reached its final value, this is $dv_1/dt > dv_2/dt, \forall t \in [t_1;t_2]$. The specimen values here mentioned are in line with the exponential function typical of capacitors charging through a resistor, where lets say a capacitor initially charges very fast, when has some charge stored it charges more slowly, and when close to being full it charges very slowly (where full means the capacitor voltage is close to power supply voltage).

\subsubsection{Voltage and current specimens for \texorpdfstring{$t\text{=}t_1$}{t=t1}}
 So looking at the schematic on left side of Fig.\ref{FIG:showNegativeCsRCOscillatorSchmittTrigger} is visible $C_1$ and $C_2$ are charging and for $t\text{=}t_1$ the charge on $C_1$ is 5[+] and on $C_2$ is 19[+], so capacitor $C_2$ is almost charged while $C_1$ is still charging.  Capacitor $C_1$ is charging through the path $V_o$$\rightarrow$$R_1$$\rightarrow$$C_1$ but mainly is charging through path $V_o$$\rightarrow$$R_2$$\rightarrow$$C_s$$\rightarrow$$C_1$, since $v_2$$>$$v_1$ then $i_s(t\text{=}t_1)<0$.\newline
For $t\text{=}t_1$, $Q_1\text{=}5[+]$, $Q_2\text{=}19[+]$, then $v_1\text{=}5[+\underline{v}]$, $v_2\text{=}19[+\underline{v}]$, since $v_s$=$v_1$$-$$v_2$ then $v_s(t\text{=}t_1)$=5[$+\underline{v}$]$-$19[$+\underline{v}$]=$-$14[$+\underline{v}$].

\subsubsection{Voltage and current specimens for \texorpdfstring{$t\text{=}t_2$}{t=t2}}
 So looking at the schematic on right side of Fig.\ref{FIG:showNegativeCsRCOscillatorSchmittTrigger} is visible $C_1$ and $C_2$ are charging and for $t\text{=}t_2$ the charge on $C_1$ is 9[+] and on $C_2$ is 20[+], so capacitor $C_2$ is fully charged while $C_1$ is still charging.   Capacitor $C_1$ is charging through the path $V_o$$\rightarrow$$R_1$$\rightarrow$$C_1$ but mainly is charging through path $V_o$$\rightarrow$$R_2$$\rightarrow$$C_s$$\rightarrow$$C_1$, since $v_2$$>$$v_1$ then $i_s(t\text{=}t_2)<0$.\par
For $t\text{=}t_2$, $Q_1\text{=}9[+]$, $Q_2\text{=}20[+]$, then $v_1\text{=}9[+\underline{v}]$, $v_2\text{=}20[+\underline{v}]$, since $v_s$=$v_1$$-$$v_2$ then $v_s(t\text{=}t_2)$=9[$+\underline{v}$]$-$20[$+\underline{v}$]=$-$11[$+\underline{v}$].

\subsubsection{\texorpdfstring{Sign of $C_s$ as calculated from $v_s$ and $i_s$ during $[t_1;t_2]$}{Sign of Cs as calculated from vs and is during [t1;t2]}}
The schematics in Fig.\ref{FIG:showNegativeCsRCOscillatorSchmittTrigger} refer to a charging cycle of the Schmitt Trigger Oscillator. Also $t_2$$>$$t_1$$\rightarrow$$\Delta$t$>$0.\par
For $t$$\in$[$t_1$;$t_2$] the capacitor $C_1$ is being charged through the path $V_o$$\rightarrow$$R_2$$\rightarrow$$C_s$$\rightarrow$$C_1$ and so $i_s(t)<0, \forall t\in[t_1;t_2] \Rightarrow \overline{i_s}<0$.\newline
Also $\Delta v_s$ between $t_1$ and $t_2$ is $\Delta v_s$=$v_s(t\text{=}t_2)$$-$$v_s(t\text{=}t_1)$ = $-$11[$+\underline{v}$]$-$($-$14[$+\underline{v}$])=3[$+\underline{v}$], so $\Delta$$v_s$$>$0 between $t_1$ and $t_2$.\par
So concluding between $t_1$ and $t_2$, $\Delta t > 0$, $\Delta v_s > 0$, $\overline{i_s}<0$ $\Rightarrow C_s < 0$ accordingly with $C_s \approx \overline{i_s} / (\Delta v_s / \Delta t)$.

\subsection{Comparison to known cases of negative capacitance}
Aspects of Multiple Sensor Interface circuit possibly related to negative capacitance phenomenon:\newline
1- Use of Schmitt-trigger 'NOT' gate which exhibits hysteresis on its $v_{o}(v_{i})$ graph.\newline
2- Multiple Sensor Interface with a capacitive sensor operates under transient(time domain), step change of voltage caused by its 'NOT' gate(Schmitt-trigger) alternating between 0V and +$V_{DDS}$ (relaxation oscillator).\par 
Negative capacitance phenomenon is reported in some scientific articles/texts, and interestingly with some coincidence to the 2 aspects mentioned above. Quotes:\newline
1- "Effective negative capacitance has been postulated in ferroelectrics because there is hysteresis in plots of polarization-electric field.", article "Towards steep slope MOSFETs using ferroelectric negative capacitance", by A. O'Neill, year 2014 \cite{SteepSlopeMOSFETFerroelectricNegCap}.\newline
2- "The phenomenon of negative capacitance, which has been reported in a variety of situations involving electrolytic as well as electronic systems,  ...  . It is suggested that the physically correct approach lies in the analysis of the corresponding time-domain behavior under step function bias, which involves a current initially falling and then rising gradually over a period of time before finally decaying to zero.", article "The physical origin of negative capacitance", by A. K. Jonscher, year 1986 \cite{PhysicalOriginNegCap}.\par

\section{Conclusions}\label{SEC:Conclusions}

The author theoretically demonstrated a more versatile design for use with sensor applications, also was provided experimental data that corroborates the presented theory. The motivation of the author was to make available an electronics design that could be more sustainable in terms of life-cycle duration, by making a design more customizable by the user and also not closed/locked to a specific application/purpose. No warranty is given that the design can provide accuracy or convenience to a specific application/use; as the article is focused on showing how a versatile design can be achieved.

\section*{Conflicts of interest}
The author declares no conflict of interest.

\section*{Acknowledgments}
I thank in general, to the Open-Source community for making available technology that everyone can use and build-on freely, thus inspiring me to also release this project as Open-Source. Also thanks to GNUplot software, that was used for drawing the plots in this article \cite{GNUplot} .\par
Also thanks to Wolfram Research Inc. for providing Wolfram Mathematica\textsuperscript{®} for Raspberry Pi with RaspberryPi-OS and so making their software easily available to use by everyone, it was used version 12 for obtaining/solving inverse function and algebraic manipulation of long equations \cite{WolframMathematicaForRaspberryPi} .

\appendix

\titleformat{\section}{\normalfont\fontsize{10}{12}\bfseries}{\appendixname~\thesection .}{0.5em}{}

\begin{appendices}

\normalsize

\section[\appendixname~\thesection]{Experimental Datasets}\label{SEC:AppxExpDatasets}
\normalsize
Experimental data obtained (Fig.\ref{FIG:showTestEquipmentMultipleSensor}) by using fixed value components connected as the sensor on the device. Were used arrays(PCBs) with inductors, resistors, capacitors that allow to obtain various different values just by changing a jumper/switch, and also single components (including in series or parallel association).
The tolerance of the PCB components is: $\pm$5$\%$ for R1, R2, and $\pm$10$\%$ for C1, C2.

\subsection[\appendixname~\thesubsection]{Frequency measurement by Multi-Sensor}
The Multiple-Sensor device measures frequency using a counter inside the microcontroller and has some accuracy and range limitations, it can measure up to 3MHz (higher frequency causes counter overflow). The Multiple-Sensor device was tested with a square wave signal from the signal generator JDS6600 (by Joy-IT, frequency accuracy: $\pm{20}$ppm).\par
The Multiple-Sensor device measurement accuracy (percentage error) of frequency, is worst at low frequencies with 9\% error at 100Hz and 0.7\% error at 1kHz, above 5kHz the error was always smaller than 0.2\% (ignoring any accuracy error by JDS6600 used as reference). The Multiple-Sensor device measurement precision (variation) for frequency was worst at low frequencies with 5\% variation at 300Hz, above 1500Hz was always smaller than 1\%, and above 15kHz was always smaller than 0.1\%.

\subsection[\appendixname~\thesubsection]{Experimental data on Multi-Sensor Int}
\textbf{- Reference instruments:} \hfill \phantom{'} \newline
The measurements of inductance($L_s$) and capacitance($C_s$) were obtained using the LCR meter TH2821A (by Tonghui, basic accuracy 0.3\%), configured to 10kHz test signal (for $L_{s}$>202mH was used 1kHz test signal).\par
The measurements of resistance($R_s$) were obtained using the meter UT603 (by UNI-T, accuracy: 0.8\% for R$\leqslant$2M$\Omega$; 2\% for R$>$2M$\Omega$).\par
\textbf{- Units:} \hspace{3mm} Hz$=$hertz, H$=$henry,  $\Omega$$=$ohm, F$=$farad.\par
\textbf{- Jumper Configurations:}\par
 \qquad \textsuperscript{a}(JPA on, JPB off): $C_1$=2.2nF; $C_2$=21.8pF.\par
 \qquad \textsuperscript{b}(JPA off, JPB on): $C_1$=21.8pF; $C_2$=2.2nF.\par
 \qquad \textsuperscript{c}(JPA on, JPB on): $C_1$=2.2nF; $C_2$=2.2nF .\par

\vspace{0.5mm}

Here is made available, the sets of experimental data that were used for drawing the plots of $L_{s}(f)$, $R_{s}(f)$, $C_{s}(f)$, these are the measured values of inductance, resistance, capacitance paired with measured frequency on the Multiple Sensor Interface device.

On \mbox{Appendixes \ref{SEC:AppxExpDatasets} and \ref{SEC:AppxAdditionalExpDatasets}} the symbols $R_{s}$, $C_{s}$, $L_{s}$ usually are/mean the same as $R_{sensor}$, $C_{sensor}$, $L_{sensor}$ and refer to measured values by the reference instruments. So on \mbox{Ap. \ref{SEC:AppxExpDatasets}} and {Ap. \ref{SEC:AppxAdditionalExpDatasets}}, unless an explicit reference to a theoretical function is made, then $R_{s}$=$R_{sensor}$, $C_{s}$=$C_{sensor}$, $L_{s}$=$L_{sensor}$.

\vspace{2mm}

\begin{table}[!htbp]
\caption{Set of experimental data for \texorpdfstring{$C_s$}{Cs} vs. frequency}
\label{TAB:tabCsvsFreq_c}
\setlength{\tabcolsep}{3pt}
\setlength{\extrarowheight}{0pt}
\begin{tabular}{|l|l|}
\hline
\thead{$C_{s}$[nF]\\Capacitance}& 
\thead{f[Hz]$^{\mathrm{c}}$\\JPA on, JPB on}\\
\hline
0&229\\
0.152&321\\
0.310&458\\
0.568&2614\\
0.615&54570\\
0.689&110286\\
0.776&140178\\
1.015&186522\\
1.34&231460\\
1.58&255526\\	
1.79&271015\\
2.00&285020\\
2.56&314530\\
2.99&329820\\
3.98&356012\\
\hline
\end{tabular}
\hspace{5mm}
\begin{tabular}{|l|l|}
\hline
\thead{$C_{s}$[nF]\\Capacitance}& 
\thead{f[Hz]$^{\mathrm{c}}$\\JPA on, JPB on}\\
\hline
4.97&373045\\
5.97&385032\\
6.95&394283\\
7.94&400750\\
8.98&405766\\
10.07&411246\\
12.04&418028\\
15.02&426040\\
20.02&433089\\
24.97&437446\\
29.68&440382\\
34.62&442584\\
39.43&444235\\
44.38&445688\\
49.38&446712\\
\hline
\end{tabular}\newline
Precision error(maximum frequency variation):\newline
$\pm$3kHz (600pF$\leqslant$$C_{s}$$<$1.6nF); $\pm$1kHz ($C_{s}$$\geqslant$21nF);\newline
$\pm$2kHz (1.6nF$\leqslant$$C_{s}$$<$21nF); $\pm$50Hz ($C_{s}$$<$600pF).\newline
\end{table}

\begin{table}[!htbp]
\caption{Set of experimental data for \texorpdfstring{$L_s$}{Ls} vs. frequency}
\label{TAB:tabLsvsFreq_abc}
\setlength{\tabcolsep}{3pt}
\setlength{\extrarowheight}{0pt}
\begin{tabular}{|l|l|l|l|}
\hline
\thead{$L_{s}[{\mu}H]$\\Inductance}& 
\thead{f[Hz]$^{\mathrm{a}}$\\JPA on, JPB off}& 
\thead{f[Hz]$^{\mathrm{b}}$\\JPA off, JPB on}&
\thead{f[Hz]$^{\mathrm{c}}$\\JPA on, JPB on}\\
\hline
1.21&834161&879205&446590\\
1.40&833855&885138&476589\\
1.65&833350&888501&473393\\
1.85&833014&887049&458577\\
2.51&831302&881055&494096\\
3.09&829880&892844&474739\\
3.80&828045&898700&467354\\
4.10&827433&891116&466711\\
4.70&826011&881239&496405\\
5.87&823121&873380&519263\\
7.32&819681&915412&577595\\
8.76&813947&864236&591646\\
9.70&813397&858548&615743\\
11.77&808321&936894&1432920\\
12.84&805905&934723&1357220\\
15.76&799009&895932&1251070\\
21.39&785691&931940&1090760\\
24.49&778123&978330&1011830\\
31.80&760632&1056660&889633\\
38.61&745112&1193360&809559\\
46.70&726580&1369450&736014\\
53.44&712070&2853370&690970\\
61.30&695603&2662840&639014\\
78.30&655604&2332760&568099\\
95.34&620697&2103380&509845\\
117.60&583023&1899190&464617\\
142.28&541388&1722220&420062\\
173.50&502398&1575330&382571\\
201.50&471222&1464370&354070\\
271.46&396546&1201810&292482\\
341.8&368641&1112820&271168\\
360.6&357938&1085180&261856\\
438.7&327053&994721&239762\\
558.1&287880&868105&212362\\
660.7&262804&794697&194290\\
777.6&241031&737803&178480\\
921.2&218738&677591&163587\\
1491&161768&606844&124338\\
2171&130897&499264&102978\\
2976&109912&446452&87856\\
3170&105439&406011&84614\\
3640&97779&383305&79110\\
4646&84110&319423&69783\\
6880&68162&282360&57750\\
10140&54034&224090&47230\\
15040&43377&192088&38790\\
20375&36588&171981&33683\\
\hline
\end{tabular}\newline
Precision error(maximum frequency variation):\newline
$\pm$2kHz (at high 'f[Hz]'); $\pm$300Hz (at low 'f[Hz]');\newline
$\pm$5kHz (2.85MHz$\leftrightarrow$1.36MHz; at $^{\mathrm{b}}$ JPA off, JPB on).
\end{table}

\begin{table*}[!htbp]
\caption{Set of experimental data for \texorpdfstring{$R_s$}{Rs} vs. frequency}
\label{TAB:tabRsvsFreq_c}
\setlength{\tabcolsep}{3pt}
\setlength{\extrarowheight}{0pt}
\begin{tabular}{|l|l|}
\hline
\thead{$R_{s}$[$\Omega$]\\Resistance}& 
\thead{f[Hz]$^{\mathrm{c}}$\\JPA on, JPB on}\\
\hline
0&456284\\
1.2&454678\\
2.2&453975\\
3.2&453149\\
4.2&452400\\
5.2&451605\\
6.2&450718\\
7.2&449923\\
8.2&449067\\
9.2&448272\\
10.2&447385\\
15.2&443226\\
20.1&439144\\
25.1&435183\\
30.1&431269\\
35.1&427447\\
40.0&423655\\
45.0&420016\\
50.0&416407\\
55.0&412845\\
60.0&409359\\
65.0&406010\\
69.9&402601\\
74.9&399328\\
79.9&396209\\
84.9&393060\\
89.9&390017\\
94.8&387005\\
99.9&383396\\
119.7&372387\\
139.7&361700\\
159.6&351608\\
179.5&342205\\
199.5&333306\\
219&325050\\
239&317236\\
259&309836\\
279&302879\\
299&296289\\
319&290112\\
339&284225\\
359&278614\\
378&273308\\
398&268278\\
418&263462\\
438&258829\\
458&254471\\
478&250251\\
498&246214\\
518&242361\\
538&238646\\
557&235144\\
\hline
\end{tabular}
\hspace{5mm}
\begin{tabular}{|l|l|}
\hline
\thead{$R_{s}$[$\Omega$]\\Resistance}& 
\thead{f[Hz]$^{\mathrm{c}}$\\JPA on, JPB on}\\
\hline
577&231735\\
597&228447\\
617&225252\\
637&222178\\
657&219228\\
677&216353\\
697&213647\\
717&210971\\
736&208402\\
756&205910\\
776&203464\\
796&201139\\
816&198877\\
836&196659\\
856&194504\\
876&192409\\
896&190406\\
916&188418\\
936&186476\\
955&184611\\
975&182776\\
996&180865\\
1096&172578\\
1195&165208\\
1295&158603\\
1394&152594\\
1494&147150\\
1593&142135\\
1693&137548\\
1792&133298\\
1892&129353\\
1992&125699\\
2090&122258\\
2190&119017\\
2290&115974\\
2390&113115\\
2490&110409\\
2590&107855\\
2690&105424\\
2790&103115\\
2890&100914\\
2990&98819\\
3090&96831\\
3190&94905\\
3290&93070\\
3390&91327\\
3490&89645\\
3580&88024\\
3680&86464\\
3780&84966\\
3880&83529\\
3980&82153\\
\hline
\end{tabular}
\hspace{5mm}
\begin{tabular}{|l|l|}
\hline
\thead{$R_{s}$[$\Omega$]\\Resistance}& 
\thead{f[Hz]$^{\mathrm{c}}$\\JPA on, JPB on}\\
\hline
4180&79553\\
4380&77107\\
4580&74783\\
4780&72612\\
4980&70593\\
5180&68682\\
5380&66863\\
5580&65135\\
5770&63514\\
5970&61985\\
6170&60517\\
6370&59111\\
6570&57765\\
6770&56511\\
6970&55288\\
7170&54126\\
7370&53010\\
7570&51940\\
7760&50915\\
7960&49952\\
8460&47643\\
8960&45548\\
9460&43622\\
9960&41879\\
10460&40243\\
10960&38760\\
11460&37353\\
11960&36053\\
12450&34861\\
12960&33714\\
13450&32659\\
13950&31680\\
14450&30732\\
14950&29861\\
15440&29035\\
15940&28255\\
16440&27522\\
16940&26818\\
17430&26145\\
17930&25503\\
18430&24907\\
18930&24326\\
19420&23775\\
19940&23271\\
20900&22292\\
21900&21375\\
22900&20549\\
23900&19769\\
24900&19066\\
25900&18393\\
26900&17782\\
27900&17201\\
\hline
\end{tabular}
\hspace{5mm}
\begin{tabular}{|l|l|}
\hline
\thead{$R_{s}$[$\Omega$]\\Resistance}& 
\thead{f[Hz]$^{\mathrm{c}}$\\JPA on, JPB on}\\
\hline
28900&16666\\
29900&16176\\
31900&15244\\
33900&14418\\
35900&13669\\
37800&13027\\
39800&12430\\
41800&11880\\
43800&11375\\
45800&10917\\
47800&10488\\
49800&10091\\
54800&9235\\
59800&8531\\
64800&7904\\
69700&7385\\
74700&6926\\
79700&6513\\
84600&6146\\
89600&5825\\
94600&5550\\
99400&5320\\
109400&4862\\
119300&4479\\
129300&4158\\
139300&3883\\
149300&3623\\
159200&3440\\
169200&3256\\
179100&3073\\
189100&2935\\
199100&2798\\
299000&1972\\
398000&1544\\
498000&1284\\
597000&1100\\
697000&978\\
796000&886\\
896000&825\\
995000&779\\
1495000&596\\
1993000&504\\
2490000&458\\
3090000&412\\
4080000&366\\
5080000&351\\
6070000&336\\
7070000&321\\
8050000&321\\
9040000&305\\
 & \\
 & \\
\hline
\end{tabular}\newline
Precision error(maximum frequency variation):\newline
$\pm$1kHz(at low $R_s$); \quad $\pm$300Hz(at 30k$\Omega$); \quad $\pm$100Hz(at high $R_s$).
\end{table*}

\normalsize

\newpage

\section[\appendixname~\thesection]{Additional Experimental Datasets}\label{SEC:AppxAdditionalExpDatasets}

\normalsize

Here are additional experimental datasets, these are measured values of inductance, resistance, capacitance paired with measured frequency by the Multi-Sensor device. These additional experimental datasets were not used on any previous plot/graph displayed along the article, and are here made available to give additional evidence, also with graphs comparing them against the respective theoretical results. The symbols $R_{s}$, $C_{s}$, $L_{s}$ are/mean the same as on \mbox{Appendix \ref{SEC:AppxExpDatasets}}.\par
\textbf{- Reference instruments:} Same as on \mbox{Appendix \ref{SEC:AppxExpDatasets}}.\par
\textbf{- Jumper Configurations/Capacitor Values:}\newline
  \textsuperscript{d}(JPA on, JPB off): $C_1$=93nF ($\approx$91nF+2.2nF); $C_2$=21.8pF.\newline
  \textsuperscript{e}(JPA off, JPB on): $C_1$=21.8pF; $C_2$=93nF ($\approx$91nF+2.2nF).\newline
  \textsuperscript{f}(JPA on, JPB on): $C_1$=93nF; $C_2$=93nF ($\approx$91nF+2.2nF).\par

\textbf{- Note:} The configurations/designs where the references $C_1$ and/or $C_2$ are 93nF($\approx$91nF//2.2nF=91nF+2.2nF), were obtained by connecting a 91nF capacitor in parallel with the 2.2nF capacitor already on the Multi-Sensor PCB.\par

\begin{figure}[H]
	\centering
		\includegraphics[width=\columnwidth,keepaspectratio]{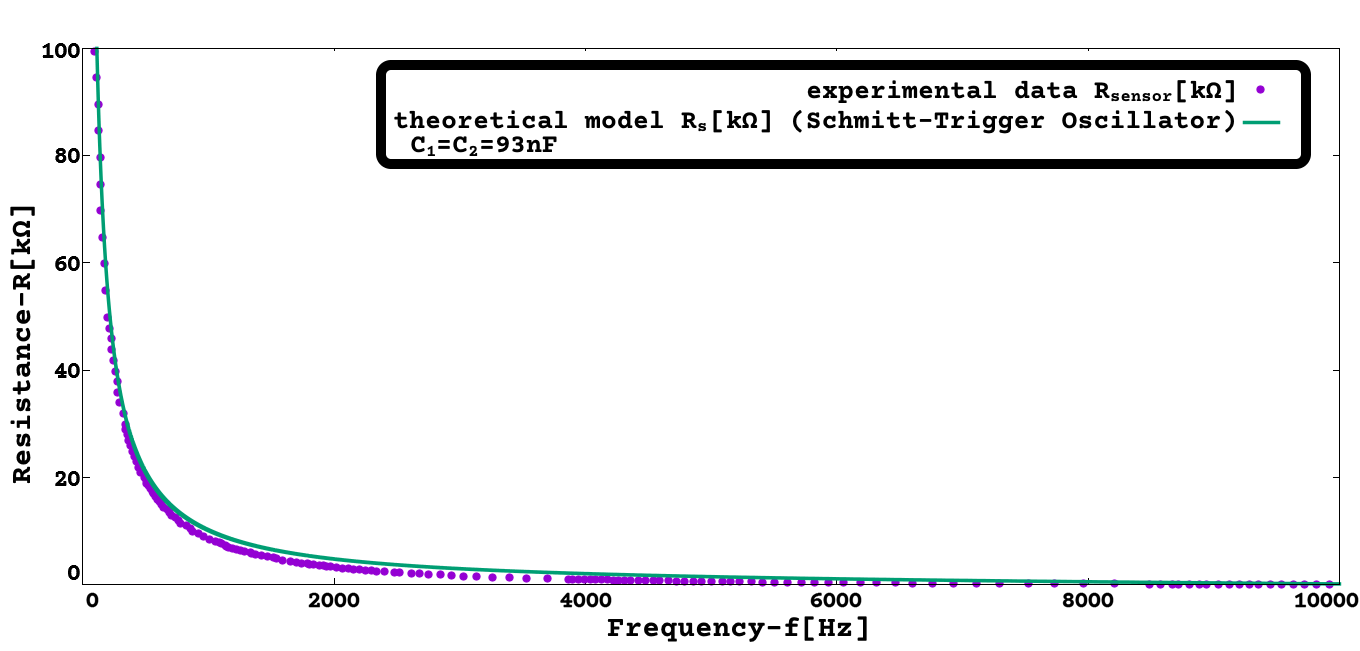}
	\caption{${R_s}(f)$ [k$\Omega$], with ${C_1}\text{=}{C_2}\text{=}93nF$, frequency in [0Hz, 10kHz] (Schmitt-trigger osc., Multiple-Sensor Int. with resistive sensor).}\label{FIG:RSensorVersusFrequencyOscillation93nFC1C2SchmittTrigger}
\end{figure}

The Fig.\ref{FIG:RSensorVersusFrequencyOscillation93nFC1C2SchmittTrigger} shows experimental data for Multi-Sensor device with various resistance values connected as the sensor and also $R_{s}(f)$ using (\ref{EQ:RSensorVersusFrequencyOscillationSchmittTrigger}) with $C_1\text{=}C_2\text{=}93nF$, $R_1\text{=}2M\Omega$, $R_2\text{=}500\Omega$, ${V_T^{-}}\text{=}1.2V$, ${V_T^{+}}\text{=}2.2V$, ${V_{DDS}}\text{=}4.18V$, $H\text{=}1.01496$.

\begin{figure}[H]
	\centering
		\includegraphics[width=\columnwidth,keepaspectratio]{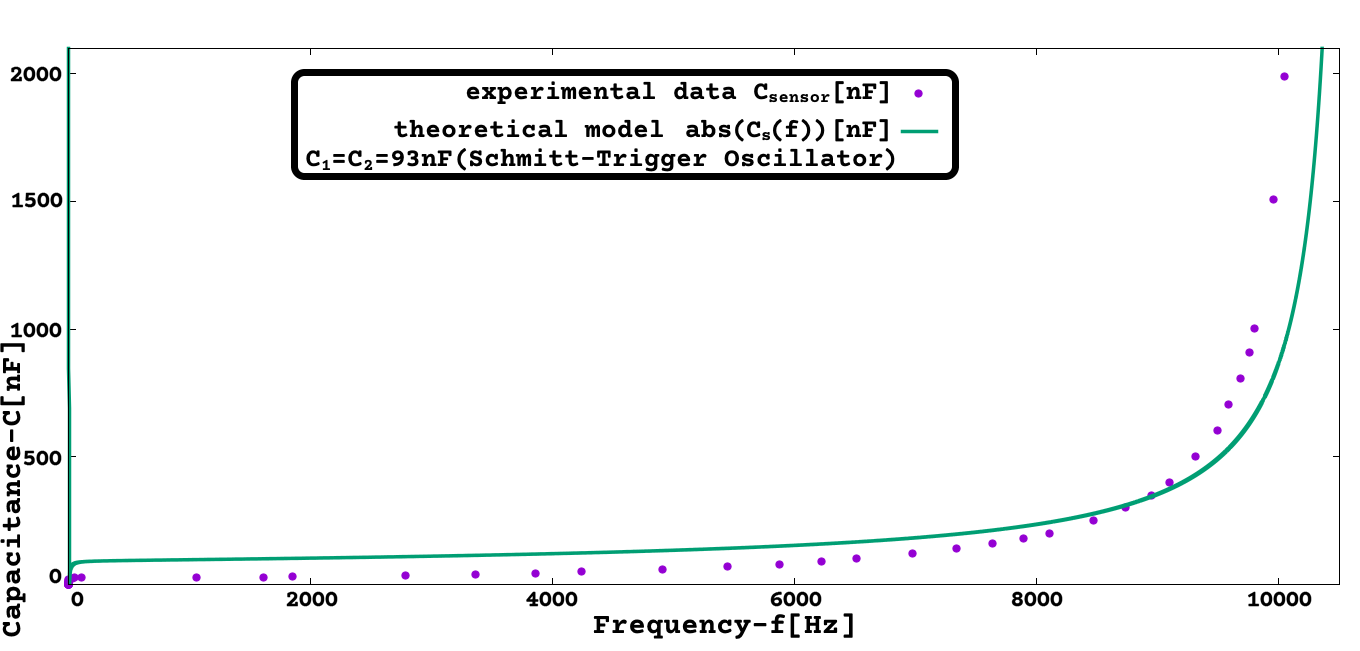}
	\caption{$|{C_s}(f)|$ [nF], with ${C_1}\text{=}{C_2}\text{=}93nF$, frequency in [0Hz, 10500Hz] (Schmitt-trigger osc., Multi-Sensor with capacitive sensor)}\label{FIG:CSensorVersusFrequencyOscillation93nFC1C2SchmittTrigger1}

	\vspace{0.1mm}

	\centering
		\includegraphics[width=\columnwidth,keepaspectratio]{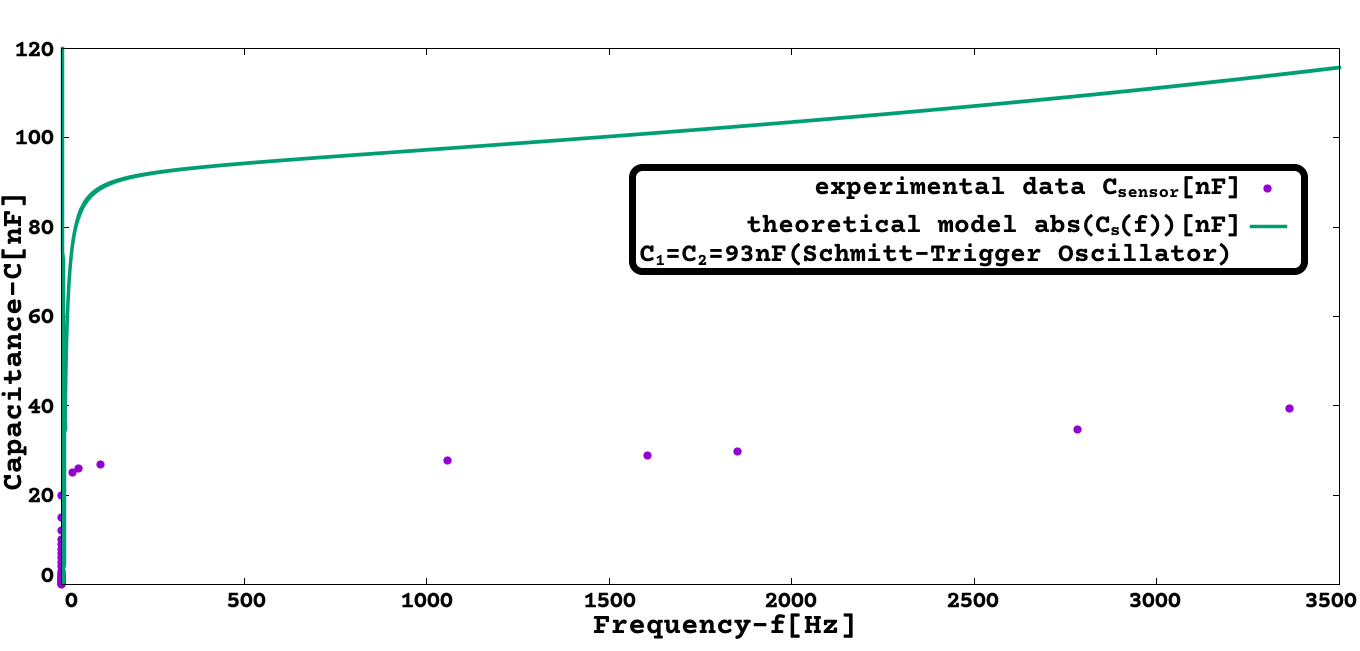}
	\caption{$|{C_s}(f)|$ [nF], with ${C_1}\text{=}{C_2}\text{=}93nF$, frequency in [0Hz, 3500Hz] (Schmitt-trigger osc., Multi-Sensor with capacitive sensor)}\label{FIG:CSensorVersusFrequencyOscillation93nFC1C2SchmittTrigger2}
\end{figure}

The Fig.\ref{FIG:CSensorVersusFrequencyOscillation93nFC1C2SchmittTrigger1} and Fig.\ref{FIG:CSensorVersusFrequencyOscillation93nFC1C2SchmittTrigger2} shows experimental data for Multi-Sensor device with various capacitance values connected as sensor and $|{C_{s}}(f)|$ using (\ref{EQ:CSensorVersusFrequencyOscillationSchmittTrigger}) with $C_1\text{=}C_2\text{=}93nF$, ${V_T^{-}}\text{=}1.2V$, ${V_T^{+}}\text{=}2.2V$, ${V_{DDS}}\text{=}4.18V$, $R_1\text{=}2M\Omega$, $R_2\text{=}500\Omega$, $H\text{=}1.01496$.

\begin{figure}[H]
	\centering
		\includegraphics[width=\columnwidth,keepaspectratio]{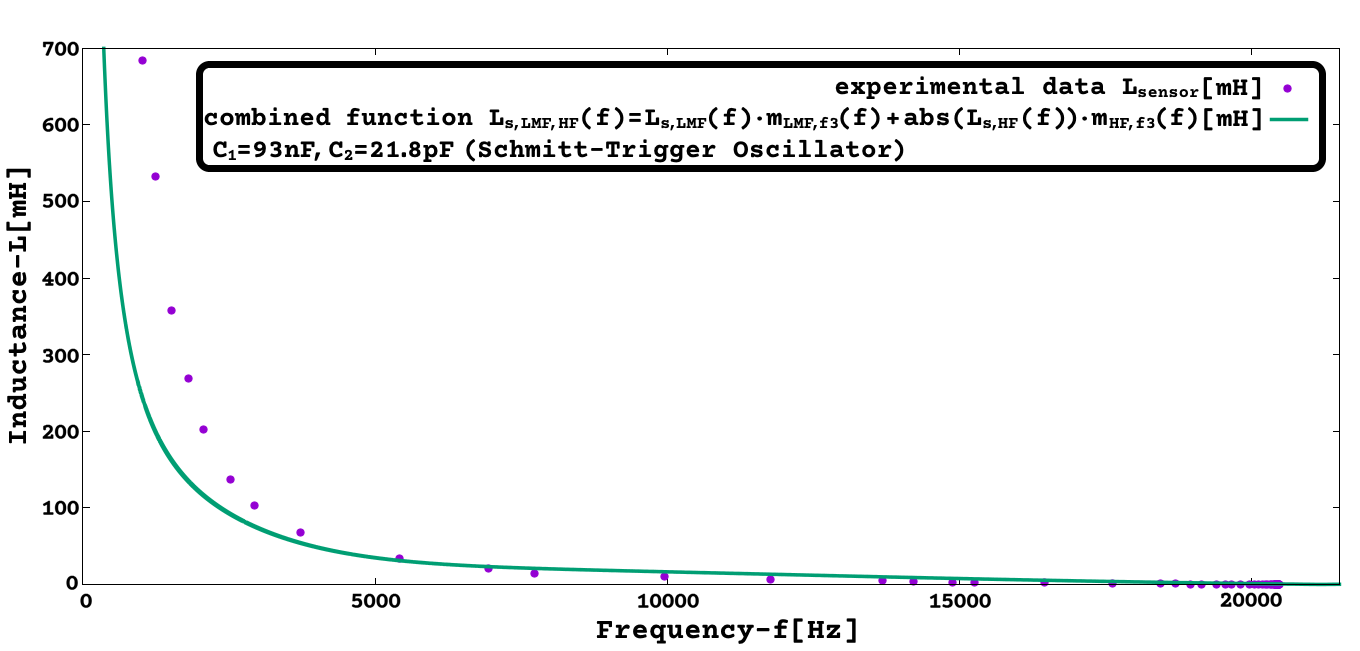}
	\caption{$L_{s,LMF,HF}(f)$ [mH], with $C_1$=93nF, $C_2$=21.78pF, frequency in [0Hz, 21500Hz] (Sch-trig. osc., Multi-Sensor w. inductive sensor)}\label{FIG:LSensorVersusFrequency93nFC1And22pFC2SchmittTriggerLMFAndHFCombined1}

	\vspace{0.1mm}
	
	\centering
		\includegraphics[width=\columnwidth,keepaspectratio]{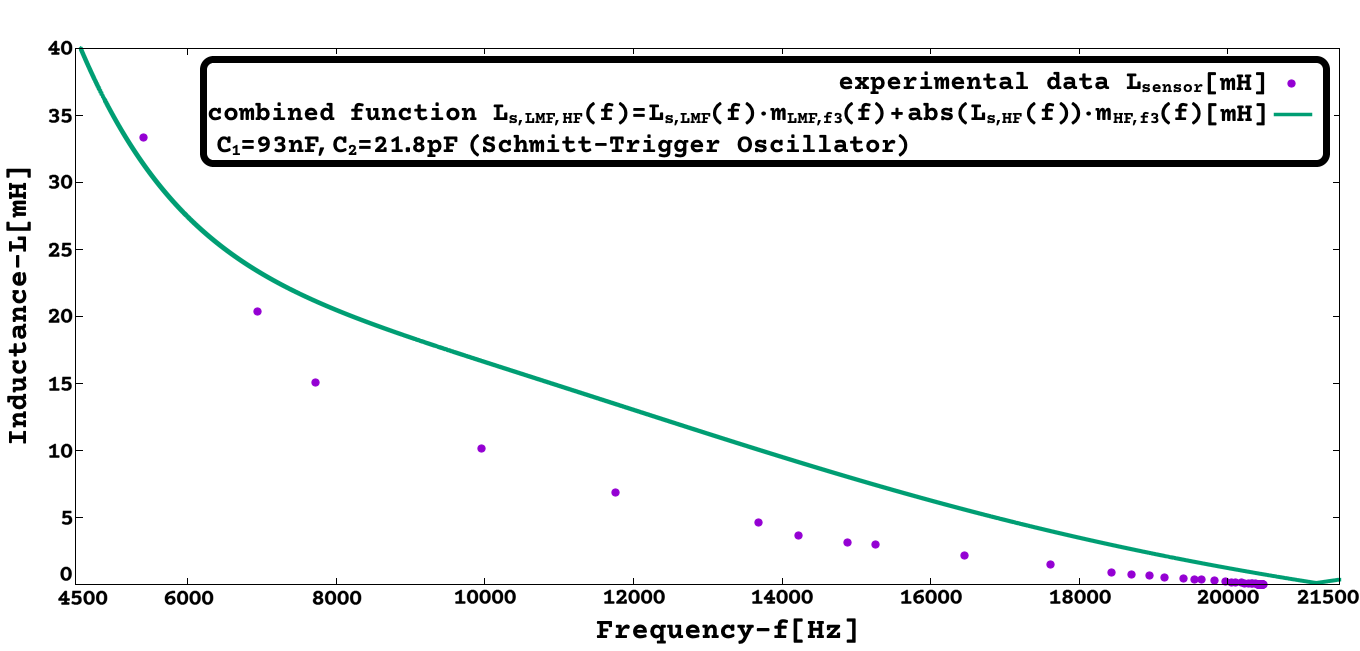}
	\caption{$L_{s,LMF,HF}(f)$ [mH], with $C_1$=93nF, $C_2$=21.78pF, frequency in [4500Hz, 21500Hz] (Sch-trig. osc., Multi-Sensor w. inductive sensor)}\label{FIG:LSensorVersusFrequency93nFC1And22pFC2SchmittTriggerLMFAndHFCombined2}
	\vspace{1mm}
\end{figure}

The Fig.\ref{FIG:LSensorVersusFrequency93nFC1And22pFC2SchmittTriggerLMFAndHFCombined1} and Fig.\ref{FIG:LSensorVersusFrequency93nFC1And22pFC2SchmittTriggerLMFAndHFCombined2} shows experimental data for Multiple-Sensor device with various inductance values connected as the sensor, and also $L_{s,LMF,HF}(f)$ using (\ref{EQ:LSensorVersusFrequencyOscillationSchmittTriggerLMFAndHFModel}) with $C_1$=93nF, $C_2$=21.78pF, $R_1\text{=}2M\Omega$, $R_2\text{=}500\Omega$, $H\text{=}1.01496$, ${m_{HF,f3}}(f$=$2{\cdot}f_{0}/3)$=0.5, ${m_{LMF,f3}}(f$=$f_{0})$=0.01, ${a_{HF}}=3.548{\cdot}{10^{-13}}$ $s^3$, ${a_{LMF}}=1.041{\cdot}10^{-11}$ $s^3$.

\begin{table}[H]
\caption{Additional set of experimental data for \texorpdfstring{$C_s$}{Cs} vs. frequency}
\label{TAB:tabCsvsFreq_f}
\setlength{\tabcolsep}{3pt}
\setlength{\extrarowheight}{0pt}
\begin{tabular}{|l|l|}
\hline
\thead{$C_{s}$[nF]\\Capacitance}& 
\thead{{f[Hz]$^{\mathrm{f}}$}\\JPA on, JPB on\\$C_1{\text{=}}C_2{\text{=}}93nF$}\\
\hline
0.152&0\\
0.568&0\\
1.015&0\\
1.58&0\\
2.00&0\\
2.99&0\\
3.98&0\\
4.97&0\\
5.97&0\\
7.94&0\\
10.07&0\\
12.04&0\\
15.02&0\\
20.02&0\\
24.97&30\\
25.89&45\\
26.87&107\\
27.86&1055\\
28.87&1605\\
29.68&1850\\
34.62&2782\\
39.43&3363\\
44.38&3853\\
\hline
\end{tabular}
\hspace{5mm}
\begin{tabular}{|l|l|}
\hline
\thead{$C_{s}$[nF]\\Capacitance}& 
\thead{{f[Hz]$^{\mathrm{f}}$}\\JPA on, JPB on\\$C_1{\text{=}}C_2{\text{=}}93nF$}\\
\hline
49.38&4235\\
59.39&4908\\
69.62&5443\\
79.96&5871\\
90.38&6223\\
100.38&6513\\
120.27&6972\\
139.62&7339\\
159.66&7629\\
180.22&7889\\
201.13&8103\\
250.42&8470\\
300.0&8730\\
349.3&8944\\
400.2&9097\\
502.9&9311\\
603.7&9495\\
706.7&9586\\
806.5&9678\\
909.1&9755\\
1003.5&9800\\
1507.3&9953\\
1991.7&10045\\
\hline
\end{tabular}\newline
Precision error(maximum frequency variation):\newline
$\pm$15Hz ($C_{s}$$\leqslant$34.62nF); $\pm$50Hz (100.38nF$<$$C_{s}$$\leqslant$180.22nF);\newline
$\pm$30Hz (34.62nF$<$$C_{s}$$\leqslant$100.38nF); $\pm$100Hz ($C_{s}$$>$180.22nF).\newline
\end{table}

\begin{table*}[!htbp]
\caption{Additional set of experimental data for \texorpdfstring{$R_s$}{Rs} vs. frequency}
\label{TAB:tabRsvsFreq_f}
\setlength{\tabcolsep}{3pt}
\setlength{\extrarowheight}{0pt}
\begin{tabular}{|l|l|}
\hline
\thead{$R_{s}$[$\Omega$]\\Resistance}& 
\thead{{f[Hz]$^{\mathrm{f}}$}\\JPA on, JPB on\\$C_1{\text{=}}C_2{\text{=}}93nF$}\\
\hline
0.1&10198\\
1.2&10183\\
2.2&10167\\
3.2&10152\\
4.2&10137\\
5.2&10106\\
6.2&10091\\
7.2&10076\\
8.2&10045\\
9.2&10030\\
10.2&10014\\
15.2&9923\\
20.1&9816\\
25.1&9724\\
30.1&9632\\
35.1&9540\\
40.0&9449\\
45.0&9357\\
50.0&9281\\
55.0&9204\\
60.0&9128\\
65.0&9036\\
69.9&8944\\
74.9&8883\\
79.9&8807\\
84.9&8715\\
89.9&8669\\
94.8&8577\\
99.9&8485\\
119.7&8210\\
139.7&7966\\
159.6&7736\\
179.5&7522\\
199.5&7293\\
219&7109\\
239&6926\\
259&6758\\
279&6605\\
299&6467\\
319&6314\\
339&6192\\
359&6054\\
378&5932\\
398&5825\\
418&5718\\
438&5611\\
458&5504\\
478&5412\\
498&5320\\
518&5229\\
538&5152\\
557&5091\\
\hline
\end{tabular}
\hspace{5mm}
\begin{tabular}{|l|l|}
\hline
\thead{$R_{s}$[$\Omega$]\\Resistance}& 
\thead{{f[Hz]$^{\mathrm{f}}$}\\JPA on, JPB on\\$C_1{\text{=}}C_2{\text{=}}93nF$}\\
\hline
577&4999\\
597&4923\\
617&4862\\
637&4785\\
657&4724\\
677&4663\\
697&4587\\
717&4541\\
736&4479\\
756&4418\\
776&4357\\
796&4311\\
816&4265\\
836&4220\\
856&4174\\
876&4128\\
896&4082\\
916&4036\\
936&3990\\
955&3944\\
975&3898\\
996&3868\\
1096&3700\\
1195&3531\\
1295&3394\\
1394&3256\\
1494&3134\\
1593&3027\\
1693&2935\\
1792&2843\\
1892&2752\\
1992&2675\\
2090&2614\\
2190&2522\\
2290&2476\\
2390&2400\\
2490&2339\\
2590&2293\\
2690&2247\\
2790&2201\\
2890&2155\\
2990&2110\\
3090&2064\\
3190&2018\\
3290&1972\\
3390&1941\\
3490&1911\\
3580&1880\\
3680&1834\\
3780&1804\\
3880&1788\\
3980&1743\\
\hline
\end{tabular}
\hspace{5mm}
\begin{tabular}{|l|l|}
\hline
\thead{$R_{s}$[$\Omega$]\\Resistance}& 
\thead{{f[Hz]$^{\mathrm{f}}$}\\JPA on, JPB on\\$C_1{\text{=}}C_2{\text{=}}93nF$}\\
\hline
4180&1697\\
4380&1651\\
4580&1590\\
4780&1544\\
4980&1513\\
5180&1467\\
5380&1421\\
5580&1376\\
5770&1345\\
5970&1330\\
6170&1284\\
6370&1253\\
6570&1223\\
6770&1192\\
6970&1162\\
7170&1146\\
7370&1131\\
7570&1100\\
7760&1085\\
7960&1055\\
8460&1009\\
8960&963\\
9460&917\\
9960&871\\
10460&856\\
10960&825\\
11460&779\\
11960&764\\
12450&733\\
12960&703\\
13450&688\\
13950&672\\
14450&642\\
14950&626\\
15440&611\\
15940&596\\
16440&581\\
16940&565\\
17430&550\\
17930&535\\
18430&519\\
18930&504\\
19420&504\\
19940&489\\
20900&458\\
21900&443\\
22900&428\\
23900&412\\
24900&397\\
25900&382\\
26900&366\\
27900&351\\
\hline
\end{tabular}
\hspace{5mm}
\begin{tabular}{|l|l|}
\hline
\thead{$R_{s}$[$\Omega$]\\Resistance}& 
\thead{{f[Hz]$^{\mathrm{f}}$}\\JPA on, JPB on\\$C_1{\text{=}}C_2{\text{=}}93nF$}\\
\hline
28900&336\\
29900&336\\
31900&321\\
33900&290\\
35900&275\\
37800&275\\
39800&259\\
41800&244\\
43800&229\\
45800&229\\
47800&214\\
49800&198\\
54800&183\\
59800&168\\
64800&152\\
69700&137\\
74700&137\\
79700&137\\
84600&122\\
89600&122\\
94600&107\\
99400&91\\
109400&91\\
119300&91\\
129300&76\\
139300&76\\
149300&61\\
159200&61\\
169200&45\\
179100&45\\
189100&45\\
199100&45\\
299000&30\\
398000&15\\
498000&15\\
597000&0\\
697000&0\\
796000&0\\
896000&0\\
995000&0\\
1495000&0\\
1993000&0\\
2490000&0\\
3090000&0\\
4080000&0\\
5080000&0\\
6070000&0\\
7070000&0\\
8050000&0\\
9040000&0\\
 & \\
 & \\
\hline
\end{tabular}\newline
Precision error(maximum frequency variation):\newline
$\pm$300Hz(at low $R_s$); \quad $\pm$50Hz(at 996$\Omega$); \quad $\pm$30Hz(at high $R_s$).
\end{table*}

\begin{table}[H]
\caption{Additional set of experimental data for \texorpdfstring{$L_s$}{Ls} vs. frequency}
\label{TAB:tabLsvsFreq_def}
\setlength{\tabcolsep}{3pt}
\setlength{\extrarowheight}{0pt}
\begin{tabular}{|l|l|l|l|}
\hline
\thead{$L_{s}[{\mu}H]$\\Inductance}& 
\thead{f[Hz]$^{\mathrm{d}}$\\JPA on, JPB off\\$C_1{\text{=}}93nF$,\\$C_2{\text{=}}21.8pF$}&
\thead{f[Hz]$^{\mathrm{e}}$\\JPA off, JPB on\\$C_1{\text{=}}21.8pF$,\\$C_2{\text{=}}93nF$}&
\thead{f[Hz]$^{\mathrm{f}}$\\JPA on, JPB on\\$C_1{\text{=}}93nF$,\\$C_2{\text{=}}93nF$}\\
\hline
1.21&20458&20503&10259\\
1.85&20412&20519&10259\\
3.09&20412&20488&10259\\
4.70&20442&20503&10259\\
7.32&20458&20503&10259\\
9.70&20442&20519&10244\\
15.76&20458&20503&10259\\
21.39&20472&20503&10244\\
24.49&20458&20503&10274\\
38.61&20396&20503&10229\\
46.70&20366&20503&10274\\
53.44&20335&20503&10320\\
61.30&20320&20503&10091\\
78.30&20274&20503&10473\\
95.34&20228&20503&9984\\
117.60&20182&20534&10473\\
142.28&20106&20503&10488\\
173.50&20045&20503&9892\\
201.50&19968&20534&9907\\
271.46&19815&20427&10917\\
341.8&19647&20503&10229\\
360.6&19555&20595&10045\\
438.7&19403&20565&10657\\
558.1&19143&20519&10366\\
660.7&18944&20626&9724\\
777.6&18699&20503&9128\\
921.2&18439&20503&8669\\
1491&17614&20687&18072\\
2171&16452&20763&14877\\
2976&15259&20595&12751\\
3170&14877&20779&12109\\
3640&14219&20549&11284\\
4646&13684&20870&9953\\
6880&11758&20779&8394\\
10140&9953&20213&6972\\
15040&7721&20962&5764\\
20375&6941&20580&4923\\
33370&5412&20794&3853\\
68050&3715&23286&2752\\
102950&2935&63744&2247\\
136900&2522&62276&1941\\
202650&2064&43408&1651\\
269650&1804&37888&1452\\
358300&1513&21100&1162\\
532200&1238&21008&1009\\
684100&1024&20870&825\\
\hline
\end{tabular}\newline
Precision error(maximum frequency variation):\newline
$\pm$300Hz (at high 'f[Hz]'); $\pm$100Hz (at low 'f[Hz]');\newline
$\pm$2kHz ([37888Hz;63744Hz]; at $^{\mathrm{e}}$ JPA off, JPB on).
\end{table}

\section[\appendixname~\thesection]{LDR Sensor Dataset}\label{SEC:LDRSensorDataset}
\normalsize
The A906013 LDR sensor has the following electrical parameters specified by the manufacturer(PerkinElmer/Excelitas) \cite{ExcelitasLDRSensorDatasheet}:\newline
R(Illum=10[lx]) is between 27k$\Omega$ (min.) and 94k$\Omega$ (max.); \quad R(Illum=100[lx]) = 8k$\Omega$ (typical value) ;\newline $\lambda_{peak}$=600nm ($\lambda_{peak}$ is the wavelength of light the LDR sensor is most sensitive to);\quad R(Illum=0, after 1s)>0.5M$\Omega$ ;\newline R(Illum=0, after 5s)>1.5M$\Omega$; \space \space ${\gamma_{10/100}}$ =  0.8 (typical value).\par
\textbf{- Reference instruments:} \hfill \phantom{'} \par
For obtaining the calibration table of the A906013 LDR sensor on the Multiple-Sensor device it was used as a reference device the Mastech MS6610 luxmeter; the MS6610 luxmeter has the maximum sensitivity at 570nm wavelength, it has an accuracy of  $\pm( 5 \% + 2 digits )$, with a resolution (value per digit) of: [0; 1999 [lx]]: 1 [lx]; \quad [2000 [lx]; 19990 [lx]]: 10 [lx]; \quad [20000 [lx]; 50000 [lx]]: 100 [lx].\par
\textbf{- Jumper Configurations:}\par
 \qquad \textsuperscript{c}(JPA on, JPB on): $C_1$=2.2nF; $C_2$=2.2nF .\par
\textbf{- Units:} \hspace{3mm} Hz$=$hertz, lx$=$lux.\par
Here is made available the calibration of a LDR sensor (Ref: A906013) connected on the Multiple-Sensor Interface.\par
\vspace{2mm}
\begin{table}[!htbp]
\caption{Set of exp. data of LDR sensor for Illuminance[lx] vs. frequency}
\label{TAB:tabIlluminancevsFreq_LDR_c}
\setlength{\tabcolsep}{3pt}
\setlength{\extrarowheight}{0pt}
\begin{tabular}{|l|l|}
\hline
\thead{$Illum$[lx]\\ref. MS6601}& 
\thead{f[Hz]$^{\mathrm{c}}$ (LDR)\\ JPA on, JPB on}\\
\hline
0&244\\
1&2507\\
3&6284\\
10&11788\\
15&16345\\
25&23026\\
30&25320\\
48&32644\\
77&44172\\
100&49768\\
122&56404\\
145&63835\\
173&67346\\
205&75456\\
251&81847\\
300&90180\\
398&102580\\
450&107641\\
512&113375\\
590&120378\\
650&124032\\
715&129124\\
820&135591\\
925&141218\\
1072&147105\\
1200&155682\\
1365&162196\\
\hline
\end{tabular}
\hspace{5mm}
\begin{tabular}{|l|l|}
\hline
\thead{$Illum$[lx]\\ref. MS6601}& 
\thead{f[Hz]$^{\mathrm{c}}$ (LDR)\\ JPA on, JPB on}\\
\hline
1485&165972\\
1660&173663\\
1880&189030\\
1985&193433\\
2160&198678\\
2400&205650\\
2930&217530\\
3950&232041\\
5000&244640\\
6000&252651\\
6950&261810\\
8100&267620\\
9650&275877\\
11000&281121\\
12050&286595\\
13350&291702\\
15500&297130\\
17250&300846\\
19000&305205\\
22600&311227\\
25600&315707\\
28600&319469\\
30300&320891\\
33400&325050\\
37000&328750\\
39800&333459\\
48000&340600\\
\hline
\end{tabular}\newline
Precision error(maximum frequency variation):\newline
$\pm$3kHz (1k[lx]$\leqslant$$Illum$$<$10k[lx]); $\pm$6kHz ($Illum$$\geqslant$10k[lx]);\newline
$\pm$1kHz (10[lx]$\leqslant$$Illum$$<$1k[lx]); $\pm$200Hz ($Illum$$<$10[lx])
\end{table}

\section[\appendixname~\thesection]{LDR Sensor Model Fitting and Error}\label{SEC:LDRSensorCalibAndError}
\normalsize
Here is shown how the theoretical model of ${R_s}(f)$, can be combined with a theoretical model of ${R_{LDR}}(Illum_{LDR})$ of the LDR sensor, to obtain a theoretical function of ${Illum_{LDR}}(f)$ of the LDR sensor plus the Multiple-Sensor interface, that can be fitted to a small experimental dataset, then obtaining a function that can generate values of ${Illum_{LDR}}$ versus frequency that are close/similar to the experimental results.

The used model for the LDR sensor is the same type of model mentioned on the datasheet of the LDR sensor (by PerkinElmer/Excelitas), where the relation between the sensor resistance ($R_{LDR}$) and the Illuminance ($\textit{E}_{v_LDR}$) is approximated by a straight line on a logarithmic scale plot (that is, ${log}(R_{LDR})$ vs ${log}(\textit{E}_{v\_LDR})$) is approximately a straight line), the slope of the straight line can be defined by a constant $\gamma$ where the manufacturer provides an aproximated value of by calculating its value on 2 reference values of Illuminance, typically at 10lx and 100lx.\par
The equation for the LDR sensor model used here is:
\quad \begin{equation}\label{EQ:ResistanceLDRVersusIllum} {R_{LDR}}(E_{v\_LDR}) = {R_{LDR\_0}}{\left(\frac{E_{v\_REF}}{{E_{v\_LDR}}+{E_{v\_REF}}}\right)^{\gamma}}\end{equation} \par
\qquad $\leftrightarrow$ \quad ${\gamma}=\frac{log({R_{LDR}}/{R_{LDR\_0}})}{log({E_{v\_REF}}/({E_{v\_LDR}}+{E_{v\_REF}}))}$ \quad \par
 \quad $ \rightarrow E_{v\_LDR}({R_{LDR}}) = {E_{v\_REF}}\left({\left(\frac{R_{LDR}}{{R_{LDR\_0}}}\right)^{-1/\gamma}}-1\right) $ .\par
\phantom{'}
\par
The ${\gamma}$ equation becomes equal to what is mentioned in the LDR sensor datasheet if ${E_{v\_LDR}}\gg{E_{v\_REF}}$, then:\par 
\quad ${R_{LDR}}(E_{v\_LDR}) \approx {R_{LDR\_0}}{\left(\frac{E_{v\_REF}}{{E_{v\_LDR}}}\right)^{\gamma}}$ \par
\qquad $\leftrightarrow$ \quad  ${\gamma} \approx \frac{log({R_{LDR}}/{R_{LDR\_0}})}{log({E_{v\_REF}}/{E_{v\_LDR}})}$ \quad .\par

The equation (\ref{EQ:ResistanceLDRVersusIllum}) of $E_{v\_LDR}({R_{LDR}})$ can be combined with equation (\ref{EQ:RSensorVersusFrequencyOscillationSchmittTrigger}) of ${R_s}(f)$ thus obtaining: 
\begin{equation}\label{EQ:IllumLDRVersusFrequencyOscillationSchmittTrigger} \resizebox{\hsize}{!}{$E_{v\_LDR}(f) {\approx} {E_{v\_REF}}\left( {\left(\frac{(C_1+C_2){R_2}{R_1}Hf-R_2-R_1}{{R_{LDR\_0}}({C_2}{R_2}Hf-1)({C_1}{R_1}Hf-1)}\right)^{-1/\gamma}} - 1\right)$}\end{equation}

\begin{figure}[H]
	\centering
		\includegraphics[width=\columnwidth,keepaspectratio]{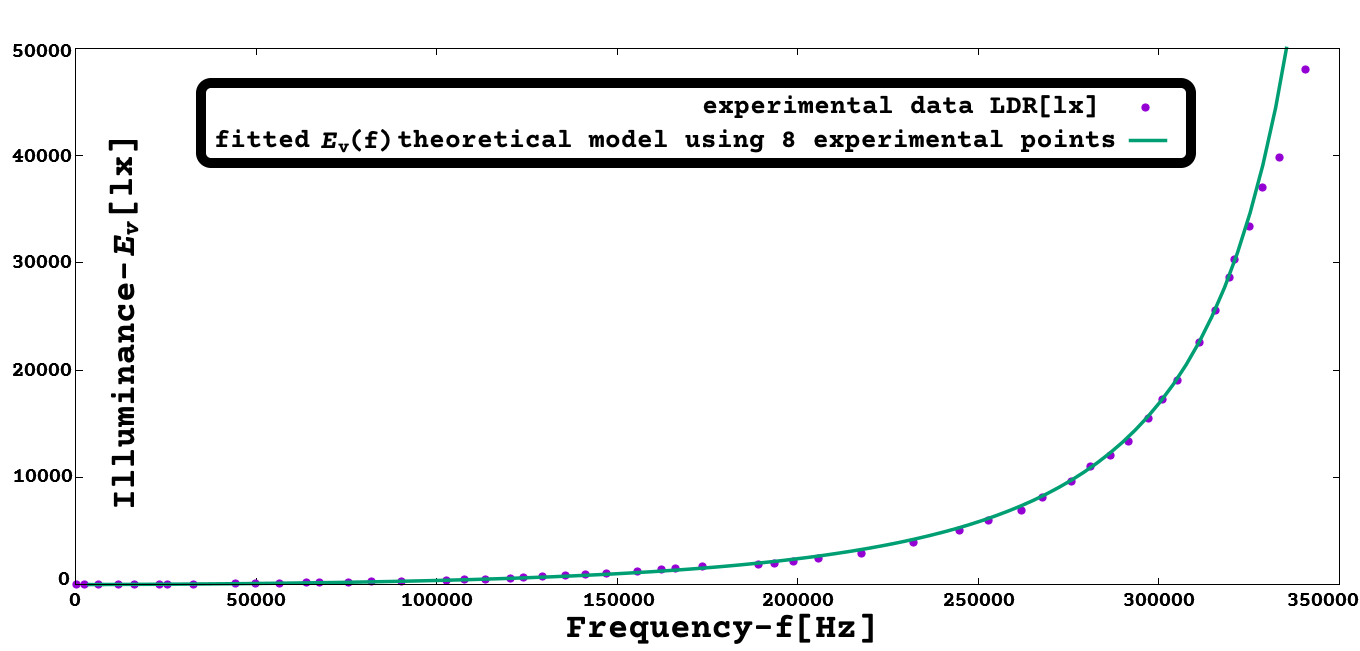}
	\caption{$E_{v\_LDR}(f)$ [lx], with $C_1$=2.2nF, $C_2$=2.2nF, frequency in [0Hz, 350kHz] (Sch-trig. osc., Multi-Sensor w. LDR sensor)}\label{FIG:LDRSensorEvVersusFrequencyRefMS6610FitTheoModel1}

	\vspace{0.1mm}
	
	\centering
		\includegraphics[width=\columnwidth,keepaspectratio]{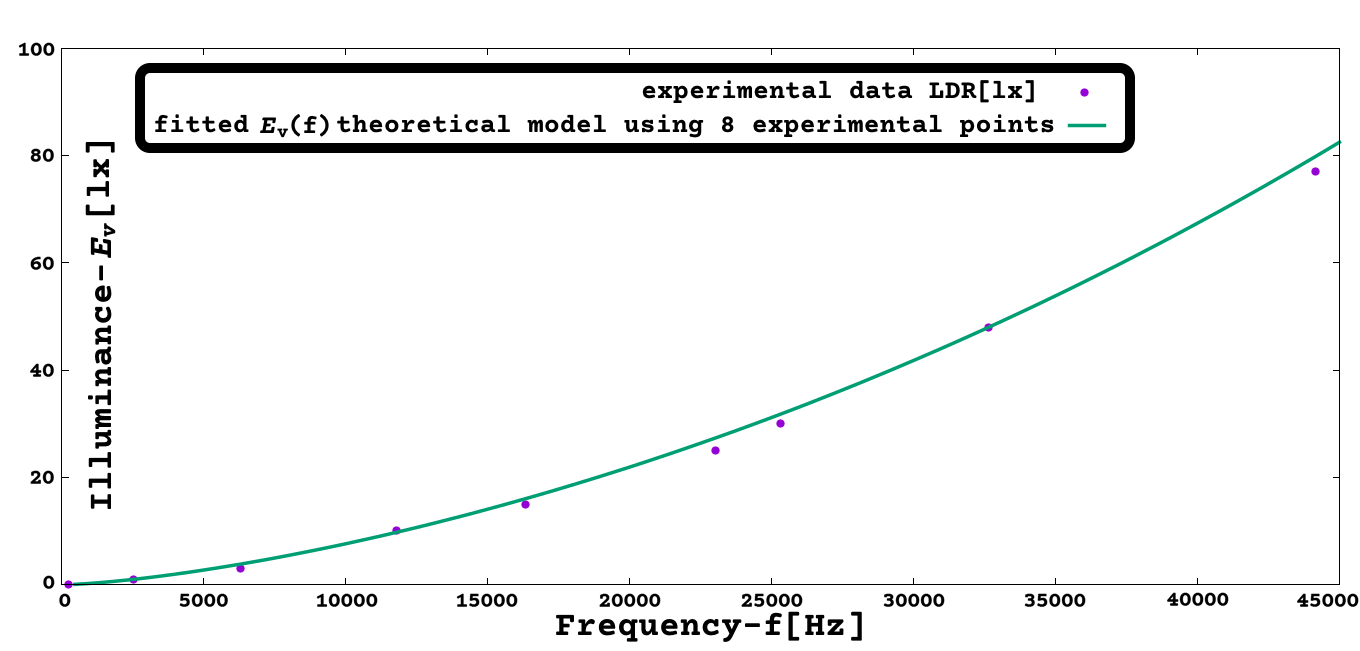}
	\caption{$E_{v\_LDR}(f)$ [lx], with $C_1$=2.2nF, $C_2$=2.2nF, frequency in [0Hz, 45kHz] (Sch-trig. osc., Multi-Sensor w. LDR sensor)}\label{FIG:LDRSensorEvVersusFrequencyRefMS6610FitTheoModel2}
	\vspace{1mm}
\end{figure}

The figures Fig. \ref{FIG:LDRSensorEvVersusFrequencyRefMS6610FitTheoModel1}, Fig. \ref{FIG:LDRSensorEvVersusFrequencyRefMS6610FitTheoModel2}
 are the result of fitting the model function $E_{v\_LDR}(f)$, to the points: (244Hz, 0.01 [lx]), (11788Hz, 10 [lx]), (25320Hz, 30 [lx]), (49768Hz, 100 [lx]), (155682Hz, 1200 [lx]), (232041Hz, 3950 [lx]), (281121Hz, 11000 [lx]), (320891Hz, 30300 [lx]); obtaining the fitted parameter values: \mbox{$\gamma$$=$0.705486}; \mbox{${R_{LDR\_0}}$$=$559200} [$\Omega$]; \mbox{$E_{v\_REF}$$=$0.0362818} [lx]; \mbox{$C_1$$=$4.90986${\cdot}$$10^{-9}$} [F]; \mbox{$C_2$$=$-2.174${\cdot}$$10^{-9}$} [F]; \mbox{$R_1$$=$2.4932${\cdot}$$10^{6}$} [$\Omega$]; \mbox{$R_2$$=$636.074} [$\Omega$]; \mbox{$H$$=$1.52273}.
The point at 244Hz was changed from 0[lx] to 0.01[lx] as it may facilitate/improve the model function fit.

\begin{figure}[H]
	\centering
		\includegraphics[width=\columnwidth,keepaspectratio]{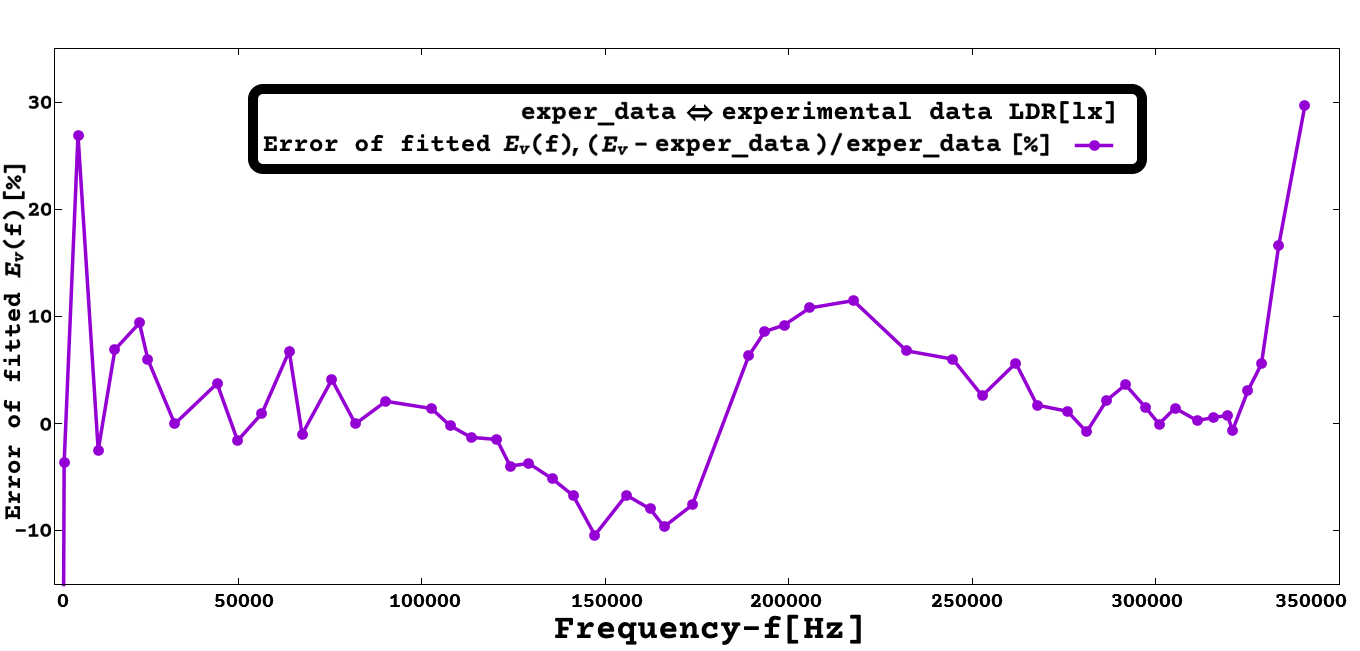}
	\caption{Error of fitted theoretical model $E_{v\_LDR}(f)$, with $C_1$=2.2nF, $C_2$=2.2nF}\label{FIG:ErrorOfFitTheoModelEvVersusFrequencyRefMS6610}
\end{figure}

On figure Fig. \ref{FIG:ErrorOfFitTheoModelEvVersusFrequencyRefMS6610} is shown the relative error in percentage [\%] of the fitted model function ${E_{v\_LDR}}(f)$ when compared with the full experimental dataset of the LDR sensor plus Multiple-Sensor interface; the relative error is calculated by subtracting the value of fitted theoretical function (${E_{v\_LDR}}(f)$) from the experimental value of ${E_{v\_LDR}}$, and then dividing by the experimental value of ${E_{v\_LDR}}$.\par
On Fig. \ref{FIG:ErrorOfFitTheoModelEvVersusFrequencyRefMS6610} is visible that for the point (244Hz, 0.01 [lx]) the relative error is very large (if calculated is about -200\%), that occurs because for that point where the Illuminance is 0 [lx] (or 0.01 [lx] as used for the model fitting) the calculation of the relative error on that experimental data point is dividing the absolute error by a number that is very close to zero, thus the relative error will be larger regardless if the absolute error has a reasonable/acceptable value for a function of ${E_{v\_LDR}}(f)$ that was fitted to both very small and very large values of Illuminance on a LDR sensor. If using the fitted model ${E_{v\_LDR}}(f)$, for example, to generate a calibration table of a LDR sensor with Multiple-Sensor interface, it would be more pratical to set the values of f<1000Hz as 0 [lx].
On Fig. \ref{FIG:ErrorOfFitTheoModelEvVersusFrequencyRefMS6610} is visible that for all the experimental dataset of LDR sensor the error of ${E_{v\_LDR}}(f)$ is always smaller than 30\% (except for ${E_{v}}$=0); and the error of ${E_{v\_LDR}}(f)$ is smaller that 10\% for the majority of the experimental dataset.
 
\section[\appendixname~\thesection]{Fluid (ex: water) Level Sensor Dataset}\label{SEC:FluidLevelSensorDataset}
\begin{figure}[b!]
	\centering
		\includegraphics[width=\columnwidth,keepaspectratio]{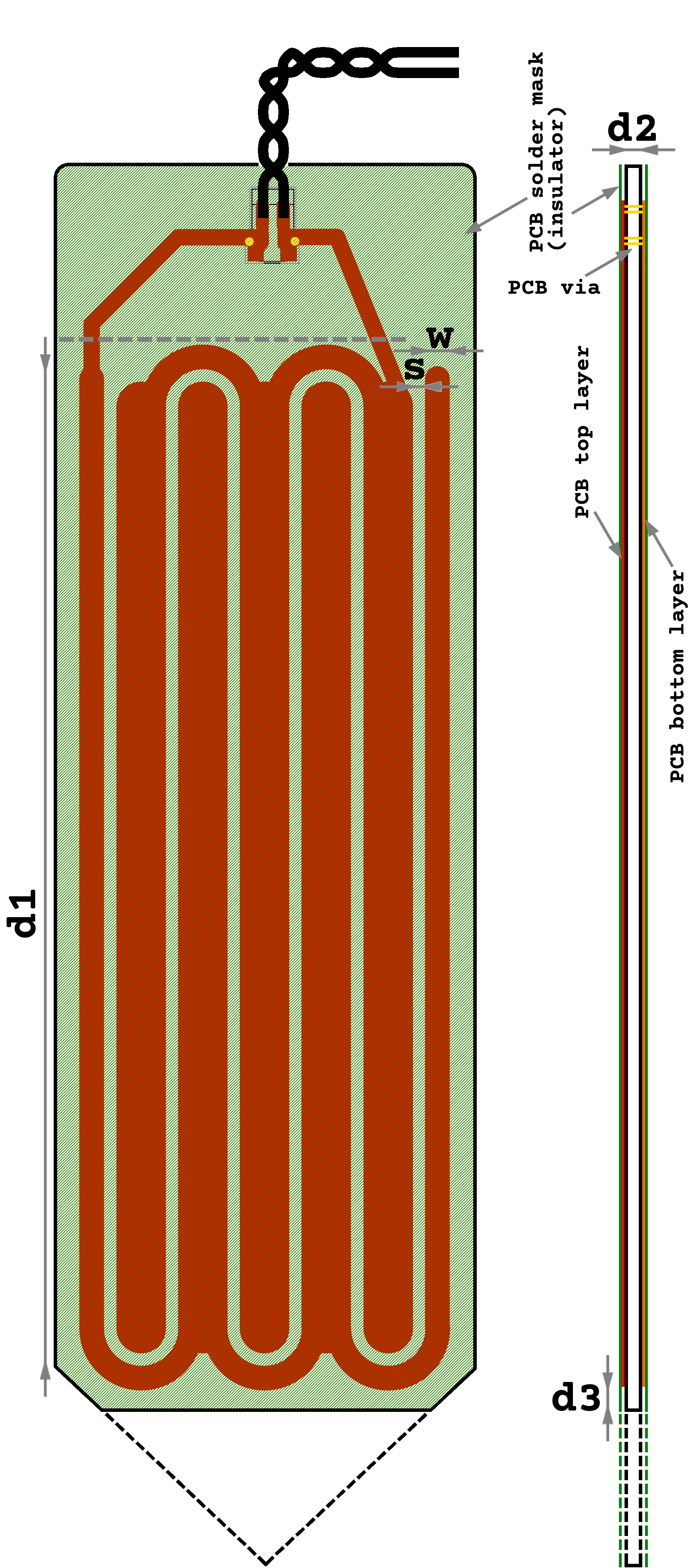}
	\caption{Diagram of 2-sided PCB (both layers have the same shape) of water (or fluid) level sensor or soil moisture sensor (with dashed line edge).}\label{FIG:FluidLevelAndSoilMoistureSensor2sidedPCB}
\end{figure}

\normalsize
It was made a custom water level sensor (that also may be used for other fluids) that is a planar capacitive sensor made as a double-sided PCB, the sensor may also be used as a soil moisture sensor. For making the sensor as a soil moisture sensor the only physical difference is the shape of the PCB at one of its edges that should be a pointy/triangle shape to make it easier to be buried in the ground, as visible on figure Fig.\ref{FIG:FluidLevelAndSoilMoistureSensor2sidedPCB} where the pointy/triangle shaped edge is represented by the dashed line.\par

For adjusting the capacitance of the sensor ($C_{sensor}$) to the region of $C_s$(f) of the device that provides a good steady change of frequency versus varying sensor capacitance, was added a 'bias' capacitor ($C_{bias}$) connected in parallel with the sensing PCB ($C_{sense}$) that measures fluid level / soil moisture.\par

 The fluid (ex: water) level / soil moisture sensor is a 2 layer PCB (top and bottom layer tracks have exactly the same shape, overlapping PCB tracks are the same circuit node, solder mask acts as insulator), with 6 parallel coplanar capacitive PCB tracks ($N_{t}$=6), with dimensions:  ${d_1}$=120mm, ${d_2}$=1.5mm, ${d_3}$=2mm, $s$=1.5mm, $w$=3mm; and electrical parameters: $C_{sense}$=45pF (capacitance with air on both sides), if using a bias capacitor of $C_{bias}$$\approx$1.8nF, then the sensor capacitance is $C_{sensor}$$\approx$1.845nF (as seen/measured by the interface device, sensor with air on both sides).\par
\textbf{- Reference instruments:} \hfill \phantom{'} \par
 For obtaining the calibration table of the fluid level sensor on the Multiple-Sensor device it was used as a reference a ruler scale laser-printed into transparent plastic film/sheets.\par
\textbf{- Jumper Configurations:}\par
 \qquad \textsuperscript{c}(JPA on, JPB on): $C_1$=2.2nF; $C_2$=2.2nF .\par
\textbf{- Units:} \hspace{3mm} Hz$=$hertz, m$=$meter.\par
Here is made available a calibration table for the fluid level sensor connected on the Multiple-Sensor Interface and tested with water (fluid), the table shows the water level ($h_{water}$) versus frequency(f).

\begin{table}[!htbp]
\caption{Set of exp. data for fluid level sensor (plus $C_{bias}$$\approx$1.8nF ), testing water level [m] vs. frequency [Hz]}
\label{TAB:tabWaterLevelvsFreq_WaterLevelSensor_c}
\setlength{\tabcolsep}{3pt}
\setlength{\extrarowheight}{0pt}
\begin{tabular}{|l|l|}
\hline
\thead{$h_{water}$[m]\\ref. ruler}& 
\thead{f[Hz]$^{\mathrm{c}}$ (sensor)\\ JPA on, JPB on}\\
\hline
0&269816\\
0.005&271244\\
0.010&274638\\
0.015&281397\\
0.020&287686\\
0.025&292650\\
0.030&296947\\
0.035&301044\\
0.040&305066\\
0.045&308980\\
0.050&312420\\
0.055&315799\\
0.060&319071\\
0.065&322282\\
\hline
\end{tabular}
\hspace{5mm}
\begin{tabular}{|l|l|}
\hline
\thead{$h_{water}$[m]\\ref. ruler}& 
\thead{f[Hz]$^{\mathrm{c}}$ (sensor)\\ JPA on, JPB on}\\
\hline
0.070&325310\\
0.075&328123\\
0.080&330875\\
0.085&333413\\
0.090&336043\\
0.095&338505\\
0.100&340722\\
0.105&342954\\
0.110&345049\\
0.115&347083\\
0.120&348703\\
0.125&350125\\
0.130&351058\\
 &  \\
\hline
\end{tabular}\newline
Precision error(maximum frequency variation):\newline
$\pm$400Hz ($h_{water}$$\geqslant$0.125m); $\pm$200Hz ($h_{water}$$<$0.11m) \newline
$\pm$300Hz (0.115m$\leqslant$$h_{water}$$<$0.125m) .
\end{table}

\section[\appendixname~\thesection]{Fluid (ex: water) Level Sensor Model Fitting and Error}\label{SEC:FluidLevelSensorCalibAndError}
\normalsize
The capacitance of the water level / moisture sensor shown on Fig. \ref{FIG:FluidLevelAndSoilMoistureSensor2sidedPCB} can be estimated by the formula of coplanar capacitance; where the copper tracks on the PCB  can be described as various straight line coplanar capacitors that are connected at the top and bottom by a curved track.\par
Also, the straight line coplanar capacitors in the water level / moisture sensor, was modelled as if there was no PCB substrate under the tracks since the overlapping tracks on both sides of the PCB are the same circuit node (they are connected by PCB via/hole), and so overlapping tracks have the same voltage (and so is assumed that between overlapping tracks the electric field is zero); also for the modelling objectives of the sensor made by coplanar capacitors was also assumed that any electric field inside the PCB substrate in the areas not located between overlapping tracks is not relevant compared to the electric field located/concentrated on the air/water/soil over the PCB that is between adjacent copper tracks. Thus the formula of coplanar capacitance \cite{CoplanarCapacitanceFormulaClaytonP} (as applied here to straight track pairs surrounded by air/water/soil) can be used (eq. \ref{EQ:CoplanarCapacitanceFormulaHomogeneousMedium1}), here will be used the approximation that the total track length (${d_s}$) is approximately the length of one of the parallel track segments (${d_1}$) (on the copper zone/area of the PCB) times the number of track segment pairs (${N_t}$) on the PCB (each track segment pair includes top and bottom layer that have the same voltage, and includes + and - terminals of the capacitor), so ${d_s} \approx {N_t}{d_1}$ is an approximation to the total length of the entire copper track from one extremity to the opposite extremity of the same capacitor terminal (+ or -); for example the capacitive sensor on figure Fig. \ref{FIG:FluidLevelAndSoilMoistureSensor2sidedPCB} has ${N_t}$=6.\par

\begin{multline}\label{EQ:CoplanarCapacitanceFormulaHomogeneousMedium1} {c_s}/{d_s} = \frac{ {\epsilon_{r}} \left( ln\left(\frac{2 \left(1 + \sqrt[4]{1 - (s/(s + 2 w))^2}\right)}{1 - \sqrt[4]{1 - (s/(s + 2 w))^2}}\right) \right)} {377 \pi {{\nu}_0}} \quad , \\ \phantom{....} \quad  0 \leqslant s/(s + 2 w) \leqslant  (1/{\sqrt{2}}) , \quad {{\nu}_0} = (1/(\sqrt{{\epsilon_0} {\mu_0}})) , \\ \text{Units of '377' constant is [$\Omega$].} \end{multline}\par

Modelling the fluid level sensor as 2 capacitors in parallel that have different dielectric medium (one capacitor uses air as dielectric the other uses fluid/water as dielectric), where $h_{fluid}$ is the height of the fluid that defines the dimensions of both capacitors, then using equation \ref{EQ:CoplanarCapacitanceFormulaHomogeneousMedium1}, is possible to write the following equation that relates $C_{sense}$ with $h_{fluid}$:

\begin{equation}
\label{EQ:CsenseVsHeightOfWaterFuncSensorModel}
    \resizebox{\hsize}{!}{$ {C_{sense}}({h_{fluid}}) \approx  \frac { ( {\epsilon_{r\_fluid}}{N_t}{h_{fluid}}  + ({\epsilon_{r\_air}}{N_t}({d_1}-{h_{fluid}}))) ln((2 (1 + \sqrt{k_2}))/(1 - \sqrt{k_2}))} {377 \pi {{\nu}_0}} $}
\end{equation}

The equation \ref{EQ:CsenseVsHeightOfWaterFuncSensorModel} can be combined with equation \ref{EQ:CSensorVersusFrequencyOscillationSchmittTrigger} to obtain the following equation, that relates fluid(ex: water) level versus frequency:

\begin{equation}
\label{EQ:HeightOfWaterVsFrequencyFuncSensorModel}
\begin{alignedat}{1}
   & \resizebox{\hsize}{!}{$ {h_{fluid}} (f) \approx   \frac {-377 \pi {{\nu}_0} {C_{sense}}(f) + ({N_t}{d_1}{\epsilon_{r\_air}} ln((2 (1 + \sqrt{k_2}))/(1 - \sqrt{k_2})))} {{N_t}({\epsilon_{r\_air}} - {\epsilon_{r\_fluid}}) ln((2 (1 + \sqrt{k_2}))/(1 - \sqrt{k_2})) } $} , \\  &{C_{sense}}(f) = {C_{sensor}}(f)  - {C_{bias}} , \\ &{C_{sensor}}(f) {\approx} \frac{({C_1}{R_1}+{C_2}{R_2})Hf-1-{C_1}{C_2}{R_1}{R_2}{H^2}{f^2}}{Hf((C_1+C_2){R_1}{R_2}Hf-R_1-R_2)} , \\ &{{\nu}_0} = (1/(\sqrt{{\epsilon_0} {\mu_0}})) , \\  &k_2 = \sqrt{1 - (s/(s + 2 w))^2}
\end{alignedat}
\end{equation}

\begin{figure}[b!]
	\centering
		\includegraphics[width=\columnwidth,keepaspectratio]{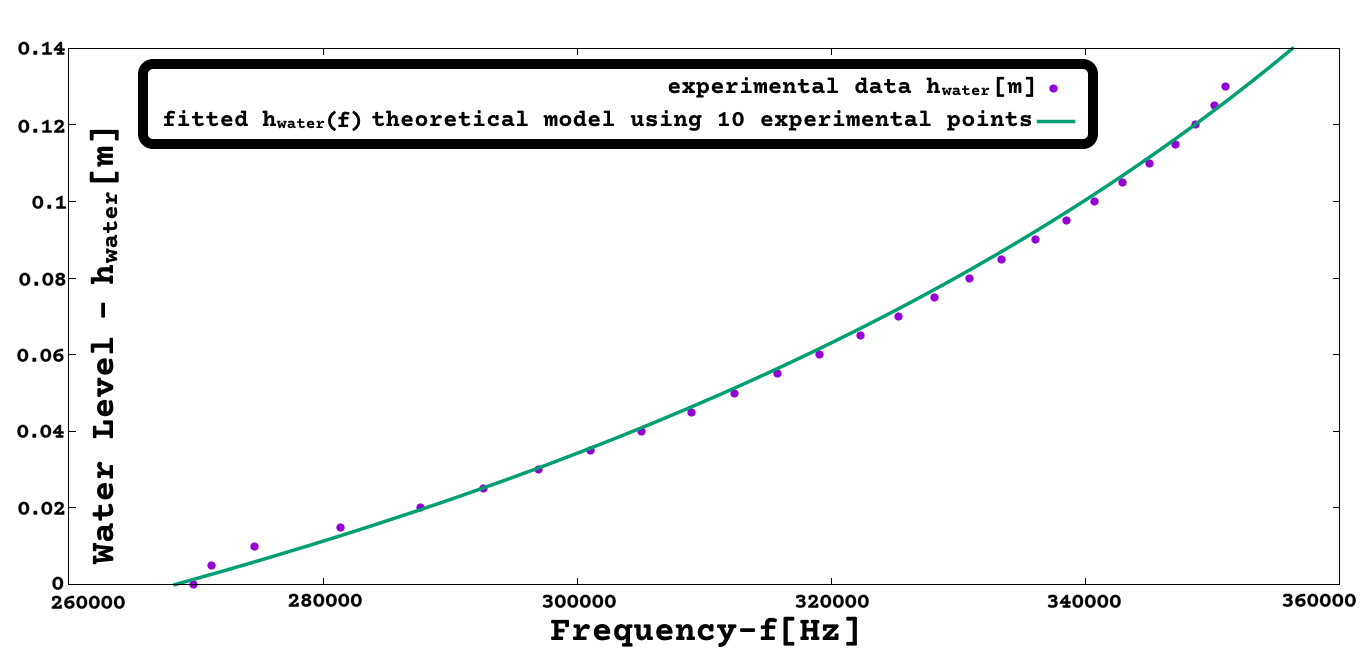}
	\caption{${h_{water}}(f)$ [m], with $C_1$=2.2nF, $C_2$=2.2nF, frequency in [260kHz, 360kHz] (Sch-trig. osc., Multi-Sensor w. water level sensor)}\label{FIG:WaterLevelSensorVsFrequencyRefRulerFitTheoModel1}

	\vspace{0.1mm}
	
	\centering
		\includegraphics[width=\columnwidth,keepaspectratio]{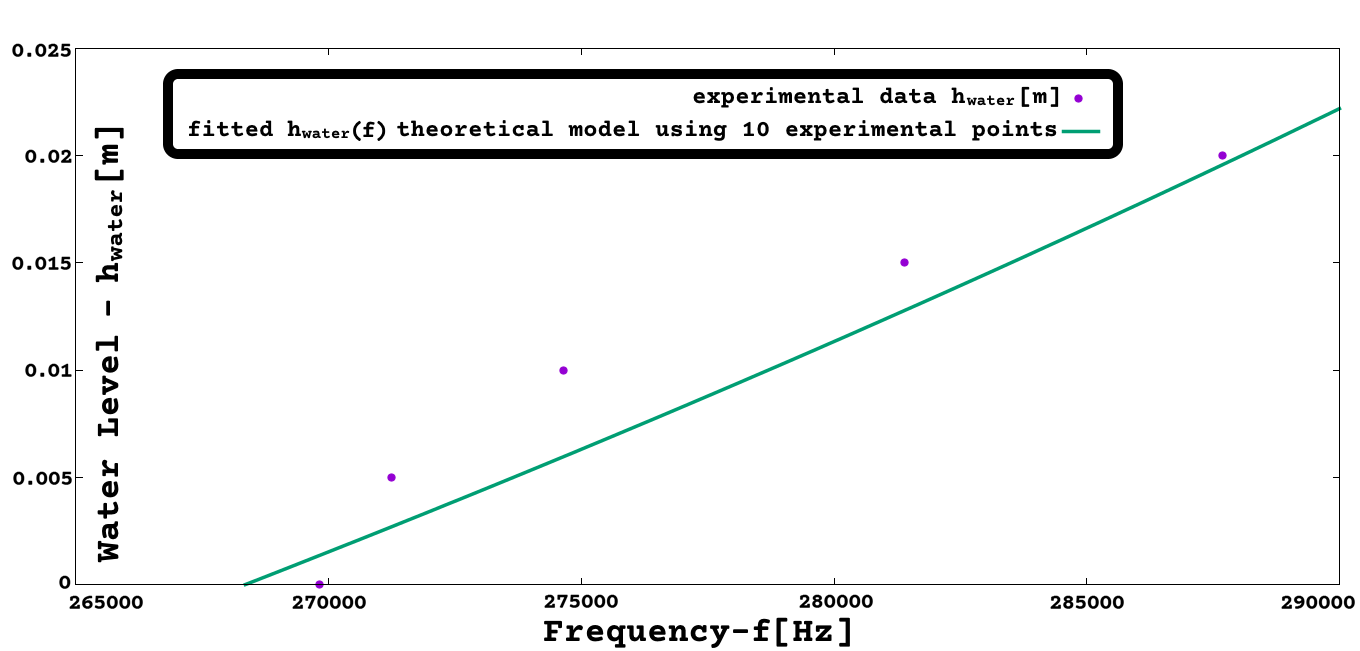}
	\caption{${h_{water}}(f)$ [m], with $C_1$=2.2nF, $C_2$=2.2nF, frequency in [265kHz, 290kHz] (Sch-trig. osc., Multi-Sensor w. water level sensor)}\label{FIG:WaterLevelSensorVsFrequencyRefRulerFitTheoModel2}
	\vspace{1mm}
\end{figure}

\begin{figure}[b!]
	\centering
		\includegraphics[width=\columnwidth,keepaspectratio]{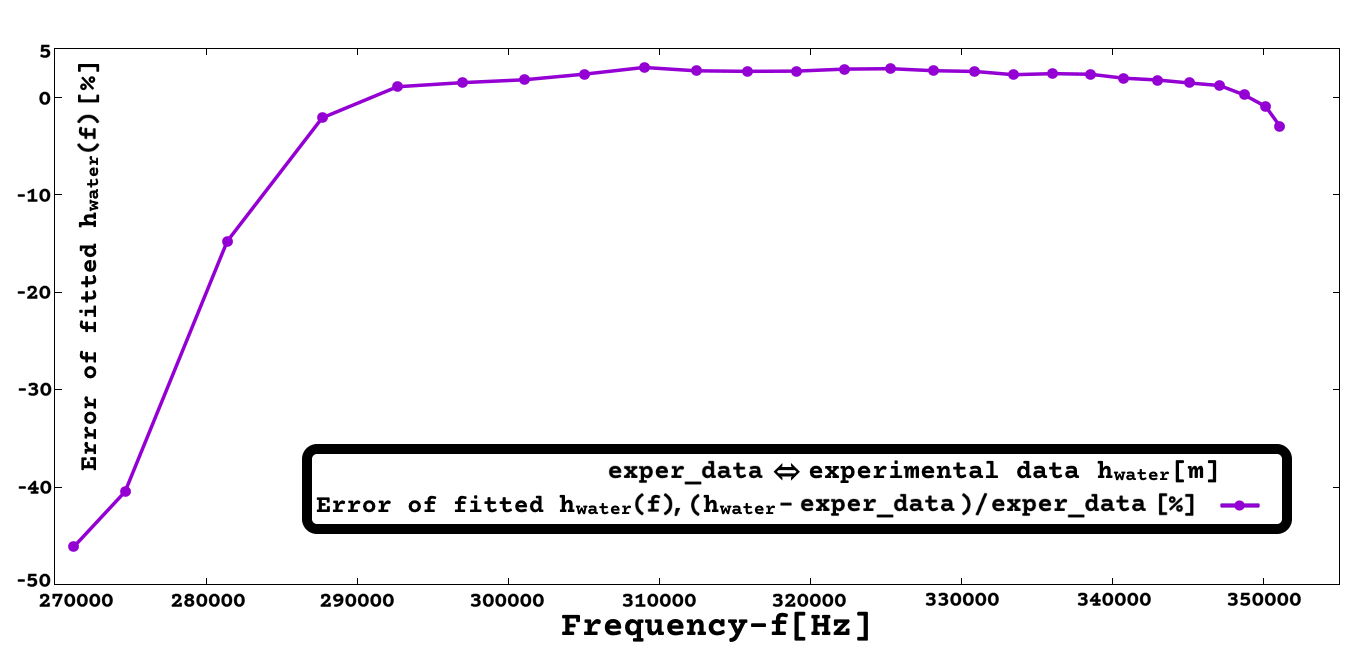}
	\caption{Error of fitted theoretical model ${h_{water}}(f)$, with $C_1$=2.2nF, $C_2$=2.2nF}\label{FIG:ErrorOfFitTheoModelWaterLevelVersusFrequencyRefRuler}
\end{figure}

The figures Fig. \ref{FIG:WaterLevelSensorVsFrequencyRefRulerFitTheoModel1}, Fig. \ref{FIG:WaterLevelSensorVsFrequencyRefRulerFitTheoModel2}
 are the result of fitting the model function \mbox{${h_{water}}(f)$ = ${h_{fluid}}(f,{\epsilon_{r\_fluid}}$=${\epsilon_{r\_water}})$}, to the points: (269816Hz, 0m), (271244Hz, 0.005m), (274638Hz, 0.01m), (287686Hz, 0.02m), (305066Hz, 0.04m), (319071Hz, 0.06m), (330875Hz, 0.08m), (340722Hz, 0.1m), (348703Hz, 0.12m), (351058Hz, 0.13m),  and constant (not fitted) parameters $d_{1}$=0.12 m, ${N_t}$=6 ;  obtaining the fitted parameter values: \mbox{$s$=0.001284 m}; \mbox{$w$=0.003457 m}; \mbox{${\epsilon_{r\_air}}$=-1.00307}; \mbox{${\epsilon_{r\_water}}$=-125.039}; \mbox{$H$=0.946237}; \mbox{$C_{bias}$=-3.76339${\cdot}$$10^{-9}$ F}; \mbox{$C_{1}$=2.32215${\cdot}$$10^{-9}$ F}; \mbox{$C_{2}$=2.45097${\cdot}$$10^{-9}$ F}; \mbox{$R_1$=1.99478${\cdot}$$10^{6}$ $\Omega$}; \mbox{$R_2$=465.095 $\Omega$}.\par

The fitted values of ${\epsilon_{r\_air}}$, ${\epsilon_{r\_water}}$, $C_{bias}$ are negative not because of some model fitting procedure, but were already expected to be negative in accordance with the theoretical analysis where is stated that the capacitive sensor $C_{sensor}$ (or  $C_{s}$) behaves as negative capacitance when the interface device is operating as Schmitt-Trigger oscillator.

On figure Fig. \ref{FIG:ErrorOfFitTheoModelWaterLevelVersusFrequencyRefRuler} is shown the relative error in percentage [\%] of the fitted model function ${h_{water}}(f)$ when compared with the full experimental dataset of the water level sensor connected on the Multiple-Sensor interface; the relative error is calculated by subtracting the value of fitted theoretical function (${h_{water}}(f)$) from the experimental value of ${h_{water}}$, and then dividing by the experimental value of ${h_{water}}$. Also is visible that for all the experimental dataset of the water level sensor the error of ${h_{water}}(f)$ is always smaller than 50\%; and the error of ${h_{water}}(f)$ is smaller that 5\% for the majority of the experimental dataset.\par

\section[\appendixname~\thesection]{Abbreviations}\label{SEC:AppxAbbreviations}
\small
\begin{tabular}{>{\hspace{-2.5mm}}p{22mm}<{\dotfill}@{}p{60mm}}
ADC & Analog to Digital Converter\\
CC BY-NC-SA & Creative Commons, Attribution - NonCommercial - ShareAlike\\
CERN & Conseil Europeen pour la Recherche Nucleaire\\
CERN-OHL-W & CERN Open Hardware Licence\newline - \mbox{Weakly reciprocal}\\
CMOS & Complementary Metal Oxide Semiconductor\\
EEPROM & Electrically Erasable Programmable Read-Only Memory\\
ESD-safe foam & Electrostatic Sensitive Device safe foam\\
FSR & Force Sensitive Resistor (sensor)\\
GND & Ground (voltage reference)\\
GPIO & General-Purpose Input/Output\\
LDR & Light Dependent Resistor (sensor)\\
$ln$(x) & Natural Logarithm, ${log_{\emph{e}}}$(x). \quad $ln({\emph{e}}^{x})=x$\\
OCR & Optical Character Recognition\\
PCB & Printed Circuit Board\\
PWM & Pulse-Width Modulation\\
QR-code & Quick Response code\\
RS-485 & Recommended Standard 485\newline \mbox{(aka. EIA/TIA-485)}\\
RTD & Resistance Temperature Detector (sensor)\\
UART & Universal Asynchronous Receiver-Transmitter\\
USB & Universal Serial Bus\\
VDD & Voltage Supply (Voltage Drain Drain)\\
VDDS & VDD Stabilized
\end{tabular}

\normalsize

\end{appendices}

\end{document}